\documentclass[12pt]{iopart}
\usepackage{graphicx}
\usepackage{rotating}

\begin{document}
\title[Spin-observables in pseudoscalar meson photoproduction]{On the spin-observables in pseudoscalar meson photoproduction}

\author{K. Nakayama}

\address{Department of Physics and Astronomy, University of Georgia,
Athens, GA 30602, USA}
\ead{nakayama@uga.edu}

\begin{abstract}
Spin-observables in pseudoscalar meson photoproduction is discussed. This work is complementary to the earlier works on this topic. Here, the reaction amplitude is expressed in Pauli-spin basis which allows to
calculate all the observables straightforwardly. We make use of the fact that the underlying reflection symmetry about the reaction plane can be most conveniently and easily exploited in this basis to help finding the non-vanishing and independent observables in this reaction.
The important issue of complete experiments is reviewed. By expressing the reaction amplitude in Pauli-spin basis, many sets of eight observables - distinct from those found in earlier works from the amplitude in transversity basis - capable of determining the reaction amplitude up to an overall phase are found.    
It is also shown that some of the combinations of the spin observables are suited for studying certain aspects of the reaction dynamics. We, then, carry out a (strictly) model-independent partial-wave analysis, in particular, of the peculiar angular behavior of the beam asymmetry observed in $\eta'$ photoproduction very close to threshold [P. Levi Sandri \textit{et al.} 2015 Eur. Phys. J. A 51, 77].
This work should be useful, especially, for newcomers in the field of baryon spectroscopy, where the photoproduction reactions are a major tool for probing the baryon spectra.
\end{abstract}


\maketitle

\normalsize


\section{Introduction}

The model-independent aspects of pseudoscalar meson photoproduction reactions have been discussed by a number of authors in the past \cite{
BDS75,Mo84,FTS92,KW96,ChT97,ARS07,SHKL11}. For example, Fasano, Tabakin and Saghai \cite{FTS92} have given a detailed account on the general structure of the single- and double-spin observables in this reaction, with emphasis on the low energy behavior of these observables. 

Chiang and Tabakin \cite{ChT97}, on the other hand, have shown that, for a given total energy of the system and meson production angle,  one requires a minimum of eight carefully chosen independent observables -- the so-called \textit{complete experiment} -- to determine the full reaction amplitude up to an arbitrary overall phase. 
This problem has been revisited quite recently in Ref.~\cite{Nak18}, where an explicit derivation of the completeness condition has been provided, covering all possible cases. In particular, 
it is shown that the completeness condition of a set of four observables to resolve the phase ambiguity of the photoproduction amplitude in transversity representation holds only when the magnitudes of the relative phases are different from each other. In situations where the equality of the magnitudes occurs, it has been shown that additional one or two or even three chosen observables are required to resolve the phase ambiguity, depending on the particular set of four observables considered. This increases the minimum number of observables required for a complete experiment. 
A way of gauging  when the equal-magnitudes situation occurs has been also provided in Ref.~\cite{Nak18}.

In Ref.~\cite{ChT97}, also the Fierz transformations of the gamma matrices have been used to develop useful linear and
nonlinear relationships between the spin observables. These relations not only help finding the complete sets of experiments, but also yield important constraints among the 16 non-redundant observables for this reaction. 
Also, Moravcsik \cite{Mo84} has discussed  the quadratic constraints among bilinear products of reaction amplitudes in terms of the experimental observables for a reaction with arbitrary spins.
Artru, Richard and Soffer \cite{ARS07} have derived various positivity constraints among the spin observables which is very useful in determining the allowed domain of physical observables, in addition to testing the consistency of various available measurements and also the validity of some dynamical assumptions in theoretical models. 

Sandorf \textit{et al.} \cite{SHKL11} have given a thorough consideration of the spin observables in pseudoscalar photoproduction  reaction, especially, related to the issues of practical purposes. Among other things, they have concluded that, while a
mathematical solution to the problem of determining an amplitude free of ambiguities
may require eight observables \cite{ChT97}, experiments with realistically achievable uncertainties will require a significantly larger number of observables. Further detailed studies along this line have been carried out by the Gent group \cite{VRCV13,NVR15,NRIG16}, where a statistically more sound analysis is performed to constrain the photoproduction amplitude \cite{NRIG16}.
Recently, with the advances in experimental techniques, many spin-observables in photoproduction reactions became possible to be measured and this has attracted much renewed interest in constraints on partial-wave analysis in the context of complete experiments \cite{SHKL11,Workman11,WPBTSOK11,DMIM11,WBT14,WTWDH17,TWDH17,WSWTB17,SWOHOSKNOTW18,FGRGP18,WU19} (for some of the related earlier works, see, e.g., Refs.~\cite{Ome81,GSLR83,Gru89} ).  
Of particular interest in this connection is the issue of whether the baryon resonances can be extracted model independently or with minimal model requirements. Efforts along this line is currently in progress  \cite{WBT14,WTWDH17,SWOHOSKNOTW18,FGRGP18}.

In this work, we consider those same issues mentioned above; as such, it is complementary to the earlier works on this topic \cite{
BDS75,FTS92,KW96,ChT97,ARS07,SHKL11}. In contrast to those studies, we consider the photoproduction amplitude in Pauli-spin basis which allows us to calculate the spin-observables, in general, in a straightforward and quite pedestrian way.  We will make used of the underlying reflection symmetry about the reaction plane - which can be conveniently and easily exploited in this basis - to help find the non-vanishing observables in this reaction.  Also, the Pauli-spin  amplitude is simply related to that of the usual Chew-Goldberger-Low-Nambu (CGLN) form \cite{CGLN57} due to this symmetry. 
The important  and non-trivial issue of the complete experiments is reviewed. In particular, by using Pauli-spin representation of the reaction amplitude instead of transversity representation, we identify other sets of observables, not reported in \cite{ChT97,Nak18}, for complete experiments.
In addition, certain combinations of the observables are shown to be useful in isolating (low) partial-wave states to learn about the reaction dynamics. Motivated by this, a model-independent analysis of the 
$\eta'$ photoproduction off the proton close to threshold is performed in connection to the
peculiar angular behavior exhibited by the beam asymmetry in this reaction \cite{GRAAL15}. 
This problem of the beam asymmetry close to threshold has been addressed recently by the Bonn-Gatchina \cite{BnGa18} and Mainz \cite{TGKNOHOOSS18} groups within their respective model calculations. The former group  \cite{BnGa18} describes the measured angular and energy dependences through the $P_{13}-D_{13}$ interference with a narrow $D_{13}(1900)$ resonance, while the latter group \cite{TGKNOHOOSS18}  describes through the $S_{11}-F_{15}$ interference with a narrow $S_{11}(1900)$ resonance. These findings, although still require to be confirmed, suggest the existence of narrow resonances in the 2 GeV energy region.   
Apart from the model-independent analysis of the $\eta'$ photoproduction mentioned above, we also show that, at energies close to threshold, even for cases where the number of partial-waves are restricted to $l \le 2$, 
the unpolarized cross section and single-polarization observables alone are not sufficient to constrain the reaction dynamics model independently. For this, double-polarization observables are required.

This note is organized as follows. In Sec.~\ref{sec:GeneralForm}, we give the general form of the pseudoscalar meson photoproduction amplitude. It is a variant of the well-known CGLN form of the amplitude \cite{CGLN57} and has a simple relation to the amplitude in Pauli-spin basis. In Sec.~\ref{sec:GSX} the definitions of various spin asymmetries are given.   
In Sec.~\ref{sec:BasicSpinObserv}, the basic formulas for the spin observables in general are derived in terms of the Pauli-spin amplitudes. Using these basic formulas, the spin observables for linearly and circularly polarized photons are derived in Secs.~\ref{sec:SpinObservLinear} and \ref{sec:SpinObservCircular}, respectively. Section~\ref{sec:Dictionary} gives a dictionary for different notations used in the literature for double-spin observables and Sec.~\ref{sec:NonRedundantObserv} summarizes the non-redundant observables.
In Sec.~\ref{sec:ComExp}, the issue of the complete experiments is reviewed.
 In Sec.~\ref{sec:SpinObserv-F} the observables are expressed in terms of the amplitudes \`a la CGLN given in Sec.~\ref{sec:GeneralForm}. Observables in the rotated frame are given in Sec.~\ref{sec:SpinObservRot}. A partial-wave analysis, especially, of the $\eta'$ photoproduction is given in Sec.~\ref{sec:PW}. Finally, Sec.~\ref{sec:Summary} contains the summary. Details of the analysis on complete experiments using the Pauli-spin representation of the reaction amplitude are given in Appendices A and B.   
In \ref{app:3}, the beam-target asymmetry E is shown to be a measure of the helicity asymmetry.

\section{General form of the pseudoscalar meson photoproduction amplitudes} \label{sec:GeneralForm}

The most general form of the pseudoscalar meson photoproduction amplitude can be written as \cite{CGLN57} 
\begin{equation}
 \hat{M}  = i f_1 \vec\sigma\cdot\vec\epsilon 
    +   f_2 \vec\sigma\cdot\hat q \vec\sigma\cdot (\hat k \times \vec \epsilon)
    +  i f_3 \vec\sigma\cdot\hat k \hat q\cdot\vec\epsilon
    +  i f_4 \vec\sigma\cdot\hat q \hat q\cdot\vec\epsilon 
\label{Jampl-S}
\end{equation}
where $\vec q$ and $\vec k$ denote the meson and photon momentum, respectively. 
$\vec{\sigma} = (\sigma_x, \sigma_y, \sigma_z)$ stands for the usual Pauli-spin matrix. 
Throughout this work, the notation $\hat{a}$ indicates the unit vector, i.e., $\hat{a} \equiv \vec{a}/ |\vec{a}|$ for an arbitrary vector $\vec{a}$. $f_i (i=1-4)$ is a function of the total energy 
of the system, $W \equiv \sqrt{s}$, and the meson production angle $\theta$ given by $x \equiv \cos\theta = \hat q\cdot\hat k$. $\vec\epsilon$ denotes the (real) photon 
polarization vector.

\vskip 0.5cm
For further convenience, following Ref.~\cite{NL05},  we re-write Eq.~(\ref{Jampl-S}) as
\begin{equation}
\hat M = F_1\, \vec\sigma\cdot\vec\epsilon
    + i F_2\, \vec\epsilon \cdot (\hat k \times \hat q)
    +   F_3\, \vec\sigma\cdot\hat k \hat q\cdot\vec\epsilon 
    +   F_4\, \vec\sigma\cdot\hat q \hat q\cdot\vec\epsilon 
\label{Jampl-NL}
\end{equation}
where, apart from an irrelevant overall constant phase,
\begin{equation}
F_1 \equiv f_1 - x f_2 \ , \ \ \ \ \ F_3   \equiv   f_2 + f_3 \ , \ \ \ \ 
F_i \equiv f_i \  (i=2,4)\ .  
\label{FJ}
\end{equation}

Note from the above equation that only the coefficient
$F_1$ contains the final $S$ partial-wave state. Also, the coefficient $F_4$ contains no partial waves lower than the $D$-wave in the final state.

\vskip 0.5cm
For (isovector) pion photoproduction the full reaction amplitude, $\hat{\cal M}$, contains also the isospin structure. Its general form can be obtained by forming a scalar out of the available operators  in isospin space for this reaction, viz., 
\begin{eqnarray}
\hat{{\cal M}} & = \hat{\pi} \cdot \vec{\tau} (\hat{M}_a + \hat{M}_b \tau_3) +  
(\hat{M}_{a'} + \hat{M}_{b'} \tau_3) \hat{\pi} \cdot \vec{\tau} \nonumber \\
& =  \hat{M}_A\, \hat{\pi} \cdot \vec{\tau} + \frac12 \hat{M}_B\, \hat{\pi} \cdot [\vec{\tau}, \tau_0]  + \hat{M}_C\, \hat{\pi}_0 \ ,
\label{Isospin}
\end{eqnarray}
where $\hat{\pi}_j\ (j=0, \pm1)$ stands for the produced pion state in isospin space. $\vec{\tau} = (\tau_{-1}, \tau_0, \tau_{+1})$ denotes the usual Pauli matrix in isospin space. 
$\hat{M}_A \equiv \hat{M}_a + \hat{M}_{a'}$, $\hat{M}_B \equiv \hat{M}_b - \hat{M}_{b'}$ and  $\hat{M}_C \equiv \hat{M}_b + \hat{M}_{b'}$. 
Each coefficient $\hat{M}_X\ (X=A,B,C)$ in the above equation is an independent amplitude $\hat{M}$ in spin space given by Eq.~(\ref{Jampl-NL}).

\vskip 0.5cm
It is well-known that the coefficients $f_i$ in Eq.~(\ref{Jampl-S}), or $F_i$ in Eq.~(\ref{Jampl-NL}), can be expanded in partial waves \cite{CGLN57,NL05,BDW67} which allows for partial-wave analyses. In photon-induced reactions, it is customary to
use the multipole decomposition instead of partial-wave decomposition. In electroproduction, which of the two decomposions to use is just a matter of choice. In photoproduction, however, the multipole decomposion is preferred over the partial-wave decomposition, since in the former decomposition, the scalar (or longitudinal) multipoles 
vanish identically due to the transversality of the real photon \cite{BDW67}, while in the latter decomposition, only a certain combinations of the partial-wave matrix elements are probed. 
 Explicitly, the multipole decomposition of the photoproduction amplitude in Eq.~(\ref{Jampl-NL}) is given by \cite{BDW67}
\begin{eqnarray}
\fl iF_1 = N \sum_{l=0}^\infty \Big\{ P'_{l+1}\, E_{l+} + P'_{l-1}\, E_{l-} + \big[ l P'_{l+1} - x (l+1) P'_l \big] M_{l+} 
+ \big[ (l+1) P'_{l-1} - x l P'_l \big] M_{l-} \Big\} \ , \nonumber \\
\fl iF_2  =  N \sum_{l=1}^\infty \Big\{ (l+1) P'_l\, M_{l+} +  l P'_l\, M_{l-} \Big\} \ , \nonumber \\
\fl iF_3  =  N \sum_{l=1}^\infty \Big\{ P''_{l+1}\, E_{l+} + P''_{l-1}\, E_{l-} + \big[ (l+1) P'_l - P''_{l+1} \big] M_{l+} 
+ \big[ P''_{l-1} + l P'_l \big] M_{l-} \Big\} \ , \nonumber \\
\fl iF_4  =  N \sum_{l=2}^\infty \Big\{ - P''_l\, E_{l+} - P''_l\, E_{l-} + P''_l\, M_{l+} 
- P''_l \, M_{l-} \Big\} \ , \nonumber \\
\label{Mult-ampl}
\end{eqnarray}
with $P^{'}_l\equiv P^{'}_l(x)$ and $P^{''}_l\equiv P^{''}_l(x)$ denoting, respectively, 
the derivative and the double-derivative of the Legendre Polynomial of first kind, 
$P_l \equiv P_l(x)$, with respect to $x$. $N \equiv  4\pi W/ m_N$, where $m_N$ denotes the nucleon mass.

The above equation  may be inverted to yield
\begin{eqnarray}
\fl E_{l+}   =  N^+_l \int^{+1}_{-1} dx \left[  P_l\, F_1  + \big[x P_l + \frac{1}{2l+1} \left(l P_{l+1} - (l+1) P_{l-1} \right) \big] F_2 \right. \nonumber \\
 \left.  \hspace{4cm} - \frac{l}{2l+1} \big(P_{l+1} -P_{l-1} \big) F_3 +  \frac{l+1}{2l+3}\big(P_l - P_{l+2} \big) F_4 \right] \ , \nonumber \\
\fl E_{l-}   =  N^-_l  \int^{+1}_{-1} dx  \left[ P_l\, F_1 + \big[x P_l - \frac{1}{2l+1} \left((l+1) P_{l+1} - l P_{l-1} \right) \big] F_2  \right. \nonumber \\
\left. \hspace{4cm} + \frac{l+1}{2l+1}\big(P_{l+1} -P_{l-1} \big) F_3 + \frac{l}{2l-1}  \big(P_l - P_{l-2} \big) F_4 \right] \ , \nonumber \\
\fl M_{l+}  =  N^+_l \int^{+1}_{-1} dx \left[ P_l\, F_1 + \big[x P_l - \frac{1}{2l+1} \left( 2(l+1) P_{l+1} - P_{l-1} \right) \big] F_2 \right. \nonumber \\
\left. \hspace{4cm} +  \frac{1}{2l+1}  \big(P_{l+1} -P_{l-1} \big) F_3 \right] \ ,  \nonumber \\
\fl M_{l-}  =  - N^-_l \int^{+1}_{-1} dx  \left[P_l\, F_1 + \big[x P_l - \frac{1}{2l+1} \left( P_{l+1} + 2 l P_{l-1} \right) \big] F_2  \right.  \nonumber \\
\left. \hspace{4cm} +  \frac{1}{2l+1} \big(P_{l+1}-P_{l-1}\big) F_3 \right] \ , \nonumber \\
\label{eq:Mult-ampl}
\end{eqnarray}
with $N^+_l \equiv i/[2(l+1)N]$ and $N^-_l \equiv i/[2l N]$.

For the relationship between the multipole and helicity amplitudes, see Ref.~\cite{BDW67}. This reference treats both the pion photo- and electro-production within a dispersion relation theory. For early description of the pion photo- and electro-production, see, e.g., Refs.~\cite{GG93,BKLM94}.     
For a more recent review of meson electroproduction, see \cite{BL04,A13}.

\section{General cross section} \label{sec:GSX}

The general spin-dependent cross section can be expressed in terms of the cross sections with specified spin polarization states of the particles in the initial and final states in a given reaction process. These cross sections are referred to as the spin observables. For pseudoscalar meson photoproduction off the nucleon, described by the amplitude given by Eq.~(\ref{Jampl-NL}), the associated general cross section can be written as
\begin{equation}
\frac{d\bar{\sigma}}{d\Omega} = \Tr\left[ \hat{M} \left(\frac{1 + \vec{{\cal P}}^\gamma \cdot \vec{\sigma}^\gamma}{2} \right)   
\left(\frac{1 + \vec{{\cal P}}^N \cdot \vec{\sigma}}{2} \right) \hat{M}^\dagger \left(\frac{1 + \vec{\cal{P}}^{N'} \cdot \vec{\sigma}}{2} \right) \right] \ ,
\label{eq:GSX}
\end{equation}
The superscript $\gamma$ on the Pauli spin matrix, $\vec{\sigma}^\gamma$, indicates that this operator acts on the real photon helicity space.
 $\vec{{\cal P}}^a$ stands for the polarization vector which specifies the direction and the degree of (spin) polarization of the ensemble of particles $a$, as $a =$ photon($\gamma$), target nucleon($N$) or recoil nucleon($N'$). It is given by
 \begin{equation}
 \vec{{\cal P}}^a \equiv \sum_{i=1}^3 {\cal P}^a_i \hat{n}_i \ , 
 \label{eq:Pol}
 \end{equation}
where $\hat{n}_i$ stands for the unit vector specifying the direction of the polarization. Details on the  polarization vector for a real photon, $\vec{{\cal P}}^\gamma$ (called Stokes vector), may be found in Ref.~\cite{FTS92}. 

Equation (\ref{eq:GSX}) can be re-written as
\begin{eqnarray}
\fl \frac{d\bar{\sigma}}{d\Omega} = \frac12 \frac{d\sigma}{d\Omega} \Bigg\{ 1  
+ \sum_{i=1}^3 \bigg[ {\cal P}^\gamma_i\, B^{p_i} + {\cal P}^N_i\, T_i + {\cal P}^{N'}_i R_i \bigg] \nonumber \\
+ \sum_{i,j=1}^3 \bigg[ {\cal P}^N_i {\cal P}^{N'}_j K_{ij}
+ {\cal P}^\gamma_i {\cal P}^{N}_j\, T^{p_i}_j + {\cal P}^\gamma_i {\cal P}^{N'}_j R^{p_i}_j \bigg] 
 + \sum_{i,j,k=1}^3  {\cal P}^\gamma_i {\cal P}^N_j {\cal P}^{N'}_k K^{p_i}_{jk}
 \Bigg\} \ , \nonumber \\
\label{eq:GSX-1}
\end{eqnarray}
with 
\footnote{Here, we refer to  $B^{p_i}$, $T_i$, $R_j$, etc. as the spin-asymmetries. Strictly speaking, they are called the analyzing powers \cite{Ohlsen72}. Asymmetries are the quantities ${\cal P}^\gamma_i B^{p_i}$, ${\cal P}^N_i T_i$, ${\cal P}^{N'}_i R_j$, etc.}
\numparts
\begin{eqnarray}
 \frac{d\sigma}{d\Omega} &\equiv \frac{1}{4} \Tr\left[ \hat{M} \hat{M}^\dagger \right] \ ,   \qquad& {\rm (unpolarized\ cross\ section)} \\ 
 \frac{d\sigma}{d\Omega} B^{p_i} &\equiv \frac{1}{4} \Tr\left[ \hat{M} \sigma^\gamma_i\,  \hat{M}^\dagger \right] \ , \qquad& {\rm (beam \ asymmetry)} \\ 
 \frac{d\sigma}{d\Omega}T_i &\equiv \frac{1}{4} \Tr\left[ \hat{M} \sigma_i\, \hat{M}^\dagger \right] \ , \qquad& {\rm (target \ asymmetry)} \\ 
 \frac{d\sigma}{d\Omega} R_i &\equiv \frac{1}{4} \Tr\left[ \hat{M} \hat{M}^\dagger \sigma_i \right] \ , \qquad& {\rm (recoil\ asymmetry)} \\ 
 \frac{d\sigma}{d\Omega} K_{ij} & \equiv \frac{1}{4} \Tr\left[ \hat{M} \sigma_i \hat{M}^\dagger \sigma_j \right]  \ , \qquad&  {\rm (target-recoil\ asymmetry)}   \\
 \frac{d\sigma}{d\Omega}T^{p_i}_j &\equiv \frac{1}{4} \Tr\left[ \hat{M} \sigma^\gamma_i \sigma_j\, \hat{M}^\dagger \right] \ , \qquad& {\rm (beam-target \ asymmetry)} \\ 
 \frac{d\sigma}{d\Omega} R^{p_i}_j &\equiv \frac{1}{4} \Tr\left[ \hat{M} \sigma^\gamma_i\hat{M}^\dagger \sigma_j \right] \ , \qquad& {\rm (beam-recoil\ asymmetry)} \\ 
 \frac{d\sigma}{d\Omega} K^{p_i}_{jk} & \equiv \frac{1}{4} \Tr\left[ \hat{M} \sigma^\gamma_i \sigma_j \hat{M}^\dagger \sigma_k \right]  \ . \qquad&  {\rm (beam-target-recoil\ asymmetry)} \nonumber \\
\end{eqnarray}
\label{eq:Aa3}
\endnumparts
In the above equations, the superscript $p_i$ stands for the linear($l$) or circular($c$) polarization of the photon as we shall specify in Secs.~\ref{sec:SpinObservLinear} and \ref{sec:SpinObservCircular}.

\section{Basic observables in photoproduction} \label{sec:BasicSpinObserv}

Observables in reaction processes may be written in terms of the corresponding amplitudes expressed in different representations (or basis). Which representation to use depends on the particular physics aspect(s) one is interested in to investigate.  
In photoproduction reactions, the most commonly used representations are in terms of the coefficients $F_i$ \`a la CGLN, as given by Eq.~(\ref{Jampl-NL}),  
and the helicity representation \cite{BDW67}. Also, the transversity representation is especially convenient in connection to the problem of complete experiments \cite{BDS75,ChT97,Nak18}. 
Of course, the multipole expansion is used for partial-wave analyses.
In this work, we express the photoproduction amplitude in the Pauli-spin representation.  As we shall see in the following, this allows us to calculate the spin-observables in general in a straightforward and quite pedestrian way.  We will make used of the underlying reflection symmetry about the reaction plane - which can be most conveniently and easily exploited in this representation - to help find the non-vanishing observables in this reaction.  Also, the Pauli-spin  amplitude is simply related to that of the usual CGLN form \cite{CGLN57} due to this symmetry as given in the next section.

We first define a set of three mutually orthogonal unit vectors $\{\hat{x}, \hat{y}, \hat{z}\}$ in terms of the available momenta in the problem to be used as the reference frame. Namely,
\begin{equation}
\hat{z} \equiv \hat{k} \ , \ \ \ \ \ \ \ 
\hat{y} \equiv \frac{\hat{k} \times \hat{q}}{\hat{k} \times \hat{q}} \ , \ \ \ \ \ \ \ \ 
\hat{x} \equiv \hat{y} \times \hat{z} \ .
\label{coord}
\end{equation}

We now note that, for a given photon polarization $\vec\epsilon_\lambda$, the 
photoproduction amplitude given by  Eq.~(\ref{Jampl-NL}) can be expressed in a 
compact notation as
\begin{equation}
\hat M^\lambda \equiv \sum_{m=0}^3 M^\lambda_m\sigma_m \ ,
\label{Mampl-pol} 
\end{equation}
with $\sigma_0=1$ and $\sigma_i\ (i=1,2,3)$ denoting the usual Pauli-spin matrices. Throughout this note we use $(1, 2, 3)$ and $(x, y, z)$ indistinguishably.
The coefficients $M^\lambda_m$ are 
given explicitly in terms of $F_i$ in Secs.~\ref{sec:SpinObservLinear} and \ref{sec:SpinObservCircular} for 
a linearly and a circularly polarized photon, respectively.

Using the photoproduction amplitude in the form given by Eq.~(\ref{Mampl-pol}), any observable corresponding to the photon polarization $\vec\epsilon_\lambda$
can be calculated straightforwardly. Then, following \cite{NL05,NL04}, the cross section with the polarization of the 
photon $\vec\epsilon_\lambda$ incident on an unpolarized target is given by 
\begin{equation}
\frac{d\sigma^\lambda}{d\Omega} \equiv 
\frac{1}{2}\Tr\left[\hat M^\lambda \hat M^{\lambda\dagger}\right] 
\, = \, \sum_{m=0}^3 |M^\lambda_m|^2 \ .
\label{BX}
\end{equation}

For a given photon polarization $\vec\epsilon_\lambda$, and target-nucleon spin
in the $i$-direction ($i=x,y,z$), the corresponding spin observable, 
$T^\lambda_i$, can be expressed as 
\begin{eqnarray}
\frac{d\sigma^\lambda}{d\Omega} T^\lambda_i & \equiv &
\frac{1}{2} \Tr[\hat M^\lambda \sigma_i \hat M^{\lambda\dagger}] \nonumber \\ 
& = & 2{\it Re}[M^\lambda_0 M^{\lambda*}_i] + 2{\it Im}[M^\lambda_j M^{\lambda*}_k] \ ,
\label{BAi}
\end{eqnarray}
where the subscripts $(i, j, k)$ run cyclically, i.e., (1,2,3), (2,3,1), (3,1,2). 

Similarly, the spin observable, $R^\lambda_i$, of the outgoing nucleon in the 
$i$-direction induced by a photon beam with polarization $\vec\epsilon_\lambda$
is given by
\begin{eqnarray}
\frac{d\sigma^\lambda}{d\Omega} R^\lambda_i & \equiv & \frac{1}{2} 
\Tr[\hat M^\lambda \hat M^{\lambda\dagger} \sigma_i] \nonumber \\
   & = & 2{\it Re}[M^\lambda_0 M^{\lambda*}_i] 
- 2{\it Im}[M^\lambda_j M^{\lambda*}_k] \ ,
\label{BPi}
\end{eqnarray}
where the subscripts $(i, j, k)$ run cyclically. 

Another spin observable is the spin transfer coefficient induced by a polarized 
photon beam, $K^\lambda_{ij}$, which is given by
\begin{eqnarray}
\frac{d\sigma^\lambda}{d\Omega} K^\lambda_{ij} & \equiv & \frac{1}{2}\Tr[\hat M^\lambda \sigma_i 
\hat M^{\lambda\dagger} \sigma_j] \nonumber \\
 & = & \left(2|M^\lambda_0|^2 - \frac{d\sigma^\lambda}{d\Omega}\right) \delta_{ij} 
+ 2{\it Re}[M^\lambda_iM^{\lambda*}_j] - 2\epsilon_{ijk}{\it Im}[M^\lambda_kM^{\lambda*}_0] \ ,
\label{BKij}
\end{eqnarray}
where $\epsilon_{ijk}$ denotes the Levi-Civita antisymmetric tensor and $(i,j,k)$ may 
take any of the values $(1,2,3)$. 
The diagonal terms reduce to
\begin{equation}
\frac{d\sigma^\lambda}{d\Omega} K^\lambda_{jj} = |M^\lambda_0|^2 + |M^\lambda_j|^2 
- \sum_{k\ne j} |M^\lambda_k|^2 \ . 
\label{BKii}
\end{equation}

Equations~(\ref{BX},\ref{BAi},\ref{BPi},\ref{BKij}) exhaust all the possible basic observables 
in photoproduction with a polarized photon beam.  Any other observables may be constructed 
by appropriate linear combinations of them. In the following two sections, the spin asymmetries defined in Sec.~\ref{sec:GSX} will be calculated based on these basic observables.

\section{Observables with linear photon polarization}\label{sec:SpinObservLinear}

In what follows, $\vec\epsilon_\perp \equiv \hat y$ and $\vec\epsilon_\parallel 
\equiv \cos\theta \hat  x - \sin\theta \hat{z}$ denote the photon polarization perpendicular and parallel to the 
reaction plane ($xz$-plane), respectively. Here, $\cos\theta \equiv \hat{k}\cdot\hat{q}$.  Recall that the reaction plane is 
defined as the plane containing the vectors $\vec k$ (in the $+z$-direction) and 
$\vec q$ and that $\hat{y}$ is perpendicular to the reaction plane (cf. Eq.~(\ref{coord})). 
Then, in this case, Eq.~(\ref{Mampl-pol}) becomes 
\begin{eqnarray}
 \hat M^{\perp} & = & F_1\ \sigma_y  + iF_2\sin\theta \ , \nonumber \\
 \hat M^{\parallel} & = & \left[F_1 + F_4\sin^2\theta\right] \sigma_x  
                      +   \left[F_3 + F_4\cos\theta\right]\sin\theta\ \sigma_z  \ .
\label{Mampl-L} 
\end{eqnarray}
The particular form of the polarized photon amplitudes above, in which only the 
two of the four coefficients $F_i$ enter the amplitude $\hat M^\perp$ and
the absence of $F_2$ in $\hat M^\parallel$, is a direct consequence of the 
reflection symmetry in the reaction plane and leads to many interesting features
of the resulting observables. The coefficients $M^\lambda_m$ in Eq.~(\ref{Mampl-pol}) 
can be read off of Eq.~(\ref{Mampl-L})
\begin{eqnarray}
M^\perp_0 & \equiv  M_0 \, = \, iF_2\sin\theta \ , \qquad& 
M^\parallel_1 \, \equiv \, M_1 \, = \, F_1 + F_4\sin^2\theta \ , \nonumber \\
M^\perp_2 & \equiv  M_2 \, = \, F_1 \ , \qquad& 
M^\parallel_3 \, \equiv \, M_3 \, = \, [F_3 + F_4\cos\theta]\sin\theta  \ , \nonumber \\
M^\perp_1 & =  M^\perp_3 \, = \, M^\parallel_0 \, = \, M^\parallel_2 = 0 \ . \nonumber \\
\label{MLcoeff}
\end{eqnarray}

As mentioned above, due to the reflection symmetry encoded in Eq.~(\ref{Mampl-L}), 
not all the observables are independent from each other. For example, from 
Eqs.~(\ref{BAi},\ref{BPi}), using Eq.~(\ref{MLcoeff}), we have
\begin{equation}
T^\perp_y \, = \, R^\perp_y \ , \qquad
T^\parallel_y \, = \, - R^\parallel_y \ , \qquad
T^\lambda_j \, = \, R^\lambda_j \, = \, 0 \ \ (j \ne y) \ .
\label{C-P}
\end{equation}
Also, from Eqs.~(\ref{BKij},\ref{BKii})
\begin{eqnarray}
\fl K^\perp_{yy}  = - K^\parallel_{yy} \, = \, 1 \ , \ \ \ 
K^\perp_{xx} \, = \, K^\perp_{zz} \ , \ \ \ 
K^\parallel_{xx} \, = \, - K^\parallel_{zz} \ , \ \ \ 
K^\perp_{xz} \, = \, - K^\perp_{zx} \ , \ \ \
K^\parallel_{xz} \, = \, K^\parallel_{zx} \ , \nonumber \\
\fl K^\lambda_{jy}  =  K^\lambda_{yj} \, = \, 0  \ \ (j \ne y ;\, \lambda = \perp ,\parallel) \ . \nonumber \\
\label{STRANSFER}
\end{eqnarray}

In terms of the amplitudes $M_m$, the completely unpolarized cross section becomes
\footnote{In Refs.~\cite{NL05,NL04}, the definition of $d\sigma/d\Omega$ differs from the above by a factor of 2.}
\begin{eqnarray}
 \frac{d\sigma}{d\Omega} & \equiv \frac{1}{4}\Tr\left[\hat M \hat M^\dagger\right]
= \frac12 \sum_{\lambda=\perp, \parallel} \frac{d\sigma^\lambda}{d\Omega} \nonumber \\
& = \frac12 \left(\, |M_0|^2 + |M_2|^2 + |M_1|^2 + |M_3|^2 \right) \ .
\label{LX-M}
\end{eqnarray}
Note that to obtain the proper unpolarized cross section, $d\sigma/d\Omega$ given above should be multiplied by the incident flux  and the final-state phase-space density factor, namely, 
\begin{equation}
\frac{d\sigma_o}{d\Omega} \equiv \left( \frac{m_N}{4\pi W}\right)^2\left(\frac{|\vec q |}{|\vec k |}\right)
\frac{d\sigma}{d\Omega} 
\label{xsc1}
\end{equation}
in the center-of-mass frame of the system. Here,  $m_N$ denotes the nucleon mass.

\vskip 0.5cm
The (linear) beam asymmetry is ($p_x \equiv l$)
\begin{eqnarray}
\frac{d\sigma}{d\Omega}B^l  \equiv \frac{d\sigma}{d\Omega}\Sigma & \equiv 
\frac{1}{4}\Tr\left[\hat M \sigma_x^\gamma\hat M^\dagger\right] \nonumber \\
& = \frac{1}{4}\Tr\left[\hat M (\hat{P}^{\perp} - \hat{P}^{\parallel}) \hat M^\dagger\right] 
 = \frac12 \left( \frac{d\sigma^{\perp}}{d\Omega} - \frac{d\sigma^{\parallel}}{d\Omega} \right) \nonumber \\
& =  \frac12 \left(\, |M_0|^2 + |M_2|^2 - |M_1|^2 - |M_3|^2 \right)  \ .
\label{LS-M}
\end{eqnarray}
Recall that the operator $\sigma^\gamma_x$ acts on the real photon helicity space. In the context of the above equation, $\sigma_x^\gamma = \hat{P}^\perp - \hat{P}^\parallel$, where the projection operator $\hat{P}^\lambda$ specifies the state of the photon polarization; namely, $ \hat{P}^\lambda \vec{\epsilon} \equiv \vec{\epsilon}_\lambda$. Note that $\hat{P}^{\lambda'} \hat{P}^\lambda = \delta_{\lambda' \lambda}$ and  $\sum_\lambda \hat{P}^\lambda = 1$. 
The projection operator $\hat{P}^\lambda$  defined here is associated with the Stokes vector, $\vec{{\cal P}}\,^\gamma$, introduced in Sec.~\ref{sec:GSX} which specifies the direction and degree of polarization of the photon. 
For example,  $\hat{P}^\perp$($\hat{P}^\parallel$) corresponds to ${\cal P}^\gamma_x = +1$ (${\cal P}^\gamma_x= -1$), while
the projection operator for circular polarizations (cf.~Eq.~(\ref{vec-cpol})), $\hat{P}^\pm$ corresponds to ${\cal P}^\gamma_z = \pm 1$.

\vskip 0.5cm
The target asymmetry is given by
\begin{eqnarray}
\frac{d\sigma}{d\Omega}T_y \equiv \frac{d\sigma}{d\Omega}T & \equiv   
 \frac{1}{4} \Tr[\hat M \sigma_{y} \hat M^\dagger] 
    =  \frac12 \sum_\lambda \frac{d\sigma^\lambda}{d\Omega} T^\lambda_{y}  \nonumber \\
& =  Re\left[M_0 M_2^*\right] +  Im\left[M_1 M_3^*\right] \ ,
\label{T-M}
\end{eqnarray}
and $T_x=T_z=0$.

\vskip 0.5cm
Similarly, we have for the recoil asymmetry 
\begin{eqnarray}
 \frac{d\sigma}{d\Omega}R_y \equiv \frac{d\sigma}{d\Omega}P & \equiv  \frac{1}{4} \Tr[\hat M M^\dagger \sigma_{y}] 
    =  \frac12 \sum_\lambda \frac{d\sigma^\lambda}{d\Omega} R^\lambda_{y} \nonumber \\
& = Re\left[M_0 M_2^*\right] -  Im\left[M_1 M_3^*\right] \ ,
\label{P-M}
\end{eqnarray}
and $R_x=R_z=0$.

\vskip 0.5cm
The target-recoil asymmetry using an unpolarized photon beam is given by
\begin{eqnarray}
\fl \frac{d\sigma}{d\Omega} K_{ij} \equiv  
\frac{1}{4}\Tr[\hat M \sigma_i \hat M^\dagger \sigma_j] 
  =  \frac12 \sum_\lambda \frac{d\sigma^\lambda}{d\Omega}K^\lambda_{ij} \nonumber \\
\fl \qquad\ \ \ \, =  \frac12  \sum_\lambda \bigg\{
\left(2|M^\lambda_0|^2 - \frac{d\sigma^\lambda}{d\Omega}\right) \delta_{ij} 
  +  2{\it Re}[M^\lambda_iM^{\lambda *}_j] 
- 2\epsilon_{ijk}{\it Im}[M^\lambda_k M^{\lambda *}_0]\bigg\} \ ,
\label{Kij}
\end{eqnarray}
where $(i,j,k)$ may take any of the values $(1,2,3)$ as in Eq.(\ref{BKij}). 
The diagonal terms reduce to
\begin{equation}
\frac{d\sigma}{d\Omega} K_{jj} = \frac12 \left\{
|M_0|^2 + |M_j|^2 - \sum_{k\ne j} |M_k|^2 \right\} \ , 
\label{Kii}
\end{equation}
with $|M_i|^2 = \sum_\lambda |M^\lambda_i|^2$. 

Explicitly, the diagonal target-recoil asymmetries $K_{ii}$ become
\begin{eqnarray}
\frac{d\sigma}{d\Omega}K_{xx} & = & 
\frac12 \left(\, |M_0|^2 + |M_1|^2 - |M_2|^2 - |M_3|^2 \right)   \ , \nonumber \\
\frac{d\sigma}{d\Omega}K_{yy} & = &  
\frac12 \left(\, |M_0|^2 - |M_1|^2 + |M_2|^2 - |M_3|^2 \right)  \ , \nonumber \\
\frac{d\sigma}{d\Omega}K_{zz} & = &  
\frac12 \left(\, |M_0|^2 - |M_1|^2 - |M_2|^2 + |M_3|^2 \right)  \ , \nonumber \\
\label{Kii-M}
\end{eqnarray}
and the non-vanishing off-diagonal ones
\begin{eqnarray}
\frac{d\sigma}{d\Omega}K_{xz} & = Re\left[M_1 M_3^*\right] - Im\left[M_0 M_2^*\right] \ , \nonumber \\
\frac{d\sigma}{d\Omega}K_{zx} & = Re\left[M_1 M_3^*\right] + Im\left[M_0 M_2^*\right] \ , \nonumber \\
\label{Kij-M}
\end{eqnarray}

It is interesting to note that, using Eq.~(\ref{MLcoeff}) in Eqs.~(\ref{BAi},\ref{BPi}), 
one obtains
\begin{equation}
\frac{d\sigma}{d\Omega}P = \frac12 \left( \frac{d\sigma^\perp}{d\Omega} T^\perp_y 
- \frac{d\sigma^\parallel}{d\Omega} T^\parallel_y \right)   \equiv \frac{d\sigma}{d\Omega}T^l_y \ ,
\label{P-BAy}
\end{equation}
which shows that the recoil-nucleon asymmetry, $P$, can be determined by 
measuring the beam-target asymmetry, $T^l_y$, 
involving the polarization of the photon and nucleon in the initial state. 
This may be relevant experimentally for photoproduction reactions where the recoil particle is not self-analyzing.   
Furthermore, from Eqs.~(\ref{LS-M},\ref{Kii-M}),
\begin{equation}
K_{yy} = \Sigma \ .
\label{Kyy-S}
\end{equation}

As pointed out in Ref.~\cite{NL04}, these results are all consequences of the
reflection symmetry in the reaction plane.

\vskip 1cm
The linear photon polarization, $\vec\epsilon_\perp$ and $\vec\epsilon_\parallel$,
discussed above can be generalized to any direction in the plane containing these vectors by rotating them
about the axis along the meson momentum $\vec{q}$ by an angle $\phi$:
\begin{eqnarray}
\vec\epsilon_{\parallel'}  \equiv  \vec\epsilon_{x'} & \equiv & \cos\phi\, \vec\epsilon_\parallel +\sin\phi\, \vec\epsilon_\perp \ , \nonumber \\
\vec\epsilon_{\perp'}  \equiv \vec\epsilon_{y'} & \equiv & -\sin\phi\, \vec\epsilon_\parallel +\cos\phi\, \vec\epsilon_\perp \ . \nonumber \\
\label{trans1}
\end{eqnarray}
For the linear polarizations defined above, Eq.~(\ref{Mampl-pol}) becomes
\begin{eqnarray}
 \hat M^{x'} & = & \cos\phi\, \hat M^\parallel + \sin\phi\, \hat M^\perp 
\, = \, \sum_{m=0}^3 M^{x'}_m\sigma_m \ , \nonumber \\
 \hat M^{y'} & = & -\sin\phi\, \hat M^\parallel + \cos\phi\, \hat M^\perp 
\, = \, \sum_{m=0}^3 M^{y'}_m\sigma_m \ , \nonumber \\
\label{Mampl-LG} 
\end{eqnarray}
with
\begin{eqnarray}
M^{x'}_0 & =  \sin\phi\, M_0 \ , \qquad& M^{y'}_0 \, = \,  \cos\phi\, M_0 \ , \nonumber \\  
M^{x'}_1 & =  \cos\phi\, M_1 \ , \qquad& M^{y'}_1 \, = \, -\sin\phi\, M_1 \ , \nonumber \\  
M^{x'}_2 & =  \sin\phi\, M_2 \ , \qquad& M^{y'}_2 \, = \,  \cos\phi\, M_2 \ , \nonumber \\  
M^{x'}_3 & =  \cos\phi\, M_3 \ , \qquad& M^{y'}_3 \, = \, -\sin\phi\, M_3 \ . \nonumber \\
\label{Mampl-LG1}
\end{eqnarray}

We, then, have from Eqs.~(\ref{BX},\ref{BAi}), 
\begin{eqnarray}
 \frac{d\sigma^{x'}}{d\Omega}  & = \sin^2\phi \left(|M_0|^2 + |M_2|^2\right) +
\cos^2\phi \left(|M_1|^2 + |M_3|^2\right) \ , \nonumber \\
  \frac{d\sigma^{y'}}{d\Omega}  & = \cos^2\phi \left(|M_0|^2 + |M_2|^2\right) +
\sin^2\phi \left(|M_1|^2 + |M_3|^2\right) \ , 
\nonumber \\
 \frac{d\sigma^{x'}}{d\Omega}T^{x'}_x  & =  \left(Re\left[M_0 M_1^*\right] + Im\left[M_2M_3^*\right]\right)\sin(2\phi) \ ,  \nonumber \\
 \frac{d\sigma^{y'}}{d\Omega}T^{y'}_x  & = -\left(Re\left[M_0 M_1^*\right] + Im\left[M_2M_3^*\right]\right)\sin(2\phi) \ , 
\nonumber \\
 \frac{d\sigma^{x'}}{d\Omega}T^{x'}_y  & =  2\left(Re\left[M_0 M_2^*\right]\sin^2\phi + Im\left[M_1M_3^*\right]\cos^2\phi \right) \ , \nonumber \\
  \frac{d\sigma^{y'}}{d\Omega}T^{y'}_y  & = 2\left(Re\left[M_0 M_2^*\right]\cos^2\phi + Im\left[M_1M_3^*\right]\sin^2\phi \right) \ , 
\nonumber \\
 \frac{d\sigma^{x'}}{d\Omega}T^{x'}_z  & = \left(Re\left[M_0 M_3^*\right] + Im\left[M_1M_2^*\right]\right)\sin(2\phi) \ , \nonumber \\
 \frac{d\sigma^{y'}}{d\Omega}T^{y'}_z & =  -\left(Re\left[M_0 M_3^*\right] + Im\left[M_1M_2^*\right]\right)\sin(2\phi) \ . 
\nonumber \\
\label{MLG-obs1}
\end{eqnarray}
The polarization observable $R^\lambda_i$ are given by the expressions for $T^\lambda_i$ above with 
the sign of $Im[M_jM_k^*]$ changed. 

\vskip 0.3cm
For the triple-polarization observables we have, from Eq.~(\ref{BKij}), 
\begin{eqnarray}
\frac{d\sigma^{x'}}{d\Omega}K^{x'}_{xx} & =  \sin^2\phi \left(|M_0|^2 - |M_2|^2\right) +
\cos^2\phi \left(|M_1|^2 - |M_3|^2\right) \ , \nonumber \\
\frac{d\sigma^{y'}}{d\Omega}K^{y'}_{xx} & =  \cos^2\phi \left(|M_0|^2 - |M_2|^2\right) +
\sin^2\phi \left(|M_1|^2 - |M_3|^2\right) \ , 
\nonumber \\
\frac{d\sigma^{x'}}{d\Omega}K^{x'}_{yy} & = \sin^2\phi \left(|M_0|^2 + |M_2|^2\right) -
\cos^2\phi \left(|M_1|^2 + |M_3|^2\right) \ , \nonumber \\
\frac{d\sigma^{y'}}{d\Omega}K^{y'}_{yy} & =  \cos^2\phi \left(|M_0|^2 + |M_2|^2\right) -
\sin^2\phi \left(|M_1|^2 + |M_3|^2\right) \ , 
\nonumber \\
\frac{d\sigma^{x'}}{d\Omega}K^{x'}_{zz} & = \sin^2\phi \left(|M_0|^2 - |M_2|^2\right) -
\cos^2\phi \left(|M_1|^2 - |M_3|^2\right) \ , \nonumber \\
\frac{d\sigma^{y'}}{d\Omega}K^{y'}_{zz} & =  \cos^2\phi \left(|M_0|^2 - |M_2|^2\right) -
\sin^2\phi \left(|M_1|^2 - |M_3|^2\right) \ , 
\nonumber \\
\frac{d\sigma^{x'}}{d\Omega}K^{x'}_{xy} & =  \left(Re\left[M_1 M_2^*\right] + Im\left[M_0M_3^*\right]\right)\sin(2\phi) \ , \nonumber \\
\frac{d\sigma^{y'}}{d\Omega}K^{y'}_{xy} & =  - \left(Re\left[M_1 M_2^*\right] + Im\left[M_0M_3^*\right]\right)\sin(2\phi) \ , 
\nonumber \\
\frac{d\sigma^{x'}}{d\Omega}K^{x'}_{xz} & =  2\cos^2\phi\, Re\left[M_1 M_3^*\right] - 2\sin^2\phi\, Im\left[M_0M_2^*\right] \ , \nonumber \\
\frac{d\sigma^{y'}}{d\Omega}K^{y'}_{xz} & = 2\sin^2\phi\, Re\left[M_1 M_3^*\right] - 2\cos^2\phi\, Im\left[M_0M_2^*\right] \ , 
\nonumber \\
\frac{d\sigma^{x'}}{d\Omega}K^{x'}_{yz} & = \left(Re\left[M_2 M_3^*\right] + Im\left[M_0M_1^*\right]\right)\sin(2\phi) \ , \nonumber \\
\frac{d\sigma^{y'}}{d\Omega}K^{y'}_{yz} & = -\left(Re\left[M_2 M_3^*\right] + Im\left[M_0M_1^*\right]\right)\sin(2\phi) \ . 
\nonumber \\
\label{MLG-obs2}
\end{eqnarray}
For $i \ne j$, $K^\lambda_{ji}$ is given by $K^\lambda_{ij}$ given above with the sign of  $Im[M_0M_k^*]$ changed. 

\vskip 0.3cm
From Eqs.~(\ref{MLG-obs1}), the beam asymmetry is given by
\begin{eqnarray}
\frac{d\sigma}{d\Omega} B^{l'} & \equiv \frac{1}{4}\Tr\left[\hat M \sigma_{x'}^\gamma \hat M^\dagger\right] \nonumber \\
& =  \frac{1}{4}\Tr\left[\hat M (\hat{P}^{\perp'} - \hat{P}^{\parallel'}) \hat M^\dagger\right] 
 = \frac12 \left( \frac{d\sigma^{\perp'}}{d\Omega} - \frac{d\sigma^{\parallel'}}{d\Omega} \right)  = \cos(2\phi) \frac{d\sigma}{d\Omega}\Sigma \nonumber \\
& = \frac12 \cos(2\phi) \left( |M_0|^2 - |M_1|^2  + |M_2|^2 - |M_3|^2\right) \ . \nonumber \\
\label{MLG-S}
\end{eqnarray}

Also, from Eq.~(\ref{MLG-obs1}), the double-polarization asymmetries (beam-target ($T^l_i$) and beam-recoil ($R^l_i$)) are  
\begin{eqnarray}
\fl \frac{d\sigma}{d\Omega}T^{l'}_x  \equiv \frac12 \left( \frac{d\sigma^{y'}}{d\Omega}T^{y'}_x - \frac{d\sigma^{x'}}{d\Omega}T^{x'}_x \right) =  - \left(Re\left[M_0 M_1^*\right] + Im\left[M_2M_3^*\right]\right)\sin(2\phi) \ ,  \nonumber \\
\fl \frac{d\sigma}{d\Omega}T^{l'}_y  \equiv \frac12 \left( \frac{d\sigma^{y'}}{d\Omega}T^{y'}_y - \frac{d\sigma^{x'}}{d\Omega}T^{x'}_y \right) =  \left(Re\left[M_0 M_2^*\right] + Im\left[M_1M_3^*\right] \right) \cos(2\phi) 
= \cos(2\phi) \frac{d\sigma}{d\Omega}P \ ,  \nonumber \\
\fl \frac{d\sigma}{d\Omega}T^{l'}_z  \equiv \frac12 \left( \frac{d\sigma^{y'}}{d\Omega}T^{y'}_z - \frac{d\sigma^{x'}}{d\Omega}T^{x'}_z \right) =  - \left(Re\left[M_0 M_3^*\right] + Im\left[M_1M_2^*\right]\right)\sin(2\phi) \ ,  \nonumber \\
\fl \frac{d\sigma}{d\Omega}R^{l'}_x  \equiv \frac12 \left( \frac{d\sigma^{y'}}{d\Omega}P^{y'}_x - \frac{d\sigma^{x'}}{d\Omega}P^{x'}_x \right) =  - \left(Re\left[M_0 M_1^*\right] - Im\left[M_2M_3^*\right]\right)\sin(2\phi) \ ,  \nonumber \\
\fl \frac{d\sigma}{d\Omega}R^{l'}_y  \equiv \frac12 \left( \frac{d\sigma^{y'}}{d\Omega}P^{y'}_y - \frac{d\sigma^{x'}}{d\Omega}P^{x'}_y \right) =  \left(Re\left[M_0 M_2^*\right] - Im\left[M_1M_3^*\right] \right) \cos(2\phi) 
= \cos(2\phi) \frac{d\sigma}{d\Omega}T \ ,  \nonumber \\
\fl \frac{d\sigma}{d\Omega}R^{l'}_z  \equiv \frac12 \left( \frac{d\sigma^{y'}}{d\Omega}P^{y'}_z - \frac{d\sigma^{x'}}{d\Omega}P^{x'}_z \right) =  - \left(Re\left[M_0 M_3^*\right] - Im\left[M_1M_2^*\right]\right)\sin(2\phi) \ . \nonumber \\
\label{MLG-BT}
\end{eqnarray}

\vskip 0.3cm
Finally, from Eq.~(\ref{MLG-obs2}), the (triple-polarization) beam-target-recoil ($K^{l'}_{ij}$) asymmetries are
\begin{eqnarray}
 \frac{d\sigma}{d\Omega}K^{l'}_{xx} \equiv  \frac12 \left( \frac{d\sigma^{y'}}{d\Omega}K^{y'}_{xx} - \frac{d\sigma^{x'}}{d\Omega}K^{x'}_{xx} \right) \nonumber \\
 \qquad\ \ \ \ =  \frac12 \cos(2\phi) \left(|M_0|^2 - |M_1|^2 - |M_2|^2 + |M_3|^2 \right) = \cos(2\phi) \frac{d\sigma}{d\Omega}K_{zz} \ , \nonumber \\
\frac{d\sigma}{d\Omega}K^{l'}_{yy}  \equiv \frac12 \left( \frac{d\sigma^{y'}}{d\Omega}K^{y'}_{yy} - \frac{d\sigma^{x'}}{d\Omega}K^{x'}_{yy} \right) \nonumber \\
\qquad \ \ \ \ = \frac12 \cos(2\phi) \left(|M_0|^2 + |M_1|^2 + |M_2|^2 + |M_3|^2 \right)  = \cos(2\phi) \frac{d\sigma}{d\Omega}   , \nonumber \\
 \frac{d\sigma}{d\Omega}K^{l'}_{zz}  \equiv \frac12 \left( \frac{d\sigma^{y'}}{d\Omega}K^{y'}_{zz} - \frac{d\sigma^{x'}}{d\Omega}K^{x'}_{zz} \right) \nonumber \\
\qquad\ \ \ \  = \frac12 \cos(2\phi) \left(|M_0|^2 + |M_1|^2 - |M_2|^2 - |M_3|^2 \right) = \cos(2\phi) \frac{d\sigma}{d\Omega}K_{xx} \ , \nonumber \\
 \frac{d\sigma}{d\Omega}K^{l'}_{xy}  \equiv \frac12 \left( \frac{d\sigma^{y'}}{d\Omega}K^{y'}_{xy} - \frac{d\sigma^{x'}}{d\Omega}K^{x'}_{xy} \right) \nonumber \\
\qquad\ \ \ \ = -\left(Re\left[M_1 M_2^*\right] + Im\left[M_0M_3^*\right]\right)\sin(2\phi) = \sin(2\phi) \frac{d\sigma}{d\Omega}T^c_{z} \ , \nonumber \\
 \frac{d\sigma}{d\Omega}K^{l'}_{xz}  \equiv \frac12 \left( \frac{d\sigma^{y'}}{d\Omega}K^{y'}_{xz} - \frac{d\sigma^{x'}}{d\Omega}K^{x'}_{xz} \right) \nonumber \\
\qquad\ \ \ \  = -\left(Re\left[M_1 M_3^*\right] + Im\left[M_0M_2^*\right]\right)\cos(2\phi) = - \cos(2\phi) \frac{d\sigma}{d\Omega}K_{zx} \ , \nonumber \\
 \frac{d\sigma}{d\Omega}K^{l'}_{yz}  \equiv \frac12 \left( \frac{d\sigma^{y'}}{d\Omega}K^{y'}_{yz} - \frac{d\sigma^{x'}}{d\Omega}K^{x'}_{yz} \right) \nonumber \\
\qquad\ \ \ \  = -\left(Re\left[M_2 M_3^*\right] + Im\left[M_0M_1^*\right]\right)\sin(2\phi) = \sin(2\phi) \frac{d\sigma}{d\Omega}R^c_{x} \ , \nonumber \\
 \frac{d\sigma}{d\Omega}K^{l'}_{yx}  \equiv \frac12 \left( \frac{d\sigma^{y'}}{d\Omega}K^{y'}_{yx} - \frac{d\sigma^{x'}}{d\Omega}K^{x'}_{yx} \right) \nonumber \\
 \qquad\ \ \ \ = -\left(Re\left[M_1 M_2^*\right] - Im\left[M_0M_3^*\right]\right)\sin(2\phi) = -\sin(2\phi) \frac{d\sigma}{d\Omega}R^c_{z}\ , \nonumber \\
 \frac{d\sigma}{d\Omega}K^{l'}_{zx}  \equiv \frac12 \left( \frac{d\sigma^{y'}}{d\Omega}K^{y'}_{zx} - \frac{d\sigma^{x'}}{d\Omega}K^{x'}_{zx} \right) \nonumber \\
\qquad\ \ \ \  = -\left(Re\left[M_1 M_3^*\right] - Im\left[M_0M_2^*\right]\right)\cos(2\phi) = - \cos(2\phi) \frac{d\sigma}{d\Omega}K_{xz} \ , \nonumber \\
 \frac{d\sigma}{d\Omega}K^{l'}_{zy}  \equiv \frac12 \left( \frac{d\sigma^{y'}}{d\Omega}K^{y'}_{zy} - \frac{d\sigma^{x'}}{d\Omega}K^{x'}_{zy} \right) \nonumber \\
\qquad\ \ \ \  = -\left(Re\left[M_2 M_3^*\right] - Im\left[M_0M_1^*\right]\right)\sin(2\phi) = - \sin(2\phi) \frac{d\sigma}{d\Omega}T^c_{x} \ . \nonumber \\
\label{MLG-BK}
\end{eqnarray}
In the above equations, $T^c_i$ and $R^c_i$ denote the double-polarization asymmetries with circularly polarized photons as given by Eq.~(\ref{MLG-CT}) in the next section. As one can see, apart from $K_{yy}^{l'}$, which is related to the unpolarized cross section, the triple-polarization asymmetries given above are directly related to the double-polarization asymmetries.

\section{Observables with circular photon polarization}\label{sec:SpinObservCircular}

The circular polarization of the photon is defined by
\begin{equation}
\vec\epsilon_\pm \equiv \mp \frac{1}{\sqrt{2}}\left(\vec\epsilon_1 \pm i \vec\epsilon_2\right) 
\, = \, \mp \frac{1}{\sqrt{2}}\left(\vec\epsilon_\parallel \pm i \vec\epsilon_\perp\right) \ .
\label{vec-cpol}
\end{equation}
The corresponding amplitudes $\hat M^\pm$ are related to those with the linear 
photon polarization given in Eq.(\ref{Mampl-L}) by
\begin{eqnarray}
 \hat M^{\pm} & = & \mp \frac{1}{\sqrt{2}}\left(\hat M^\parallel \pm i\hat M^\perp\right) \nonumber \\ 
& = & \mp \frac{1}{\sqrt{2}}\left\{\left(M_1\sigma_x + M_3\sigma_z\right) \pm
i\left(M_2\sigma_y + M_0\right)\right\} \, \equiv \, \sum_{m=0}^3 \tilde M^\pm_m\sigma_m \ , 
\label{Mampl-C} 
\end{eqnarray}
where
\begin{equation}
\fl \tilde M^\pm_0 = - i\frac{1}{\sqrt{2}} M_0 \ , \ \ \ 
\tilde M^\pm_2 = - i\frac{1}{\sqrt{2}} M_2 \ , \ \ \ 
\tilde M^\pm_1 = \mp \frac{1}{\sqrt{2}} M_1 \ , \ \ \ 
\tilde M^\pm_3 = \mp \frac{1}{\sqrt{2}} M_3 \ ,
\label{MCcoeff}
\end{equation}
with $M_m$ given by Eq.~(\ref{MLcoeff}).

The cross section for a circularly polarized photon and an unpolarized nucleon target
is then given by
\begin{equation}
\frac{d\sigma^\pm}{d\Omega} = \frac12 \sum_{m=0}^3 |\tilde M^\pm_m|^2 
                               = \frac{1}{4}\sum_{m=0}^3 |M_m|^2 \ ,
\label{CX}
\end{equation}
which  immediately leads to the (circular) beam asymmetry ($p_z\equiv c$)
\begin{eqnarray}
\frac{d\sigma}{d\Omega}B^c  & \equiv \frac{1}{4}\Tr\left[\hat M \sigma_{z}^\gamma \hat M^\dagger\right] \nonumber \\
& =  \frac{1}{4}\Tr\left[\hat M (\hat{P}^{+} - \hat{P}^{-}) \hat M^\dagger\right] 
= \frac12 \left( \frac{d\sigma^+}{d\Omega} - \frac{d\sigma^-}{d\Omega} \right) = 0  \ .
\label{CS-M}
\end{eqnarray}

From Eqs.~(\ref{BAi},\ref{BPi}), the double-polarization observables $T^\pm_i$ 
and $R^\pm_i$  become
\begin{eqnarray}
\fl \frac{d\sigma^\pm}{d\Omega}T^\pm_x  =  
\pm Re\left[M_2 M_3^*\right] \mp Im\left[M_0 M_1^*\right] \ , \qquad&
\frac{d\sigma^\pm}{d\Omega}R^\pm_x  = 
\mp Re\left[M_2 M_3^*\right] \mp Im\left[M_0 M_1^*\right] \ , \nonumber \\
\fl \frac{d\sigma^\pm}{d\Omega}T^\pm_y =  
Re\left[M_0 M_2^*\right] - Im\left[M_1 M_3^*\right] \ , \qquad&
\frac{d\sigma^\pm}{d\Omega}R^\pm_y =  
Re\left[M_0 M_2^*\right] + Im\left[M_1 M_3^*\right] \ , \nonumber \\
\fl \frac{d\sigma^\pm}{d\Omega}T^\pm_z  =  
\mp Re\left[M_1 M_2^*\right] \mp Im\left[M_0 M_3^*\right] \ , \qquad&
\frac{d\sigma^\pm}{d\Omega}R^\pm_z  = 
\pm Re\left[M_1 M_2^*\right] \mp Im\left[M_0 M_3^*\right] \ . \nonumber \\
\label{CTiPi-M}
\end{eqnarray}

It is straightforward to show that, for the diagonal spin observables  
$K^\pm_{ii}$, one obtains
\begin{equation}
\frac{d\sigma^\pm}{d\Omega}K^\pm_{ii} = \frac{d\sigma}{d\Omega}K_{ii}
\label{CKii-M}
\end{equation}
and that the off-diagonal ones become
\begin{eqnarray}
\fl \frac{d\sigma^\pm}{d\Omega}K^\pm_{xy}  =  
\mp Re\left[M_0 M_3^*\right] \mp Im\left[M_1 M_2^*\right] \ , \quad&
\frac{d\sigma^\pm}{d\Omega}K^\pm_{yx}  = 
\pm Re\left[M_0 M_3^*\right] \mp Im\left[M_1 M_2^*\right] \ , \nonumber \\
\fl \frac{d\sigma^\pm}{d\Omega}K^\pm_{yz}  =  
\mp Re\left[M_0 M_1^*\right] \pm Im\left[M_2 M_3^*\right] \ , \quad&
\frac{d\sigma^\pm}{d\Omega}K^\pm_{zy}  =  
\pm Re\left[M_0 M_1^*\right] \pm Im\left[M_2 M_3^*\right] \ . \nonumber \\
\fl \frac{d\sigma^\pm}{d\Omega}K^\pm_{xz}  =  
Re\left[M_1 M_3^*\right] - Im\left[M_0 M_2^*\right]  = 
\frac{d\sigma}{d\Omega}K_{xz}  \ , \nonumber \\
\fl \frac{d\sigma^\pm}{d\Omega}K^\pm_{zx} =  
Re\left[M_1 M_3^*\right] + Im\left[M_0 M_2^*\right]  = 
\frac{d\sigma}{d\Omega}K_{zx}  \ ,   \nonumber \\
\label{CKij-M}
\end{eqnarray}

\vskip 0.3cm
From Eq.~(\ref{CTiPi-M}), the double-polarization asymmetries, beam-target ($T^c_i$) and beam-recoil ($R^c_i$) asymmetries, are  
\begin{eqnarray}
\frac{d\sigma}{d\Omega}T^c_x & \equiv \frac12 \left( \frac{d\sigma^{+}}{d\Omega}T^{+}_x - \frac{d\sigma^{-}}{d\Omega}T^{-}_x \right) =  Re\left[M_2 M_3^*\right] - Im\left[M_0M_1^*\right]  \ ,  \nonumber \\
\frac{d\sigma}{d\Omega}T^c_z & \equiv \frac12 \left( \frac{d\sigma^{+}}{d\Omega}T^{+}_z - \frac{d\sigma^{-}}{d\Omega}T^{-}_z \right) =  - Re\left[M_1 M_2^*\right] - Im\left[M_0M_3^*\right] \ ,  \nonumber \\
\frac{d\sigma}{d\Omega}R^c_x & \equiv \frac12 \left( \frac{d\sigma^{+}}{d\Omega}R^{+}_x - \frac{d\sigma^{-}}{d\Omega}R^{-}_x \right) =  - Re\left[M_2 M_3^*\right] - Im\left[M_0M_1^*\right]  \ ,  \nonumber \\
\frac{d\sigma}{d\Omega}R^c_z & \equiv \frac12 \left( \frac{d\sigma^{+}}{d\Omega}R^{+}_z - \frac{d\sigma^{-}}{d\Omega}R^{-}_z \right) =  Re\left[M_1 M_2^*\right] - Im\left[M_0M_3^*\right]  \ , \nonumber \\
\label{MLG-CT}
\end{eqnarray}
and $T^c_y = R^c_y = 0$.

Similarly, from Eq.~(\ref{CKij-M}), the beam-target-recoil asymmetries ($K^c_{ij}$) are
\begin{eqnarray}
\fl \frac{d\sigma}{d\Omega}K^c_{xy}  \equiv \frac12 \left( \frac{d\sigma^{+}}{d\Omega}K^{+}_{xy} - \frac{d\sigma^{-}}{d\Omega}K^{-}_{xy} \right) =  - Re\left[M_0 M_3^*\right] - Im\left[M_1M_2^*\right] = \frac{1}{\sin(2\phi)} \frac{d\sigma}{d\Omega}T^l_z \ ,  \nonumber \\
\fl \frac{d\sigma}{d\Omega}K^c_{yz} \equiv \frac12 \left( \frac{d\sigma^{+}}{d\Omega}K^{+}_{yz} - \frac{d\sigma^{-}}{d\Omega}K^{-}_{yz} \right) =  - Re\left[M_0 M_1^*\right] + Im\left[M_2M_3^*\right] = \frac{1}{\sin(2\phi)} \frac{d\sigma}{d\Omega}R^l_x \ ,  \nonumber \\
\fl \frac{d\sigma}{d\Omega}K^c_{yx}  \equiv \frac12 \left( \frac{d\sigma^{+}}{d\Omega}K^{+}_{yx} - \frac{d\sigma^{-}}{d\Omega}K^{-}_{yx} \right) =  - Re\left[M_0 M_3^*\right] + Im\left[M_1M_2^*\right] = \frac{1}{\sin(2\phi)} \frac{d\sigma}{d\Omega}R^l_z \ ,  \nonumber \\
\fl \frac{d\sigma}{d\Omega}K^c_{zy} \equiv \frac12 \left( \frac{d\sigma^{+}}{d\Omega}K^{+}_{zy} - \frac{d\sigma^{-}}{d\Omega}K^{-}_{zy} \right) =  - Re\left[M_0 M_1^*\right] - Im\left[M_2M_3^*\right] = \frac{1}{\sin(2\phi)} \frac{d\sigma}{d\Omega}T^l_x  \ , \nonumber \\
\label{MLG-CK}
\end{eqnarray}
and $K^c_{xz} = K^c_{zx} = K^c_{ii} = 0$.

\vskip 0.3cm
From Eqs.~(\ref{MLG-BK},\ref{MLG-CK}), we see that the triple-polarization asymmetries are simply redundant.

\section{Different notations} \label{sec:Dictionary}

In the literature, one uses different notations for the double-polarization observables than used in this note. Here is a dictionary:
\begin{eqnarray}
E  \equiv T^c_z  \ , \qquad&  C_{z} \equiv R^c_{z} \ , \qquad\qquad\ \ \ \ L_{z} \equiv K_{zz} \ , \nonumber \\
F  \equiv T^c_x  \ , \qquad& C_{x} \equiv R^c_{x} \ , \qquad\qquad\ \ \ \ L_{x} \equiv K_{zx} \ , \nonumber \\
G  \equiv T^{l'}_z(\phi=\frac{\pi}{4}) \ , \qquad&  O_{z} \equiv R^{l'}_{z}(\phi=\frac{\pi}{4}) \ , \qquad  T_{z} \equiv K_{xz} \ , \nonumber \\
H \equiv T^{l'}_x(\phi=\frac{\pi}{4}) \ , \qquad&  O_{x} \equiv R^{l'}_{x}(\phi=\frac{\pi}{4}) \ , \qquad  T_{x} \equiv K_{xx} \ . \nonumber \\
\label{obs-dic}
\end{eqnarray}

Also, some authors use different conventions for defining the double-polarization asymmetries. 
See, e.g., Ref.~\cite{RS12} for a list of different definitions used by some of the groups. 
The conventions in this work are consistent with those of Refs.~\cite{FTS92,SHKL11}, except for the sign of $E$ in the latter reference. Also, our $E$, $G$, $C_{x'}$, $C_{z'}$, $O_{x'}$, $O_{z'}$, and $L_x$ differ in sign from those in Ref.~\cite{Ronchen}.

\section{Non-redundant observables} \label{sec:NonRedundantObserv}

In Secs.~\ref{sec:SpinObservLinear} and \ref{sec:SpinObservCircular} we found, all together (non-vanishing),
\begin{itemize}

\item[a)]
 1 unpolarized  cross section: \qquad\qquad\ \ \ Eq.~(\ref{LX-M}) \ ,
\item[b)]
 3 single-polarization asymmetries: \qquad\  Eqs.~(\ref{LS-M} or \ref{MLG-S}), (\ref{T-M}), (\ref{P-M}) \ ,

\item[c)]
 15 double-polarization asymmetries: \ \ \ \ \ Eqs.~(\ref{Kii-M}), (\ref{Kij-M}), (\ref{MLG-BT}), (\ref{MLG-CT}) \ ,

\item[d)]
 13 triple-polarization asymmetries: \qquad  Eqs.~(\ref{MLG-BK}), (\ref{MLG-CK}) \ .

\end{itemize}

\vskip 0.2cm
Two of the double-polarization asymmetries, $T^{l'}_y$ and $R^{l'}_y$, are related to the single-polarization asymmetries, $R_y (\equiv P)$ and $T_y (\equiv T)$, respectively. A third double-polarization asymmetry, $K_{yy}$ (target-recoil asymmetry), is related to the beam asymmetry, $B^l (\equiv \Sigma)$.  All the triple-polarization asymmetries are related to the double-polarization asymmetries. 
There are, then, 16 non-redundant observables in total.   Appropriate combinations of them [cf. Eqs.~(\ref{LX-M},\ref{LS-M},\ref{T-M},\ref{P-M},\ref{Kii-M},\ref{Kij-M},\ref{MLG-BT},\ref{MLG-CT}) ] yield
(using the commonly adopted notations as given in the previous section)
\begin{eqnarray}
\frac{d\sigma}{d\Omega}\left(1+\Sigma\right) =  |M_0|^2 + |M_2|^2 \ , 
\qquad&
\frac{d\sigma}{d\Omega}\left(1-\Sigma\right)  =  |M_1|^2 + |M_3|^2 \ ,
\nonumber \\
\frac{d\sigma}{d\Omega}\left(T_x+L_z\right) =  |M_0|^2 - |M_2|^2 \ , 
\qquad&
\frac{d\sigma}{d\Omega}\left(T_x-L_z\right) = |M_1|^2 - |M_3|^2 \ , 
\nonumber \\
\frac{d\sigma}{d\Omega}\left(T+P\right) = \ \ \ 2 Re\left[M_0 M_2^*\right] \ , 
\qquad& 
\frac{d\sigma}{d\Omega}\left(T-P\right)  = - 2 Im\left[M_1 M_3^*\right] \ , 
\nonumber \\
\frac{d\sigma}{d\Omega}\left(T_z-L_x\right)  =  - 2 Im\left[M_0 M_2^*\right] \ , 
\qquad&
\frac{d\sigma}{d\Omega}\left(T_z+L_x\right) = \ \ \ 2 Re\left[M_1 M_3^*\right] \ , \nonumber \\
\frac{d\sigma}{d\Omega}\left(H+O_x\right)
 =  -2Re\left[M_0 M_1^*\right] \ ,
\qquad&
\frac{d\sigma}{d\Omega}\left(H-O_x\right) = -2Im\left[M_2 M_3^*\right] \ ,
\nonumber \\
\frac{d\sigma}{d\Omega}\left(F + C_x\right)  =  - 2 Im\left[M_0 M_1^*\right] \ , 
\qquad&
\frac{d\sigma}{d\Omega}\left(F - C_x\right) = \ \ \  2 Re\left[M_2 M_3^*\right] \ ,
\nonumber \\ 
\frac{d\sigma}{d\Omega}\left(G+O_z\right)  =  -2Re\left[M_0 M_3^*\right]\ ,
\qquad&
\frac{d\sigma}{d\Omega}\left(G-O_z\right)  =  -2Im\left[M_1 M_2^*\right]\ ,
\nonumber \\
\frac{d\sigma}{d\Omega}\left(E + C_z\right)  =  - 2 Im\left[M_0 M_3^*\right] \ ,
\qquad&
\frac{d\sigma}{d\Omega}\left(E - C_z\right)  = - 2 Re\left[M_1 M_2^*\right] \ . 
\nonumber \\
\label{M-obs}
\end{eqnarray}

\section{Complete experiments} \label{sec:ComExp}

The relations in the upper two rows in Eq.~(\ref{M-obs}) determine the magnitude 
of the amplitudes $M_m$, while the relations in the lower rows determine the
phase difference between the corresponding two amplitudes, i.e., $M_0$ and $M_2$,
and, $M_1$ and $M_3$, etc. 
Note that it requires both $Re\left[M_iM_j^*\right]$ and $Im\left[M_iM_j^*\right]$ 
for fixing the phase difference between the amplitudes $M_i$ and $M_j$ completely.

From the results in Eq.~(\ref{M-obs}), one might naively conclude that one needs at least 12 independent 
observables to determine the photoproduction amplitude completely, apart from an arbitrary
overall phase. For example, as mentioned above, the observables $\Sigma$, $T_x$, $L_z$, together with the unpolarized cross section $d\sigma/d\Omega$ (first two rows in Eq.~(\ref{M-obs})) 
determine the magnitudes of the amplitudes $M_m$, while $T$, $P$, $T_z$ and $L_x$
(the next two rows in Eq.~(\ref{M-obs})) determine the relative phase of $M_0$ and $M_2$, 
and of $M_1$ and $M_3$, respectively.  We, then, need two more constraints provided by, say, 
$F$, $C_x$, $H$, and $O_x$ (left column of the next two rows in Eq.~(\ref{M-obs}))
to determine the relative phase between $M_0$ and $M_1$. This ends up with the total of 12 
independent observables to determine the photoproduction amplitude.
However, as discussed in the following, careful considerations reveal that one actually needs less observables to determine the photoproduction amplitude.

An early account on the issue of complete experiments has been given by Baker, Donnachie and Storrow  in their classical work \cite{BDS75}. The problem is not trivial. In fact, early studies on this issue have resulted in contradictory findings. 
The situation has been cleared by the authors of Ref.~\cite{BDS75} who have derived the necessary and sufficient conditions for determining the full photoproduction amplitude up to discrete ambiguities. They also provided the rules for choosing further measurements to resolve these ambiguities. According to these authors, for a given kinematics (total energy of the system and meson production angle), one requires nine observables to determine the full reaction amplitude
up to an arbitrary overall phase.
Keaton and Workman \cite{KW96}, however, have realized that the issue is not quite that simple and that there are cases obeying the rules given in Ref.~\cite{BDS75} that still leave unsolved ambiguities. Then, as mentioned in the Introduction, in a detailed analysis, Chiang and Tabakin \cite{ChT97} have shown that the minimum number of required observables to determine the pseudoscalar meson photoproduction amplitude is eight. The issue of complete experiments has been revisited quite recently in Ref.~\cite{Nak18}, where an explicit derivation of the completeness condition is provided and it corroborates the original findings of Ref.~\cite{ChT97}. However, it is found that the argument of eight observables required for a complete experiment hols only when the magnitudes of the relative phases $\alpha_{ij}$, as defined in Ref.~\cite{Nak18},  differ from each other.

To address the problem of complete experiments, it is convenient to express the complex amplitudes $M_i$'s in the form 
\begin{equation}
M_i = B_i e^{i\phi_i} \ ,  \ \ \ \ \ \ B_i \equiv |M_i| \ \ \ \ \ (i=0, \cdots, 3) \ .
\label{eq:aux0}
\end{equation}
As mentioned before, the magnitudes $B_i$'s are determined from the four observables through the relations in the first two rows in Eq.~(\ref{M-obs}). Then, a set of carefully chosen four observables from the remaining spin-observables in Eq.~(\ref{M-obs}) serves to determine the relative phases $\{\phi_{01}, \phi_{12}, \phi_{23}\}$  ($\phi_{ij} \equiv \phi_i - \phi_j$) or
any combinations of three relative phases $\phi_{ij}$'s  that allow to determine the phases $\phi_i\ (i=0, \cdots, 3)$ of the reaction amplitude up to an overall phase.

In Refs.~\cite{ChT97,Nak18}, the above described procedure of determining the reaction amplitude, directly in terms of the magnitudes and phases, have been carried out for amplitudes expressed in transversity basis instead of Pauli-spin basis as used in the present work. 
Transversity basis is related to Pauli-spin basis by a simple rotation of the quantization axis. In the latter case, the quantization axis is along the incident photon momentum, while in the former case, it is along the axis perpendicular to the reaction plane \cite{BDS75}. In transversity basis, the magnitudes of the amplitudes $M_i$'s are determined by the unpolarized cross section $d\sigma/d\Omega$, and single-spin observables $\Sigma$, $T$ and $P$, leaving the double-spin observables to determine the phases of the amplitude \cite{ChT97,Nak18}. 
This is a nice feature of the transversity amplitudes from the experimental point of view for reactions where the baryon in the final state is self-analyzing, such as the $\Lambda$ and $\Sigma$ hyperons, so that the recoil-asymmetry $P$ can be measured without measuring the spin of the recoil baryon explicitly. Otherwise, the measurement of the recoil asymmetry is much more involved than measuring the other single-spin observables. Also, measuring the phases of the transversity amplitudes is trickier than measuring their magnitudes in general.
As can be seen from Eq.~(\ref{M-obs}), in Pauli-spin basis, the double-spin observables $T_x$ and $L_z$ enter in the determination of the magnitudes of the Pauli-spin amplitudes. We also mention that none of the single-spin observables enter in the determination of the magnitudes of the helicity amplitudes \cite{ChT97}.   
An immediate consequence of these differences is that  one can find more possible sets of minimum number of observables, distinct from those determined in Refs.~\cite{ChT97,Nak18}, that can determine the photoproduction amplitude up to an overall phase.

\vskip 0.5cm 
The sets of minimum number (four) of spin observables that resolve the phase ambiguity of the Pauli-spin amplitudes up to an overall phase can be determined in complete analogy to what has been done in Ref.~\cite{Nak18} for transversity amplitudes.  For details, see \ref{app:1}.
The results are given in Tables.~\ref{tab:ab}, \ref{tab:ac}, \ref{tab:bc}, and \ref{tab:211}.  
Note that, there, the kinematic restrictions on the relative phase angles for those sets of four observables - that resolve the phase ambiguity otherwise - are indicated. The restrictions of the type $(\beta_{ik} \pm \alpha_{kl}) \ne \pm \pi/2$ at the level of pairs of observables as discussed in \ref{app:2} are not indicated. For details, see \ref{app:2}. The
restrictions are less severe for determining the relative phases of the reaction amplitude in Pauli-spin representation than in transversity representation as discussed also in \ref{app:2}.

\begin{table*}[t!]
\caption{\label{tab:ab} Sets of two pairs of double-spin observables for $2 + 2$ (groups $a$ + $b$) case mentioned in \ref{app:1}.  $\surd =$ do resolve; $X =$ don't resolve; $^{**}=$ don't resolve if $\beta_{02}=|\alpha_{13}| \ {\rm and}\  \beta_{03}=-\alpha_{12}$; $^*=$ don't resolve if $\beta_{02}=\alpha_{13} \ {\rm and}\  \beta_{03}=-\alpha_{12}$;  $^{\dagger\dagger}=$ don't resolve if $\beta_{02}=-\alpha_{13} \ {\rm and}\  \beta_{03}=|\alpha_{12}|$; $^\dagger=$ don't resolve if $\beta_{02}=-\alpha_{13} \ {\rm and}\  \beta_{03}=\alpha_{12}$; $^\# =$ don't resolve if $\beta_{02}=-\alpha_{13} \ {\rm and}\  \beta_{03}=-\alpha_{12}$.  Restrictions of the type $(\beta_{ik} \pm \alpha_{kl}) \ne \pm \pi/2$ at the level of pairs of observables as discussed in \ref{app:2} are not indicated. }
\begin{tabular}{ccccccc}
\hline\hline
  & $(O^b_{1+}, O^b_{1-}) $ & $(O^b_{1+}, O^b_{2+}) $ & $(O^b_{1+}, O^b_{2-}) $ & $(O^b_{1-}, O^b_{2+}) $ & $(O^b_{1-}, O^b_{2-}) $ & $(O^b_{2+}, O^b_{2-}) $ \\ 
  & $(G, O_z) $                    & $(G, E) $                          & $(G, C_z) $                     & $(O_z, E) $                     & $(O_z, C_z) $                 & $(E, C_z) $                     \\ \hline
$(O^a_{1+}, O^a_{1-})  $    &       X      &      X     & $\surd\ ^{**}$  & $\surd\ ^{**}$  &      X    &   X    \\
$(T, P) $                              &    &   &   &   &   &   \\
$(O^a_{1+}, O^a_{2+}) $    &       X     &      X      & $\surd\ ^*$  & $\surd\ ^*$ &      X     &   X    \\
$(T, L_x) $                          &    &   &   &   &   &   \\
$(O^a_{1+}, O^a_{2-}) $    & $\surd\ ^{\dagger\dagger}$  & $\surd\ ^\dagger$ & $\surd\ ^*$ & $\surd\ ^*$ & $\surd\ ^\dagger$  & $\surd\ ^{\dagger\dagger}$   \\
$(T, T_z) $                         &    &   &   &   &   &   \\
$(O^a_{1-}, O^a_{2+}) $   & $\surd\ ^{\dagger\dagger}$  &  $\surd\ ^\dagger$ & $\surd\ ^*$ & $\surd\ ^*$ & $\surd\ ^\dagger$  & $\surd\ ^{\dagger\dagger}$   \\
$(P, L_x) $                        &    &   &   &   &   &   \\
$(O^a_{1-}, O^a_{2-}) $    &     X      &       X      & $\surd\ ^*$ & $\surd\ ^*$ &      X       &     X        \\
$(P, T_z) $                        &    &   &   &   &   &   \\
$(O^a_{2+}, O^a_{2-}) $  &        X    &       X      & $\surd\ ^{**}$ & $\surd\ ^{**}$ &      X       &     X          \\
$(L_x, T_z) $                   &    &   &   &   &   &   \\
\hline\hline
\end{tabular}
\end{table*}
\begin{table*}[t!]
\caption{\label{tab:ac} Sets of two pairs of double-spin observables for $2 + 2$ (groups $a$ + $c$) case mentioned in \ref{app:1}.  $\surd =$ do resolve; $X =$ don't resolve; $^{**}=$ don't resolve if $\beta_{02}=|\alpha_{13}| \ {\rm and}\  \beta_{01}=-\alpha_{23}$; $^*=$ don't resolve if $\beta_{02}=\alpha_{13} \ {\rm and}\  \beta_{01}=-\alpha_{23}$;  $^{\dagger\dagger}=$ don't resolve if $\beta_{02}=-\alpha_{13} \ {\rm and}\  \beta_{01}=|\alpha_{23}|$; $^\dagger=$ don't resolve if $\beta_{02}=-\alpha_{13} \ {\rm and}\  \beta_{01}=\alpha_{23}$; $^\# =$ don't resolve if $\beta_{02}=-\alpha_{13} \ {\rm and}\  \beta_{01}=-\alpha_{23}$. Restrictions of the type $(\beta_{ik} \pm \alpha_{kl}) \ne \pm \pi/2$ at the level of pairs of observables, as discussed in \ref{app:2}, are not indicated. }
\begin{tabular}{ccccccc}
\hline\hline
  & $(O^c_{1+}, O^c_{1-}) $ & $(O^c_{1+}, O^c_{2+}) $ & $(O^c_{1+}, O^c_{2-}) $ & $(O^c_{1-}, O^c_{2+}) $ & $(O^c_{1-}, O^c_{2-}) $ & $(O^c_{2+}, O^c_{2-}) $ \\ 
  & $(H, O_x) $                    & $(H, C_x) $                      & $(H, F) $                         & $(O_x, C_x) $                 & $(O_x, F) $                    & $(C_x, F) $                      \\ \hline
$(O^a_{1+}, O^a_{1-})  $    &       X     & $\surd\ ^{**}$ &     X      &     X       & $\surd\ ^{**}$ &       X    \\
$(T, P) $                              &    &   &   &   &   &   \\
$(O^a_{1+}, O^a_{2+}) $    & $\surd\ ^{\dagger\dagger}$ & $\surd\ ^*$ & $\surd\ ^\dagger$ & $\surd\ ^\dagger$ & $\surd\ ^*$ & $\surd\ ^{\dagger\dagger}$  \\
$(T, L_x) $                          &    &   &   &   &   &   \\
$(O^a_{1+}, O^a_{2-}) $    &       X      & $\surd\ ^*$ &       X    &      X      & $\surd\ ^*$ &      X      \\
$(T, T_z) $                         &    &   &   &   &   &   \\
$(O^a_{1-}, O^a_{2+}) $    &       X       & $\surd\ ^*$ &       X    &      X     & $\surd\ ^*$ &   X      \\
$(P, L_x) $                          &    &   &   &   &   &   \\
$(O^a_{1-}, O^a_{2-}) $     & $\surd\ ^{\dagger\dagger}$ & $\surd\ ^*$ & $\surd\ ^\dagger$ & $\surd\ ^\dagger$ & $\surd\ ^*$ & $\surd\ ^{\dagger\dagger}$   \\
$(P, T_z) $                         &    &   &   &   &   &   \\
$(O^a_{2+}, O^a_{2-}) $    &     X         & $\surd\ ^{**}$ &      X    &      X     & $\surd\ ^{**}$ &      X         \\
$(L_x, T_z) $                     &    &   &   &   &   &   \\
\hline\hline
\end{tabular}
\end{table*}
\begin{table*}[h!]
\caption{\label{tab:bc} Sets of two pairs of double-spin observables for $2 + 2$ (groups $b$ + $c$) case mentioned in \ref{app:1}.  $\surd =$ do resolve; $X =$ don't resolve; $^{**}=$ don't resolve if $\beta_{03}=|\alpha_{12}| \ {\rm and}\  \beta_{01}=-\alpha_{23}$; $^*=$ don't resolve if $\beta_{03}=\alpha_{12} \ {\rm and}\  \beta_{01}=-\alpha_{23}$;  $^{\dagger\dagger}=$ don't resolve if $\beta_{03}=-\alpha_{12} \ {\rm and}\  \beta_{01}=|\alpha_{23}|$; $^\dagger=$ don't resolve if $\beta_{03}=-\alpha_{12} \ {\rm and}\  \beta_{01}=\alpha_{23}$; $^\# =$ don't resolve if $\beta_{03}=-\alpha_{12} \ {\rm and}\  \beta_{01}=-\alpha_{23}$. Restrictions of the type $(\beta_{ik} \pm \alpha_{kl}) \ne \pm \pi/2$ at the level of pairs of observables, as discussed in \ref{app:2}, are not indicated. }
\begin{tabular}{ccccccc}
\hline\hline
  & $(O^c_{1+}, O^c_{1-}) $ & $(O^c_{1+}, O^c_{2+}) $ & $(O^c_{1+}, O^c_{2-}) $ & $(O^c_{1-}, O^c_{2+}) $ & $(O^c_{1-}, O^c_{2-}) $ & $(O^c_{2+}, O^c_{2-}) $ \\ 
  & $(H, O_x) $                    & $(H, C_x) $                      & $(H, F) $                         & $(O_x, C_x) $                 & $(O_x, F) $                    & $(C_x, F) $                      \\ \hline
$(O^b_{1+}, O^b_{1-})  $    &       X      &       X      & $\surd\ ^{**}$ & $\surd\ ^{**}$ &     X       &        X    \\
$(G, O_z) $                         &    &   &   &   &   &   \\
$(O^b_{1+}, O^b_{2+}) $    & $\surd\ ^{\dagger\dagger}$ & $\surd\ ^\dagger$ &  $\surd\ ^*$ &  $\surd\ ^*$   & $\surd\ ^\dagger$ & $\surd\ ^{\dagger\dagger}$  \\
$(G, E) $                         &    &   &   &   &   &   \\
$(O^b_{1+}, O^b_{2-}) $    &       X     &       X      & $\surd\ ^*$ & $\surd\ ^*$ &      X       &      X       \\
$(G, C_z) $                            &    &   &   &   &   &   \\
$(O^b_{1-}, O^b_{2+}) $   &       X     &       X      & $\surd\ ^*$ & $\surd\ ^*$ &      X       &      X        \\
$(O_x, E) $                   &    &   &   &   &   &   \\
$(O^b_{1-}, O^b_{2-}) $     & $\surd\ ^{\dagger\dagger}$ & $\surd\ ^\dagger$ & $\surd\ ^*$ & $\surd\ ^*$ & $\surd\ ^\dagger$  & $\surd\ ^{\dagger\dagger}$   \\
$(O_x, C_z) $                       &    &   &   &   &   &   \\
$(O^b_{2+}, O^b_{2-}) $  &     X         &     X      & $\surd\ ^{**}$ & $\surd\ ^{**}$ &       X      &      X         \\
$(E, C_z) $                      &    &   &   &   &   &   \\
\hline\hline
\end{tabular}
\end{table*}
\begin{table*}[t!]
\caption{\label{tab:211} Sets of two pairs of double-spin observables for case $2 (a) + 1 (b) + 1 (c)$. Other combinations can be obtained by appropriate permutations of the indices $a, b, c$.  $\surd =$ do resolve; $X =$ don't resolve; $^{**}=$ don't resolve if $\beta_{02}=|\alpha_{13}| \ {\rm and}\  \beta_{01}=|\alpha_{23}|$; $^*=$ don't resolve if $\beta_{02}=\alpha_{13} \ {\rm and}\  \beta_{01}=|\alpha_{23}|$.  Restrictions of the type $(\beta_{ik} \pm \alpha_{kl}) \ne \pm \pi/2$ at the level of pairs of observables, as discussed in \ref{app:2}, are not indicated. }
\begin{tabular}{ccccccc}
\hline\hline
  & $(O^a_{1+}, O^a_{1-})$ & $(O^a_{1+}, O^a_{2+})$ & $(O^a_{1+}, O^a_{2-})$ & $(O^a_{1-}, O^a_{2+})$ & $(O^a_{1-}, O^a_{2-})$  & $(O^a_{2+} , O^a_{2-})$  \\ 
  & $(T, P)$                          & $(T, L_x)$                       & $(T, T_z)$                      & $(P, L_x)$                       & $(P, T_z)$                      & $(L_x, T_z)$                     \\ \hline
$(O^b_{1+}, O^c_{1+}) $  & $\surd\ ^{**}$ & $\surd\ ^*$ & $\surd\ ^*$ & $\surd\ ^*$ & $\surd\ ^*$ &      X       \\
$(G, H) $                          &    &   &   &   &   &         \\
$(O^b_{1+}, O^c_{1-}) $   & $\surd\ ^{**}$ & $\surd\ ^*$ & $\surd\ ^*$ & $\surd\ ^*$ & $\surd\ ^*$ &      X       \\
$(G, O_x) $                      &    &   &   &   &   &         \\
$(O^b_{1+}, O^c_{2+}) $  &      X     & $\surd\ ^*$ & $\surd\ ^*$ & $\surd\ ^*$ & $\surd\ ^*$ & $\surd\ ^{**}$       \\
$(G, C_x) $                      &    &   &   &   &   &        \\
$(O^b_{1+}, O^c_{2-}) $  &      X      & $\surd\ ^*$ & $\surd\ ^*$ & $\surd\ ^*$ & $\surd\ ^*$ & $\surd\ ^{**}$      \\
$(G, F) $                          &    &   &   &   &   &         \\
  &    &   &   &   &   &         \\
$(O^b_{1-}, O^c_{1+}) $  & $\surd\ ^{**}$ & $\surd\ ^*$ & $\surd\ ^*$ & $\surd\ ^*$ & $\surd\ ^*$ &       X  \\
$(O_z, H) $                     &    &   &   &   &   &         \\
$(O^b_{1-}, O^c_{1-}) $  & $\surd\ ^{**}$ & $\surd\ ^*$ & $\surd\ ^*$ & $\surd\ ^*$ & $\surd\ ^*$ &        X  \\
$(O_z, O_x) $                 &    &   &   &   &   &         \\
$(O^b_{1-}, O^c_{2+}) $ &     X      & $\surd\ ^*$ & $\surd\ ^*$ & $\surd\ ^*$ & $\surd\ ^*$  & $\surd\ ^{**}$  \\
$(O_z, C_x) $                 &    &   &   &   &   &        \\
$(O^b_{1-}, O^c_{2-}) $  &    X       & $\surd\ ^*$ & $\surd\ ^*$ & $\surd\ ^*$ &  $\surd\ ^*$ & $\surd\ ^{**}$  \\
$(O_z, F) $                     &    &   &   &   &   &         \\
  &    &   &   &   &   &         \\
$(O^b_{2+}, O^c_{1+}) $ &      X      & $\surd\ ^*$ & $\surd\ ^*$ & $\surd\ ^*$ & $\surd\ ^*$ & $\surd\ ^{**}$   \\
$(E, H) $                          &    &   &   &   &   &         \\
 $(O^b_{2+}, O^c_{1-}) $ &      X      & $\surd\ ^*$ & $\surd\ ^*$ & $\surd\ ^*$ & $\surd\ ^*$ & $\surd\ ^{**}$   \\
$(E, O_x) $                      &    &   &   &   &   &         \\
$(O^b_{2+}, O^c_{2+}) $ & $\surd\ ^{**}$ & $\surd\ ^*$ & $\surd\ ^*$ & $\surd\ ^*$ & $\surd\ ^*$ &       X  \\
$(E, C_x) $                      &    &   &   &   &   &        \\
$(O^b_{2+}, O^c_{2-}) $  & $\surd\ ^{**}$ & $\surd\ ^*$ & $\surd\ ^*$ & $\surd\ ^*$ & $\surd\ ^*$ &       X  \\
$(E, F) $                          &    &   &   &   &   &         \\
  &    &   &   &   &   &         \\
$(O^b_{2-}, O^c_{1+}) $ &      X      & $\surd\ ^*$ & $\surd\ ^*$ & $\surd\ ^*$ & $\surd\ ^*$ & $\surd\ ^{**}$   \\
$(C_z, H) $                     &    &   &   &   &   &         \\
 $(O^b_{2-}, O^c_{1-}) $ &      X      & $\surd\ ^*$ & $\surd\ ^*$ & $\surd\ ^*$ & $\surd\ ^*$ & $\surd\ ^{**}$   \\
$(C_z, O_x) $                 &    &   &   &   &   &         \\
$(O^b_{2-}, O^c_{2+}) $ & $\surd\ ^{**}$ & $\surd\ ^*$ & $\surd\ ^*$ & $\surd\ ^*$ & $\surd\ ^*$ &        X  \\
$(C_z, C_x) $                       &    &   &   &   &   &        \\
$(O^b_{2-}, O^c_{2-}) $  & $\surd\ ^{**}$ & $\surd\ ^*$ & $\surd\ ^*$ & $\surd\ ^*$ & $\surd\ ^*$ &        X  \\
$(C_z, F) $                     &    &   &   &   &   &         \\
\hline\hline
\end{tabular}
\vskip -0.05cm
\end{table*}
Those sets of four spin observables not involving the single-spin observables $T$ and $P$  yield distinct sets of eight observables to determine the photoproduction amplitude compared to those sets of eight observables containing the sets of four double-spin observables given in Refs.~\cite{ChT97,Nak18} that do not involve $T_x$ and $L_z$. 

Anyway, these results reveal that it is experimentally extremely demanding to determine the 
photoproduction amplitude uniquely for it requires to measure both the single- and double-spin
observables for each kinematics.
Moreover, as has been pointed out in Ref.~\cite{SHKL11}, experimental data have finite accuracies which limits our ability to determine the reaction amplitude in a complete experiment. Due to these difficulties, in practice, it requires to measure more observables than those of a complete experiment to determine the amplitude. In this regard, measurements of more observables among those 16 non-redundant ones are not just desirable for consistency check purposes, but they are a necessity. Note that for isovector meson photoproduction, we see from Eq.~(\ref{Isospin}) that one requires
to measure, in principle, at least three charged  channels (e.g., $\gamma p \to \pi^0 p$, $\gamma p \to \pi^+ n$, and $\gamma n \to \pi^- p$) to determine the corresponding reaction amplitude. Measurements of the fourth channel ($\gamma n \to \pi^0 n$) may be useful, especially, if comparable level of accuracies of the data achieved in the charged channels can be attained. This is, however, an extremely difficult challenge.
The issue of the complete experiments in connection to the actual experimental data has been further addressed by the Gent group 
\cite{VRCV13,NVR15,NRIG16} in their detailed analyses, including a statistically more sound analysis to constrain the photoproduction amplitude \cite{NRIG16}. 
It should also be mentioned that there have been initiatives toward finding constraints in partial-wave analysis in the context of complete experiments \cite{SHKL11,Workman11,WPBTSOK11,DMIM11}, motivated by recent advances in experimental techniques that allow for possibilities of many spin-observables in photoproduction reactions to be  measured.
Of particular interest in this connection is the issue of whether the baryon resonances can be extracted model independently or with minimal model assumptions. Efforts in this direction are currently in progress  \cite{WBT14,WTWDH17,FGRGP18}.
Here, we remark that quantum mechanics does not allow to determine the overall phase of the reaction amplitude from experiment. For this, some physics input is required. This fact must have a strong impact on partial-wave analysis for extracting the baryon resonances in the context of complete experiments for, if the overall phase of the amplitude is unknown, the corresponding partial-wave amplitude is an ill defined quantity, unless the unknown overall phase is a constant, independent on the meson production angle.  The issues related to the unknown overall phase have been discussed earlier by several authors. In particular,
Omelaenko \cite{Ome81} mentioned the overall phase problem for photoproduction in the summary section of
his paper on discrete ambiguities in truncated partial-wave analysis. In the classic review paper by Bowcock and Burkhardt \cite{BB75}, this problem is discussed as well. 
Dean and Lee \cite{DL72} also investigated this problem mainly for the formalism of $\pi N$-scattering. Two recent publications \cite{WSWTB17,SWOHOSKNOTW18} treat the same problem, but mostly in the simpler context of spinless particle scattering.

\vskip 0.5cm
Before leaving this section, it should be mentioned that the Fierz transformation may be used to obtain various 
relationships among the 16 non-redundat observables \cite{ChT97}. They are very helpful for testing the consistency of these
observables extracted either experimentally or calculated based on theoretical models. An exhaustive number of these Fierz relations are given in Refs.~\cite{ChT97,SHKL11}. Thus, we refrain from giving them here. The conventions for the spin observables in Ref.~\cite{SHKL11} are consistent with those used in this work, except for the sign difference of $E$.

\section{Observables in terms of the amplitudes $F_i$} \label{sec:SpinObserv-F}

Of course, if one wishes, any of the observables discussed in Secs.\ref{sec:SpinObservLinear} 
and \ref{sec:SpinObservCircular}, can be expressed directly in terms of the coefficients $F_i$ 
appearing in the most general form of the photoproduction amplitude given by 
Eq.~(\ref{Jampl-NL}). For example, inserting Eq.~(\ref{MLcoeff}) into Eq.~(\ref{LX-M}), the 
unpolarized cross section becomes
\begin{equation}
\fl \frac{d\sigma}{d\Omega} =  |F_1|^2  + \frac12 \Big(|F_2|^2 + |F_3|^2 + |F_4|^2 \Big)\sin^2\theta 
+ Re\left[\left(F_1 + F_3\cos\theta\right) F_4^*\right]\sin^2\theta  \ .
\label{xsc}
\end{equation}

\vskip 0.3cm
From Eqs.~(\ref{MLcoeff},\ref{LS-M},\ref{T-M},\ref{P-M}), the single-polarization asymmetries become
\begin{eqnarray}
\fl \frac{d\sigma}{d\Omega}B^l \equiv  \frac{d\sigma}{d\Omega}\Sigma  =  \frac12 \Big( |F_2|^2 -|F_3|^2 - |F_4|^2\Big)\sin^2\theta
- Re\left[\left(F_1 + F_3\cos\theta\right)F_4^*\right]\sin^2\theta \ , \nonumber \\
\fl \frac{d\sigma}{d\Omega}T_y  \equiv  \frac{d\sigma}{d\Omega}T =  Im\left[ 
\left(- F_2 + F_3 + F_4\cos\theta\right)F_1^* 
+ F_3F_4^*\sin^2\theta\right] \sin\theta \ , \nonumber \\
\fl \frac{d\sigma}{d\Omega}R_y  \equiv  \frac{d\sigma}{d\Omega}P  =  - Im\left[ 
\left(F_2 + F_3 + F_4\cos\theta\right)F_1^* 
+ F_3F_4^*\sin^2\theta\right] \sin\theta \ . \nonumber \\
\label{asym}
\end{eqnarray}

The beam-target asymmetries, are given by  (from Eqs.~(\ref{MLcoeff},\ref{MLG-BT},\ref{MLG-CT}))
\begin{eqnarray}
\fl \frac{d\sigma}{d\Omega}T^c_z & \equiv  \frac{d\sigma}{d\Omega}E  = - |F_1|^2 - Re\left[F_2^*(F_3 + F_4\cos\theta) + F_1^*F_4 \right]\sin^2\theta  \ , \nonumber \\
\fl \frac{d\sigma}{d\Omega}T^c_x & \equiv \frac{d\sigma}{d\Omega}F  = - Re\left[ F_2^*(F_1+F_4\sin^2\theta) - F_1^*(F_3 + F_4\cos\theta) \right]\sin\theta  \ , \nonumber \\
\fl \frac{d\sigma}{d\Omega}T^{l'}_z(\phi=\frac{\pi}{4}) & \equiv \frac{d\sigma}{d\Omega}G  =  - Im\left[ F_2^*\left(F_3 + F_4\cos\theta\right) + F_1^*F_4\right] \sin^2\theta \ , \nonumber \\
\fl \frac{d\sigma}{d\Omega}T^{l'}_x(\phi=\frac{\pi}{4}) & \equiv  \frac{d\sigma}{d\Omega}H  =  - Im\left[ F_2^*\left(F_1 + F_4\sin^2\theta\right) - F_1^*\left(F_3 + F_4\cos\theta\right) \right]\sin\theta  \ .  \nonumber \\
\label{eq:EFGH}
\end{eqnarray}
In \ref{app:3}, the beam-target asymmetry $E$ is shown to correspond to the helicity asymmetry.

\vskip 0.3cm
The beam-recoil asymmetries are given by (from Eqs.~(\ref{MLcoeff},\ref{MLG-CT}))
\begin{eqnarray}
\fl \frac{d\sigma}{d\Omega}R^c_{x} & \equiv \frac{d\sigma}{d\Omega}C_{x}  =  - Re\left[ F^*_2 (F_1+F_4\sin^2\theta) + F^*_1(F_3 + F_4\cos\theta) \right]\sin\theta  \ , \nonumber \\
\fl \frac{d\sigma}{d\Omega}R^c_{z} & \equiv \frac{d\sigma}{d\Omega}C_{z}  =   |F_1|^2 - Re\left[F_2^*(F_3 + F_4\cos\theta) - F_1^*F_4 \right]\sin^2\theta  \ , \nonumber \\
\label{eq:Cn1Cn3}
\end{eqnarray}
and, from Eqs.~(\ref{MLcoeff},\ref{MLG-BT}),
\begin{eqnarray}
\fl \frac{d\sigma}{d\Omega}R^{l'}_{x}(\phi=\frac{\pi}{4}) & \equiv \frac{d\sigma}{d\Omega}O_{x}  =  -Im\left[ F^*_2(F_1+F_4\sin^2\theta) + F^*_1(F_3 + F_4\cos\theta) \right]\sin\theta  \ , \nonumber \\
\fl \frac{d\sigma}{d\Omega}R^{l'}_{z}(\phi=\frac{\pi}{4}) & \equiv \frac{d\sigma}{d\Omega}O_{z}  =   Im\left[F^*_1F_4 - (F_3 + F_4\cos\theta)F^*_2 \right]\sin^2\theta  \ , \nonumber \\
\label{eq:On1On3}
\end{eqnarray}

The target-recoil asymmetries are, from Eqs.~(\ref{Kii-M},\ref{Kij-M}),
\begin{eqnarray}
\fl \frac{d\sigma}{d\Omega} K_{xx} \equiv \frac{d\sigma}{d\Omega} T_x = 
Re\big[(F_1-F_3\cos\theta)F^*_4\big]\sin^2\theta + \frac12 \Big( |F_2|^2 - |F_3|^2 - |F_4|^2 \cos(2\theta) \Big)\sin^2\theta \ , \nonumber \\
\fl \frac{d\sigma}{d\Omega} K_{xz} \equiv \frac{d\sigma}{d\Omega} T_z = Re\big[ F_1(-F^*_2 + F^*_3 + F^*_4\cos\theta) + F_3F^*_4\sin^2\theta \big] \sin\theta  \nonumber \\
\hspace{1cm} +  \frac12 |F_4|^2 \sin(2\theta) \sin^2\theta \ , \nonumber \\
\fl \frac{d\sigma}{d\Omega} K_{zx}  \equiv \frac{d\sigma}{d\Omega} L_x = Re\big[ F_1(F^*_2 + F^*_3 + F^*_4\cos\theta) + F_3F^*_4\sin^2\theta \big] \sin\theta  \nonumber \\
\hspace{1cm} + \frac12 |F_4|^2 \sin(2\theta) \sin^2\theta  , \nonumber \\
\fl \frac{d\sigma}{d\Omega} K_{zz}  \equiv \frac{d\sigma}{d\Omega} L_z = - |F_1|^2 - Re\big[(F_1-F_3\cos\theta)F^*_4\big]\sin^2\theta  \nonumber \\
\fl \hspace{3.5cm} + \frac12\Big( |F_2|^2 + |F_3|^2 + |F_4|^2\cos(2\theta) \Big)\sin^2\theta \ , \nonumber \\
\label{eq:TniLni}
\end{eqnarray}

\vskip0.5cm
Some of the appropriate combinations of the spin observables may be useful for learning about certain aspects of the reaction dynamics. For example, 
\begin{eqnarray}
\frac{d\sigma}{d\Omega}(G - O_z) & = 2Im[F_1F^*_4]\sin^2\theta \ ,
\label{eq:OzG}
\end{eqnarray}
may tell us about the $D$-wave interferences at low energies, because the amplitude $F_4$ contains no partial-waves lower than the $D$-wave in the final state.

Another example is the case of the combinations 
\begin{eqnarray}
\frac{d\sigma}{d\Omega}(C_z+E) & = - 2Re[F^*_2(F_3+F_4\cos\theta)]\sin^2\theta \ , \nonumber \\
\frac{d\sigma}{d\Omega}(O_z+G) & = -2Im[F^*_2(F_3+F_4\cos\theta)]\sin^2\theta \ , \nonumber \\
\label{eq:CzEOzG}
\end{eqnarray}
which filter out the $S$-wave contributions since the $S$-wave is contained only in the amplitude $F_1$.

Other combinations such as
\begin{eqnarray}
\fl \frac{d\sigma}{d\Omega}\left(1+\Sigma\right) & =  |F_1|^2 + |F_2|^2\sin^2\theta \ , 
\qquad& 
\frac{d\sigma}{d\Omega}\left(T + P\right) \, = \, 2Im[F_1F^*_2] \sin\theta \ ,
\nonumber \\
\fl \frac{d\sigma}{d\Omega}\left(L_z + T_x\right) & =   -|F_1|^2 + |F_2|^2\sin^2\theta \ , 
\qquad& 
\frac{d\sigma}{d\Omega}\left(L_x -T_z\right) \, = \, 2Re[F_1F^*_2] \sin\theta \  ,  \nonumber \\
\label{M-obsF}
\end{eqnarray}
can determine the amplitudes $F_i$ (here, $i=1,2$). Note that the combinations $T+P$ and/or $L_x-T_z$ can be useful to shed light on the $S$-wave interference at low energies as the $S$-waves are contained only in $F_1$, as mentioned above. 
 
Also, note that the quantities that are proportional to the imaginary part of the product of the amplitudes should be more sensitive to the final state interaction compared to those that depend only on the real part.

\section{Observables in the rotate frame} \label{sec:SpinObservRot}

In the literature, some of the spin observables involving recoil-nucleon spin polarization are specified in the rotated coordinate system with the quantization axis along the direction of the meson momentum $\vec{q}$. 
Here we give the spin observables in this rotated frame.  The rotated reference frame, $\{\hat{x}', \hat{y}', \hat{z}'\}$, is obtained from $\{\hat{x},\hat{y},\hat{z}\}$ (cf. Eq.~(\ref{coord})) by rotating the latter by an angle $\theta$ ($\cos\theta \equiv \hat{k}\cdot\hat{q}$) counterclockwise about the $\hat{y}$-axis. Explicitly,
\begin{eqnarray}
\hat{x}' & = \cos\theta\, \hat{x} - \sin\theta\, \hat{z} \ , \nonumber \\
\hat{z}' & = \sin\theta\, \hat{x} + \cos\theta\, \hat{z} \ , \nonumber \\
\hat{y}' & = \hat{y} \ .
\label{eq:D1}
\end{eqnarray}

Note that, in terms of this rotated primed reference frame, the linear photon polarization $\vec{\epsilon}^{\, \parallel}$ and $\vec{\epsilon}^{\, \perp}$ defined in the unprimed frame can be expressed as 
\begin{eqnarray}
\vec{\epsilon}^{\, \parallel} & = \hat{x} = \cos\theta\, \hat{x}' + \sin\theta\, \hat{z}' \ , \nonumber \\
\vec{\epsilon}^{\, \perp} & = \hat{y} = \hat{y}'  \ .
\label{eq:D2}
\end{eqnarray}
Also, the photon and meson momenta, $\vec k$ and $\vec q$, respectively, are given by 
\begin{eqnarray}
\hat{k} & = \hat{z}  = -\sin\theta\, \hat{x}' + \cos\theta\, \hat{z}' \ , \nonumber \\
\hat{q} & = \sin\theta\, \hat{x} + \cos\theta\, \hat{z} = \hat{z}' \ .
\label{eq:D3}
\end{eqnarray}

Inserting Eqs.~(\ref{eq:D2},\ref{eq:D3}) into Eq.~(\ref{Jampl-NL}), we can express the photoproduction amplitude $\hat{M}$ for a given photon polarization in the primed reference frame as
\begin{equation}
\hat M^\lambda \equiv \sum_{m'=0}^3 M^\lambda_{m'}\sigma_{m'} \ ,
\label{eq:D4} 
\end{equation}
where
\begin{eqnarray}
\fl M^\perp_{0'} & \equiv  M_{0'} \, = \, iF_2\sin\theta  = M_0 \ , \qquad&
M^\parallel_{1'} \, \equiv \, M_{1'} \, = \, F_1\cos\theta  - F_3\sin^2\theta \ , \nonumber \\
\fl M^\perp_{2'} & \equiv  M_{2'} \, = \, F_1  = M_2 \ , \qquad&
M^\parallel_{3'} \, \equiv \, M_{3'} \, = \, [F_1 + F_3\cos\theta + F_4]\sin\theta  \ , \nonumber \\
\fl M^\perp_{1'} & =  M^\perp_{3'} \, = \, M^\parallel_{0'} \, = \, M^\parallel_{2'} = 0 \ .
\label{eq:D5}
\end{eqnarray}

We are now in position to express all the observables in the rotated primed frame, $\{\hat{x}', \hat{y}', \hat{z}'\}$. To this end, we simply make the substitution $M_i \to M_{i'}$ in the observables in the unprimed frame, $\{\hat{x}, \hat{y}, \hat{z}\}$, considered in the previous sections. In particular, from Eq.~(\ref{MLG-CT}), we have
\begin{eqnarray}
 R^c_{x'}  &= -Re[M_{2'}M^*_{3'}] - Im[M_{0'}M^*_{1'}] \ , \nonumber \\
R^c_{z'} & = Re[M_{1'}M^*_{2'}] - Im[M_{0'}M^*_{3'}]\ ,  \nonumber \\
 R^{l'}_{x'} & =  - ( Re[M_{0'}M^*_{1'}] - Im[M_{2'}M^*_{3'}] ) \sin(2\phi')  \ , \nonumber \\
 R^{l'}_{z'} &=  - (Re[M_{0'}M^*_{3'}] - Im[M_{1'}M^*_{2'}])\sin(2\phi') \ ,
\label{eq:D6}
\end{eqnarray}
where $\phi'$ plays the same role in the rotated frame as the angle $\phi$ introduced in Eq.~(\ref{trans1}).

Similarly, since
\begin{eqnarray}
\sigma_{x'} & = \cos\theta\, \sigma_{x} - \sin\theta\, \sigma_{z} \ , \nonumber \\
\sigma_{z'} & = \sin\theta\, \sigma_{x} + \cos\theta\, \sigma_{z} \ , \nonumber \\
\sigma_{y'} & = \sigma_{y}  \ ,
\label{eq:D7}
\end{eqnarray}
we have
\begin{eqnarray}
K_{xx'} & = \cos\theta\, K_{xx} - \sin\theta\, K_{xz} \ , \ \ \ \ \ \ \ \
K_{xz'} = \sin\theta\, K_{xx} + \cos\theta\, K_{xz} \ , \nonumber \\
K_{zx'} & = \cos\theta\, K_{zx} - \sin\theta\, K_{zz} \ , \ \ \ \ \ \ \ \
K_{zz'} = \sin\theta\, K_{zx} + \cos\theta\, K_{zz} \ .
\label{eq:D8}
\end{eqnarray}

\vskip 0.5cm
In terms of the amplitudes $F_i$ in Eq.~(\ref{Jampl-NL}), we have, from Eqs.~(\ref{eq:D5},\ref{eq:D6}),
\begin{eqnarray}
\fl \frac{d\sigma}{d\Omega}R^c_{x'} & \equiv \frac{d\sigma}{d\Omega}C_{x'}  =  -|F_1|^2\sin\theta  - Re\left[(F_2+F_3)F^*_1\cos\theta \right. \nonumber \\
& \hspace{5.5 cm}  \left. + (F^*_1F_4 - F^*_2F_3\sin^2\theta)\right] \sin\theta  \ , \nonumber \\
\fl \frac{d\sigma}{d\Omega}R^c_{z'} & \equiv \frac{d\sigma}{d\Omega}C_{z'}  =  |F_1|^2\cos\theta  - Re\left[F_1^*(F_2 + F_3) + F_2^*(F_3\cos\theta +F_4) \right]\sin^2\theta  \ , \nonumber \\
\fl \frac{d\sigma}{d\Omega}R^{l'}_{x'}(\phi'=\frac{\pi}{4}) & \equiv \frac{d\sigma}{d\Omega}O_{x'}  = -Im\left[(F_2 -F_3)F^*_1\cos\theta - (F^*_1F_4 + F_2F^*_3\sin^2\theta)\right] \sin\theta  \ , \nonumber \\
\fl \frac{d\sigma}{d\Omega}R^{l'}_{z'}(\phi'=\frac{\pi}{4}) & \equiv \frac{d\sigma}{d\Omega}O_{z'}  =  - Im\left[(F_2 - F_3)F^*_1 - (F_3\cos\theta +F_4)F^*_2 \right]\sin^2\theta  \ . \nonumber \\
\label{eq:Cn1pCn3p}
\end{eqnarray}
and, from Eqs.~(\ref{eq:TniLni},\ref{eq:D8}),
\begin{eqnarray}
\fl \frac{d\sigma}{d\Omega} K_{xx'}  \equiv \frac{d\sigma}{d\Omega} T_{x'} = \cos\theta\, K_{xx} - \sin\theta\, K_{xz} \nonumber \\
\fl \qquad\ \ \ \  = Re\big[ F_1(F^*_2 - F^*_3) - F_3F^*_4 \big] \sin^2\theta   
+ \frac12\Big(|F_2|^2 - |F_3|^2 - |F_4|^2 \Big) \cos\theta\sin^2\theta  \ , 
\nonumber \\
\fl \frac{d\sigma}{d\Omega} K_{xz'}  \equiv \frac{d\sigma}{d\Omega} T_{z'} = \sin\theta\, K_{xx} + \cos\theta\, K_{xz} \nonumber \\
\fl \qquad\ \ \ \
= - Re\big[ F_1(F^*_2 - F^*_3)\cos\theta - F_1F^*_4 \big] \sin\theta 
+ \frac12\Big(|F_2|^2 - |F_3|^2 + |F_4|^2\Big) \sin^3\theta  \ ,
\nonumber \\
\fl \frac{d\sigma}{d\Omega} K_{zx'}  \equiv \frac{d\sigma}{d\Omega} L_{x'} = \cos\theta\, K_{zx} - \sin\theta\, K_{zz} \nonumber \\
 \fl \qquad\ \ \ \
= |F_1|^2\sin\theta + Re\big[F_1(F^*_2+F^*_3+F^*_4) \big]\sin\theta -\frac12\Big(|F_2|^2 + |F_3|^2 - |F_4|^2 \Big)\sin^3\theta \ , 
\nonumber \\
\fl \frac{d\sigma}{d\Omega} K_{zz'}  \equiv \frac{d\sigma}{d\Omega} L_{z'} = \sin\theta\, K_{zx} + \cos\theta\, K_{zz} \nonumber \\
\fl \qquad\ \ \ \
= -|F_1|^2\cos\theta + Re\big[F_1(F^*_2+F^*_3) + F_3F^*_4\big]\sin^2\theta \nonumber \\
+ \frac12 \Big( |F_2|^2 + |F_3|^2 + |F_4|^2\cos\theta\Big)\sin^2\theta \ .
\nonumber \\
\label{eq:TnipLnip}
\end{eqnarray}

\section{Photoproduction near threshold} \label{sec:PW}

Consider the partial waves with the orbital angular momentum $l \le 3$ 
in the final state.
\footnote{Here, we note that, if the expansion were made in $j = l \pm \frac12 \le \frac52$ 
instead of in $l \le 3$, then, no $E_{3+}$ and $M_{3+}$ multipoles would enter in any of the 
expressions in this section.}
Then, from Eq.~(\ref{Mult-ampl}) (apart from the overall factor $-iN$),
\begin{eqnarray}
 F_1 & =  \left[E_{0+} - \frac{3}{2}E_{2+} + E_{2-} - 3\left(M_{2+}-M_{2-}\right)\right] \nonumber \\
     & + \left[3E_{1+} + M_{1+} - M_{1-} + 3E_{3-} - \frac{15}{2}E_{3+} - \frac{33}{2}\left(M_{3+}-M_{3-}\right)\right]x  \nonumber \\
& +   3\left[\frac{5}{2}E_{2+} + 2\left(M_{2+} - M_{2-}\right)\right]x^2 
 + \frac{5}{2}\left[ 7E_{3+} + 9\left(M_{3+}-M_{3-}\right)\right]x^3 \ , \nonumber \\
 F_2 & =  \left[ 2M_{1+} + M_{1-}  -\frac{9}{2}M_{3-} - 6M_{3+} \right] \nonumber \\
& + 3\left[3M_{2+}+2M_{2-}\right]x
 + 5\left[\frac{9}{2}M_{3-}+6M_{3+}\right]x^2 \ , \nonumber \\
 F_3 & =  \left[ 3E_{1+} - M_{1+} + M_{1-} - 3\left(\frac{5}{2}E_{3+}-E_{3-}\right)  + \frac{3}{2}\left(M_{3+} - M_{3-}\right) \right] \nonumber \\
 & + 3\left[5E_{2+} - 2\left(M_{2+}-M_{2-}\right) \right] x 
  + \frac{15}{2} \left[ 7E_{3+} - 3\left(M_{3+}-M_{3-}\right) \right]x^2\ , \nonumber \\
 F_4 & =  -3\left[E_{2+} + E_{2-} - M_{2+} + M_{2-}\right]
  -15\left[E_{3+}+E_{3-} -M_{3+}+M_{3-}\right]x \ . \nonumber \\
\label{NL_mult}
\end{eqnarray}

Writing now Eq.~(\ref{NL_mult}) in a short hand notation, we have
\begin{eqnarray}
F_1 & = & \left(\alpha_0 + \tilde\alpha_2\right) + \left(\alpha_1 + \tilde{\alpha}_3\right)x + \alpha_2 x^2 + \alpha_3 x^3 \ , \nonumber \\ 
F_2 & = & \left(\beta_1+\tilde{\alpha}_3\right)  + \beta_2 x + \beta_3 x^2 \ , \nonumber \\
F_3 & = & \delta_1 + x\delta_2 \ , \nonumber \\  
F_4 & = & \rho_2 + \rho_3 x \ , \nonumber \\
\label{NL_mult1}
\end{eqnarray}
where $\alpha_l$, $\tilde\alpha_l$, $\beta_l$, $\delta_l$ and $\rho_l$ 
can be read off of Eq.~(\ref{NL_mult}).

\vskip 0.5cm
First, we consider  the cross section and 
single-spin asymmetries, according to Eqs.~(\ref{xsc},\ref{asym}), taking the partial waves  $l \le 2$. We have, 
\begin{eqnarray}
\fl \frac{d\sigma}{d\Omega}  =   |\alpha_0 + \tilde\alpha_2 + \alpha_1\cos\theta + \alpha_2\cos^2\theta|^2
+ \frac12 \Big\{ |\beta_1 + \beta_2\cos\theta|^2  + |\delta_1 + \delta_2\cos\theta|^2 + |\rho_2|^2  \nonumber \\
\fl \qquad +   2Re\left[\left(\alpha_0 + \tilde\alpha_2 + \left(\alpha_1 + \delta_1\right)\cos\theta + 
\left(\alpha_2 + \delta_2\right)\cos^2\theta\right)\rho_2^*\right] \Big\}\sin^2\theta \ , \nonumber \\
\fl \frac{d\sigma}{d\Omega}\Sigma  =  \frac12 \left\{ |\beta_1 + \beta_2\cos\theta|^2 
- |\delta_1 + \delta_2\cos\theta|^2 - |\rho_2|^2 \right. \nonumber \\
\fl \qquad  -  \left. 2Re\left[\left(\alpha_0 + \tilde\alpha_2 + 
\left(\alpha_1 + \delta_1\right)\cos\theta + 
\left(\alpha_2 + \delta_2\right)\cos^2\theta\right)\rho_2^*\right] \right\}\sin^2\theta 
\ , \nonumber \\
\fl \frac{d\sigma}{d\Omega}T  =  Im\left[ 
\left(\delta_1 - \beta_1 + \left(\delta_2 - \beta_2 + \rho_2\right)\cos\theta\right)
\left(\alpha_0 + \tilde\alpha_2 + \alpha_1\cos\theta + \alpha_2\cos^2\theta\right)^* \right. 
\nonumber \\
\fl \qquad +  \left. \left(\delta_1 + \left(\delta_2 + \rho_2\right)\cos\theta\right)
\rho_2^*\sin^2\theta\right] \sin\theta 
\ , \nonumber \\
\fl \frac{d\sigma}{d\Omega}P  =  - Im\left[ 
\left(\delta_1 + \beta_1 + \left(\delta_2 + \beta_2 + \rho_2\right)\cos\theta\right)
\left(\alpha_0 + \tilde\alpha_2 + \alpha_1\cos\theta + \alpha_2\cos^2\theta\right)^* \right. 
\nonumber \\
\fl \qquad +  \left. \left(\delta_1 + \left(\delta_2 + \rho_2\right)\cos\theta\right)
\rho_2^*\sin^2\theta\right] \sin\theta \ . \nonumber \\
\label{asym1}
\end{eqnarray}

If we consider only the $S$- and $P$-waves in the final state, the above results reduce to
\begin{eqnarray}
\frac{d\sigma}{d\Omega} & = &  
 |\alpha_0 + \alpha_1\cos\theta|^2
+ \frac12 \left( |\beta_1|^2 + |\delta_1|^2 \right)\sin^2\theta  \ , \nonumber \\
\frac{d\sigma}{d\Omega}\Sigma & = & 
\frac12 \left( |\beta_1|^2 - |\delta_1|^2\right)\sin^2\theta 
\ , \nonumber \\
\frac{d\sigma}{d\Omega}T & = & Im\left[ 
\left(\delta_1 - \beta_1 \right)\left(\alpha_0 + \alpha_1\cos\theta\right)^* \right] \sin\theta
\ , \nonumber \\
\frac{d\sigma}{d\Omega}P & = & - Im\left[ 
\left(\delta_1 + \beta_1 \right)\left(\alpha_0 + \alpha_1\cos\theta\right)^* \right] \sin\theta \ . \nonumber \\
\label{asym2}
\end{eqnarray}

Another possible scenario is to keep only the $S$-wave and its interference 
with the $P$- and $D$-waves, as has been done in Ref.~\cite{TDK99}, which 
is motivated by the fact that, near threshold $\eta$ photoproduction is 
dominated by the $S$-wave final state. We then have
\begin{eqnarray}
\frac{d\sigma}{d\Omega} & = &  
 |\alpha_0|^2 + Re\left[
\alpha_0^*\left\{ \left(2\tilde\alpha_2 + \rho_2\right) + 2\alpha_1\cos\theta + 
\left(2\alpha_2 - \rho_2\right)\cos^2\theta \right\} \right]  \ , \nonumber \\
\frac{d\sigma}{d\Omega}\Sigma & = & -Re\left[\alpha_0^* \rho_2 \right]\sin^2\theta \ , 
\nonumber \\
\frac{d\sigma}{d\Omega}T & = & Im\left[\alpha_0^* 
\left\{\delta_1 - \beta_1 + \left(\delta_2 - \beta_2 + \rho_2\right)\cos\theta\right\} 
\right] \sin\theta 
\nonumber \\
\frac{d\sigma}{d\Omega}P & = & - Im\left[\alpha_0^* 
\left\{\delta_1 + \beta_1 + \left(\delta_2 + \beta_2 + \rho_2\right)\cos\theta\right\} 
\right] \sin\theta \ . \nonumber \\
\label{asym3}
\end{eqnarray}

It is interesting to note that Eqs.~(\ref{asym2}, \ref{asym3}) reveal that the two scenarios
considered above lead exactly to the same angular dependences for the corresponding 
observables considered here. However, the dynamics is very different from each other. For 
example, while the beam asymmetry, $\Sigma$, is due to the difference of the squared 
magnitudes of the $P$-wave multipoles in the former scenario (cf. Eq.~(\ref{asym2})), it is 
entirely due to the interference between the $S$- and $D$-waves in the later scenario 
(cf. Eq.~(\ref{asym3})). The $\cos\theta\sin\theta$ terms in both the target ($T$) and recoil ($P$) 
asymmetries arise from the interference among the $P$-wave multipoles
in the former scenario, while in the later scenario, they arise from the 
interference of the $S$- and $D$-wave multipoles. In other words, the assumption of the $S$-wave 
dominance alone, in conjunction with the observables considered above, does not constrain 
the presence or absence of the $D$-wave contribution. 
To do this, we require double-polarization observables. For example, the combination of the double-spin asymmetries given by Eq.~(\ref{eq:OzG}), 
\begin{equation}
\fl \frac{d\sigma}{d\Omega}(G - O_z) = \left\{
\begin{array}{ll}
0 \ , & S-P \ {\rm scenario} \\
2Im\big[ \left(  \alpha_0 + \tilde\alpha_2 + x^2\alpha_2 \right) \rho_2^* \big] \sin^2\theta\ , & S-D \ {\rm scenario}
\end{array}
\right.
\label{eq:OzG-1}
\end{equation}
will tell us about the presence or absence of the $D$-wave for, $F_4$ contains no lower partial waves than $l=2$.

The above consideration reveals that even when the partial waves are restricted to $l\le 2$, which is expected to suffice for low energies close to threshold in most cases, the unpolarized cross section and single-polarization observables alone are - strictly speaking - not sufficient to constrain the reaction amplitude model independently. For this, double-polarization observables are required.

\vskip 0.5cm
Although, in general, we do not expect the higher partial-waves to influence the results at low energies close to threshold, the issue of when the higher partial-waves start to become significant as the energy increases depends, actually, on the particular dynamics of the reaction processes. For example, 
in $\eta'$ photoproduction the measured beam asymmetry ($\Sigma$) exhibits a nearly $\cos\theta\sin^2\theta$ dependence at an excess energy of $Q \equiv W-W_{thr}=7$ MeV only \cite{GRAAL15}, which may indicate a possible contribution of even higher partial-waves than the $D$-wave. Moreover, the observed angular dependence becomes less pronounced at $Q=16$ MeV.
Although this problem requires further investigation to be settled, the Mainz group \cite{TGKNOHOOSS18}, based on their isobar model (etaMAID2018),  explained this peculiar angular and energy behavior as the $S_{11}-F_{15}$ interference with a narrow $S_{11}(1900)$ resonance.
If this interpretation holds, it means a presence of the $F$-wave even at very close to the threshold energy.  
The Bonn-Gatchina group \cite{BnGa18}, on the other hand, describes the measured beam asymmetry by a $P_{13}-D_{13}$ interference with a narrow $D_{13}(1900)$ resonance. 
Higher partial-wave contributions very close to threshold energy can be observed also in other reactions (see, e.g., Ref.\cite{JOHN15}). 

In the remaining of this section, we make a model-independent analysis of the peculiar angular behavior of the beam asymmetry in $\eta'$ photoproduction mentioned above. To this end, we consider, not only the beam asymmetry - more precisely, the unnormalized beam asymmetry $(d\sigma_o/ d\Omega) \Sigma$ - but also the cross section, $d\sigma_o/d\Omega$, and the combinations $d\sigma_o/d\Omega(1 \pm \Sigma)$.
First, we fit the available data for these quantities at $W=1903$ and $1912$ MeV with  a polynomial function of the form ($x=\cos\theta$)
\begin{equation}
y  = a_0 + a_1 x + a_2 x^2 + (a_3x + a_4x^2+a_5x^3+a_6 x^4)(1-x^2) \ ,
\label{eq:fit}
\end{equation}
which includes partial waves through $l=3$.  The purpose of the fit is to extract the coefficient $a_3$ of the term proportional to $x(1-x^2)=\cos\theta\sin^2\theta$ from each of the quantities mentioned above.
Since Ref.\cite{GRAAL15} reports only the beam asymmetry $\Sigma$ (at $W=1903$ and 1912 MeV) and not the cross section,  we have multiplied $\Sigma$ by the cross section measured by the A2 Collaboration at MAMI \cite{A2MAMI17} (at $W=1904$ and 1912 MeV, respectively) to obtain the corresponding unnormalized beam asymmetries. 
%
\begin{figure}[t] 
\includegraphics[height=0.5\textwidth,width=0.45\textwidth,clip=1]{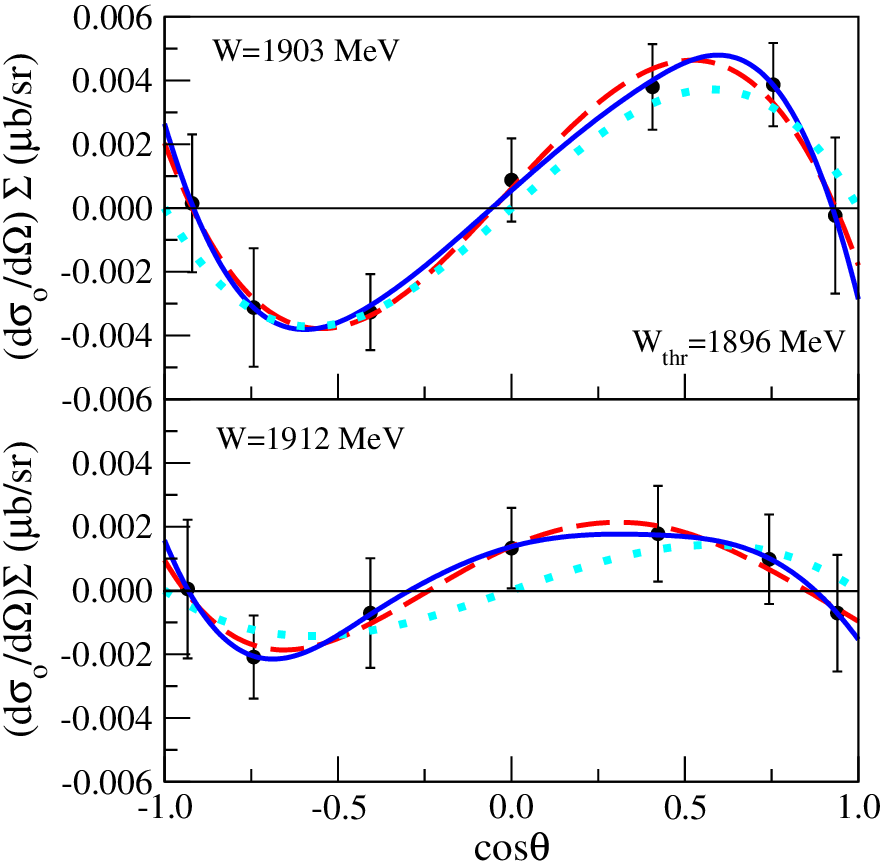} \hspace{1cm}
\includegraphics[height=0.5\textwidth,width=0.45\textwidth,clip=1]{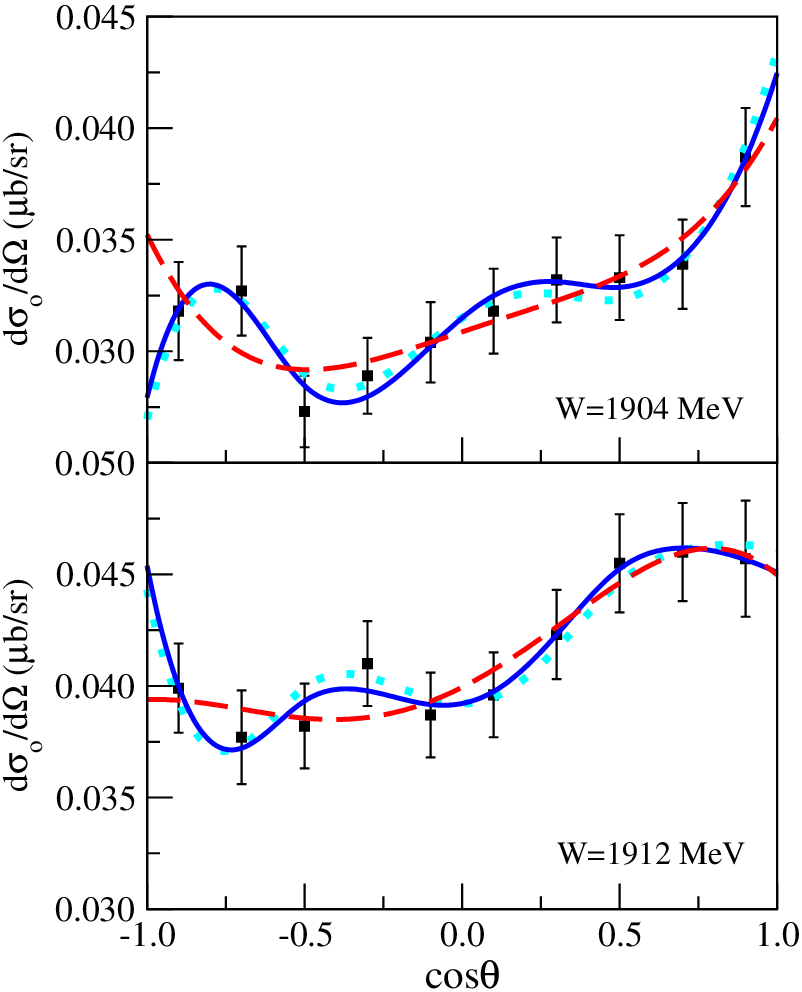} 
\caption{
Fit results for $(d\sigma_o/d\Omega) \Sigma$ and $d\sigma_o/d\Omega$ by a polynomial function of the form given by Eq.(\ref{eq:fit}).
The dashed red lines correspond to the results including through the $D$-wave ($a_5=a_6=0$). The solid blue lines include through the $F$-wave; for $(d\sigma_o/d\Omega) \Sigma$, $a_4=0$ in the fit at $W=1903$ MeV, while $a_6=0$ at $W=1912$ MeV.
The dotted cyan curves for $(d\sigma_o/d\Omega) \Sigma$ correspond to setting all the coefficients $a_i$'s in Eq.(\ref{eq:fit}) to zero, except for $a_3$; for the cross sections, they are the same as the solid blue curves, except for setting $a_3=0$.
The cross section data are from \cite{A2MAMI17}. The beam asymmetry data from Ref.\cite{GRAAL15} where multiplied by the corresponding cross section data from \cite{A2MAMI17} to obtain the unnormalized beam asymmetry  $(d\sigma_o/d\Omega) \Sigma$.}    
\label{fig:1}
\end{figure}
%
In Fig.\ref{fig:1}, we show the fit results of the data for $(d\sigma_o/ d\Omega) \Sigma$ and $d\sigma_o/ d\Omega$. 
As we can seen, $(d\sigma_o/ d\Omega) \Sigma$ is well described by the fits. The dashed red curves correspond to the fit including partial waves through $D$-wave ($a_5=a_6=0$). We mention that, at $W=1903$ MeV, a fit result with $a_4=a_5=a_6=0$ (not shown in the figure) is indistinguishable from the dashed red curve, indicating that the $a_4$ term is insignificant at this energy. The solid blue curve, at $W=1903$ MeV, corresponds to including partial waves through $F$-wave but setting $a_4=0$ while, at $W=1912$ MeV, it corresponds to setting $a_6=0$ which has a very small contribution. The dotted cyan curves correspond to the fit with only the $a_3x(1-x^2) = a_3\cos\theta\sin^2\theta$ term.  They illustrate the dominance of this term in this observable.  We see that the data cannot distinguish the different fits shown. 
For the cross sections, we also see that different fits considered describe the data very well overall, although the dashed red curves, corresponding to $l \le 2$ ($a_5=a_6=0$), miss one(two) data point(s) slightly at $W=1912$(1904) MeV. They hint for a presence of the $F$-wave as shown by the solid blue curves with $l \le 3$. 
The dotted cyan curves here are the same as the solid blue curves, except for setting $a_3=0$. They illustrate that the $\cos\theta\sin^2\theta$ term is relatively small, unlike in the beam asymmetry. 
The dashed red and solid blue curves in Fig.\ref{fig:2}  correspond to the fit results for the combinations $d\sigma_o/d\Omega(1\pm\Sigma)$ with $l \le 2$ and $l\le 3$, respectively. Again, the fits at $W=1903$ MeV hint for a presence of the $F$-wave. The extracted values of the coefficient $a_3$ from the two observables and their combinations considered are displayed in Table.\ref{tab:fit}. The values of $a_3$ corresponding to different fit results lay within the error bars quoted in that table. 
%
\begin{figure}[t] 
\hspace{3cm}
\includegraphics[height=0.5\textwidth,clip=1]{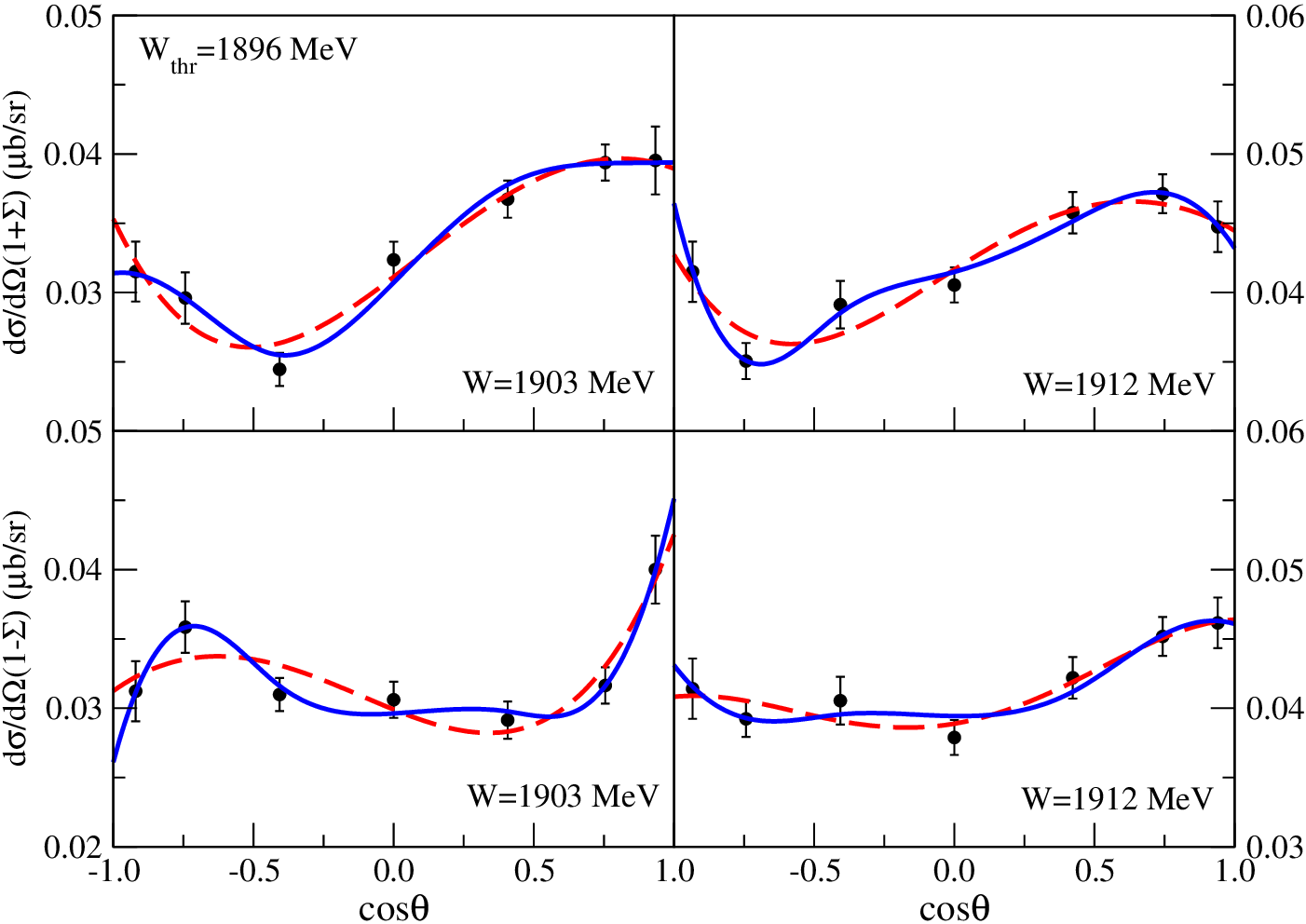} 
\caption{Same as Fig.\ref{fig:1} for the combinations $d\sigma_o/d\Omega(1 \pm\Sigma)$.} 
\label{fig:2}
\end{figure}
%

%
\begin{table}[h]
\caption{\label{tab:fit}Extracted values of the coefficient $a_3$ in Eq.(\ref{eq:fit}) in units of $[(|\vec{q}|/|\vec{k}|) nb]$ corresponding to the fit  results shown in Figs.\ref{fig:1} and \ref{fig:2}. The notations used for $a_3$ are: 
${a_3}^\sigma$ from $d\sigma_o/d\Omega$,  ${a_3}^\Sigma$  from  $(d\sigma_o/ d\Omega) \Sigma$, and  ${a_3}^{\Sigma \pm}$ from $d\sigma_o/d\Omega(1\pm \Sigma)$. The predicted results from the partial-wave interferences as explained in the text are also shown in bold face. }
\begin{indented}\item[]
\begin{tabular}{@{}cccccc}
\br
 $W$ (MeV) &  & ${a_3}^\sigma$ & ${a_3}^\Sigma$  & ${a_3}^{\Sigma +}$ & ${a_3}^{\Sigma -}$  \\
\mr
 1903  & fit & $27.1 \pm 20.8$ &  $104.1 \pm 11.5$  & $133.1 \pm 56.2$ & $-75.2\pm 35.7$ \\
          & $S_{11}-F_{15}$ & $27.1 \pm 20.8$ & $104.1 \pm 11.5$  & ${\bf 131.2 \pm 32.3}$ & ${\bf -77.0 \pm \bf 32.3}$ \\
          & $S_{11}-F_{17}$ & $27.1 \pm 20.8$ & $104.1 \pm 11.5$  & ${\bf 131.2 \pm 32.3}$ & ${\bf -217.3 \pm \bf 38.9}$ \\
          & $P_{11}-D_{15}$ & $27.1 \pm 20.8$ & $104.1 \pm 11.5$  & ${\bf 131.2 \pm 32.3}$ & ${\bf -77.0 \pm \bf 32.3}$ \\
          & $P_{13}-D_{13}$ & $27.1 \pm 20.8$ & $104.1 \pm 11.5$  & $133.1 \pm 56.2$ & ${\bf -75.1 \pm \bf 79.2}$ \\
\mr
1912 & fit  & $21.4 \pm 15.6$ &  $27.1 \pm 8.3$ & $44.7 \pm 23.8$ & $-9.9 \pm 4.7$ \\
         & $S_{11}-F_{15}$ & $21.4 \pm 15.6$ & $27.1 \pm 8.3$  & ${\bf 48.5 \pm 23.9}$ & ${\bf -5.7 \pm \bf 23.9}$ \\
         & $S_{11}-F_{17}$ & $21.4 \pm 15.6$ & $27.1 \pm 8.3$  & ${\bf 48.5 \pm 23.9}$ & ${\bf -47.6 \pm \bf 28.4}$ \\
         & $P_{11}-D_{15}$ & $21.4 \pm 15.6$ & $27.1 \pm 8.3$  & ${\bf 48.5 \pm 23.9}$ & ${\bf -5.7 \pm \bf 23.9}$ \\
         & $P_{13}-D_{13}$ & $21.4 \pm 15.6$ & $27.1 \pm 8.3$  & $44.7 \pm 23.8$ & ${\bf -9.5 \pm \bf 40.1}$ \\
\br
\end{tabular}
\end{indented}
\end{table}
%

There are a number of partial-wave interferences that can, in principle, provide the observed $\cos\theta\sin^2\theta$ behavior in the beam asymmetry. For partial waves with $l \le 3$,  the possible interferences are listed in Table.\ref{tab:ResInt}, together with the products of the corresponding multipole amplitudes contributing to the cross section, unnormalized beam asymmetry and their combinations.
These product amplitudes are obtained with the help of Eqs.(\ref{xsc},\ref{asym},\ref{NL_mult}) (cf. also Eq.(\ref{M-obsF})).
%
\hspace{5cm}
\begin{table}[h]
\caption{\label{tab:ResInt} Partial-wave interferences with $l \le 3$ leading to the $\cos\theta\sin^2\theta$ dependence in the cross section, unnormalized beam asymmetry and their combinations as indicated. These quantities are expressed here in terms of the multipole amplitudes. The $P_{11}-D_{13}$ and $D_{15}-F_{17}$ interferences do not contribute to the $\cos\theta\sin^2\theta$ term in the beam asymmetry.}
\begin{indented}\item[]
\begin{tabular}{@{}ll}
 \br
 interference & $d\sigma_o/d\Omega$   \\ 
\mr
 $S_{11}-F_{15}$  & $15 Re[E^*_{0+}(2M_{3-} - E_{3-})]$  \\
 $S_{11}-F_{17}$  & $-5Re[E^*_{0+}(6 M_{3+} + 10 E_{3+})]$  \\
 $P_{11}-D_{15}$ & $15Re[M^*_{1-}(M_{2+}+2E_{2+})]$  \\
 $P_{13}-D_{13}$ & $18Re[E^*_{1+}(2M_{2-}-E_{2-}) + M^*_{1-}M_{2-}]$  \\
 $P_{13}-D_{15}$ & $3Re[-6E^*_{1+}(2M_{2+}+E_{2+}) + 2M^*_{1+}(2M_{2+}-5E_{2+})]$  \\
 $D_{13}-F_{15}$ & $3Re[2E^*_{2-}(5M_{3-}+2E_{3-}) + 6M^*_{2-}(2E_{3-}-M_{3-})]$  \\
 $D_{13}-F_{17}$ & $-5Re[2E^*_{2-}(3M_{3+}-4E_{3+})-3M^*_{2-}(E_{3+}-3M_{3+})]$ \\
 $D_{15}-F_{15}$ & $-\frac92Re[-E^*_{2+}(11E_{3-}+5M_{3-}) + M^*_{2+}(8E_{3-}+11M_{3-})]$ \\
 \br
 interference & $(d\sigma_o/d\Omega)\Sigma$  \\
\mr
 $S_{11}-F_{15}$  & $15 Re[E^*_{0+}(M_{3-} + E_{3-})]$ \\
 $S_{11}-F_{17}$  & $-5Re[3E^*_{0+}(M_{3+}-E_{3+})]$ \\
 $P_{11}-D_{15}$ & $15Re[M^*_{1-}(M_{2+} - E_{2+})]$ \\
 $P_{13}-D_{13}$ & $18Re[E^*_{1+}E_{2-} + M^*_{1-}M_{2-}]$ \\
 $P_{13}-D_{15}$ & $3Re[-9E^*_{1+}E_{2+} + M^*_{1+}(4M_{2+} + 5E_{2+})]$ \\
 $D_{13}-F_{15}$ & $3Re[E^*_{2-}(5M_{3-}-4E_{3-}) + 9M^*_{2-}M_{3-}]$ \\
 $D_{13}-F_{17}$ & $-5Re[3E^*_{2-}(M_{3+}+5E_{3+})+18M^*_{2-}M_{3+}]$ \\
 $D_{15}-F_{15}$ & $-\frac{9}{2}Re[21(E^*_{2+}E_{3-} + M^*_{2+}M_{3-})]$ \\
 \br
interference & $d\sigma_o/d\Omega(1+\Sigma)$ \\
\mr
 $S_{11}-F_{15}$  & $45 Re[E^*_{0+} M_{3-}]$ \\
 $S_{11}-F_{17}$  & $-5Re[E^*_{0+}(9M_{3+}+7E_{3+})]$ \\
 $P_{11}-D_{15}$ & $15Re[M^*_{1-}(2M_{2+} + E_{2+})]$ \\
 $P_{13}-D_{13}$ & $18Re[2(E^*_{1+} + M^*_{1-})M_{2-}]$ \\
 $P_{13}-D_{15}$ & $3Re[-3E^*_{1+}(4M_{2+}+5E_{2+}) + M^*_{1+}(8M_{2+}-5E_{2+})]$ \\
 $D_{13}-F_{15}$ & $3 Re[15E^*_{2-}M_{3-}+ 3M^*_{2-}(M_{3-}+4E_{3-})]$ \\
 $D_{13}-F_{17}$ & $-5Re[E^*_{2-}(9M_{3+}+7E_{3+})+3M^*_{2-}(9M_{3+}-E_{3+})]$ \\
 $D_{15}-F_{15}$ & $-\frac92 Re[5E^*_{2+}(2E_{3-}-M_{3-}) + 8M^*_{2+}(4M_{3-}+E_{3-})]$ \\
 \br
 interference & $d\sigma_o/d\Omega(1-\Sigma)$  \\ 
\mr
 $S_{11}-F_{15}$  & $15 Re[E^*_{0+}(M_{3-}- 2E_{3-}]$ \\
 $S_{11}-F_{17}$  & $-5Re[E^*_{0+}(-3M_{3+} + 13E_{3+})]$ \\
 $P_{11}-D_{15}$ & $45Re[M^*_{1-} E_{2+}]$ \\
 $P_{13}-D_{13}$ & $18Re[2E^*_{1+}(M_{2-} - E_{2-})]]$ \\
 $P_{13}-D_{15}$ & $3Re[3E^*_{1+}(E_{2+}-4M_{2+})-15M^*_{1+}E_{2+}]$ \\
 $D_{13}-F_{15}$ & $3 Re[E^*_{2-}(5M_{3-}+8E_{3-}) - 3 M^*_{2-}(5M_{3-}-4E_{3-})]$ \\
 $D_{13}-F_{17}$ & $-5Re[E^*_{2-}(3M_{3+}-23E_{3+})-3M^*_{2-}(3M_{3+}+E_{3+})]$ \\
 $D_{15}-F_{15}$ & $-\frac92 Re[-E^*_{2+}(32E_{3-}+5M_{3-}) - 2M^*_{2+}(5M_{3-}-4E_{3-})]$  \\
 \br
\end{tabular}
\end{indented}
\end{table}

%

Let us begin by examining the contributions of the $S_{11}-F_{15}$ interference. We see from Table.\ref{tab:ResInt} that
\begin{eqnarray}
\frac{d\sigma_o}{d\Omega} & = 15Re[E^*_{0+}(2M_{3-} - E_{3-})] =  2a - b  = {a_3}^\sigma  \ , \nonumber \\
\left(\frac{d\sigma_o}{d\Omega}\right)\Sigma & = 15Re[E^*_{0+}(M_{3-} + E_{3-})] =  a + b  = {a_3}^\Sigma \ , \nonumber \\
\left(\frac{d\sigma_o}{d\Omega}\right)(1+\Sigma) & = 45Re[E^*_{0+}M_{3-}] =  3a  = {a_3}^{\Sigma+} \ , \nonumber \\
\left(\frac{d\sigma_o}{d\Omega}\right)(1-\Sigma) & = 15Re[E^*_{0+}(M_{3-}-2E_{3-})] =  a - 2b = {a_3}^{\Sigma-}  \ , \nonumber \\
\label{eq:S11F15}
\end{eqnarray}
where $a\equiv 15Re[E^*_{0+}M_{3-}]$ and $b\equiv 15Re[E^*_{0+}E_{3-}]$. If the contribution to the $\cos\theta\sin^2\theta$ term in each of the quantities in the above equation is dominated by the $S_{11}-F_{15}$ interference, then, 
as indicated above, $a$ and $b$ can be fixed from the extracted values from the data of the two of the coefficients, say, ${a_3}^\sigma$ and ${a_3}^\Sigma$ given in Table.\ref{tab:fit}.  Once $a$ and $b$ are determined, they can be used to predict the other two coefficients, ${a_3}^{\Sigma+}$ and ${a_3}^{\Sigma-}$. The results are shown also in Table.\ref{tab:fit}. As we can see, the predicted values are in remarkable agreement with the corresponding extracted values from the data. 
Thus, the $S_{11}-F_{15}$ interference mechanism leading to the $\cos\theta\sin^2\theta$  behavior is certainly consistent with the currently existing data close to threshold. Indeed, as mentioned before, the Mainz group \cite{TGKNOHOOSS18} describes both the angular and energy dependences of the cross sections and beam asymmetries close to threshold as an $S_{11}-F_{15}$ interference with a narrow $S_{11}(1900)$ resonance.

The analysis as described above for the $S_{11}-F_{15}$ interference can be repeated for other partial-wave interferences listed in Table.\ref{tab:ResInt}. The results for the $S_{11}-F_{17}$ interference shown in Table.\ref{tab:fit}, reveals that, although the predicted ${a_3}^{\Sigma+} $ is in excellent agreement with that extracted from the data, ${a_3}^{\Sigma-}$ is clearly at odds with its value from the data at both energies.  Admittedly, the uncertainties in the extracted values of the coefficients $a_3$'s are relatively large, except for ${a_3}^\Sigma$, but the present analysis strongly suggests that the $S_{11}-F_{17}$ interference is a unlikely mechanism for the observed angular behavior in the beam asymmetry.

 The $P_{13}-D_{13}$ interference involves three independent products of multipole amplitudes ($a \equiv 18Re[E^*_{1+} E_{2-}]$,  $b \equiv 18Re[M^*_{1+}M_{2-}]$ and $c \equiv 18Re[E^*_{1+} M_{2-}]$)  for describing the $\cos\theta\sin^2\theta$ behavior of the observables in consideration as can be seen from Table.\ref{tab:ResInt}. Thus, using three of the extracted coefficients, say, ${a_3}^\sigma$, ${a_3}^\Sigma$ and ${a_3}^{\Sigma+}$ given in Table.\ref{tab:fit}, we can predict ${a_3}^{\Sigma-}$. The result is shown in Table.\ref{tab:fit}. We see that, although the uncertainty involved is large, the agreement with the extracted centroid value is excellent.  We conclude that the $P_{13}-D_{13}$ interference is also consistent with the observed angular behavior of the beam asymmetry. The Bonn-Gatchina group \cite{BnGa18}, indeed, describes the measured beam asymmetry close to threshold as the $P_{11}-D_{13}$ interference with a narrow $D_{13}(1900)$ resonance.

An inspection of the results in Table.\ref{tab:ResInt} reveals that the $P_{11}-D_{15}$ interference exhibits the same feature as the $S_{11}-F_{15}$ interference, in that, if the latter is consistent with the extracted values of $a_3$'s in Table.\ref{tab:fit}, then, the former is also consistent with the same extracted values (cf. the corresponding numerical results in Table.\ref{tab:fit} which are identical). 
Hence, the only way to distinguish between the $S_{11}-F_{15}$ and $P_{11}-D_{15}$ interferences is to look for their respective energy dependences provided by some underlying model dynamics, such as via the widths and masses of the resonances involved. In fact, the model calculations of Refs.\cite{BnGa18,TGKNOHOOSS18} strongly suggest that one of the interfering resonances should be a narrow resonance in order to be able to reproduce the observed energy dependence of the beam asymmetry.

The remaining partial-wave interferences quoted in Table.\ref{tab:ResInt} involve four independent products of multipole amplitudes. Since we have only four coefficients $a_3$'s to constrain these products of amplitudes, the present analysis cannot provide any further insights into their roles in explaining the observed behavior of the beam asymmetry close to threshold, other than the obvious fact that they are consistent with the extracted values of $a_3$'s given in Table.\ref{tab:fit}.

\section{Summary} \label{sec:Summary}

 In this note we have discussed the model-independent aspects of the reaction amplitude in pseudoscalar meson photoproduction. 
 In contrast to the earlier works, the photoproduction amplitude has been expressed in Pauli-spin basis which allows to calculate the spin-observables straightforwardly and in a quite pedestrian way.  In doing so, we made use of the fact that the underlying reflection symmetry about the reaction plane can be most conveniently and easily exploited in this basis to help finding the non-vanishing and non-redundant observables in this reaction. 
 
The problem of complete experiments has been reviewed. In particular, by expressing the photoproduction amplitude in Pauli-spin basis, we have 
identified additional sets of eight observables that can determine the photoproduction amplitude up to an arbitrary phase compared to those found in Refs.~\cite{ChT97,Nak18} using transversity basis for expressing the reaction amplitude. In addition, we found that the kinematic restrictions required for the sets of four observables to resolve the phase ambiguity of the reaction amplitude is less severe for the case of Pauli-spin amplitudes than for transversity amplitudes in general. 
On the other hand, as has been pointed out in Ref.~\cite{SHKL11} (see, also, Refs.~\cite{VRCV13,NVR15,NRIG16}), 
in view of the experimental data having finite accuracies - which limits our ability to
determine the reaction amplitude in a complete experiment - in practice, it requires to measure more observables than those of a complete experiment to determine the photoproduction amplitude. In this regard, measurements of more observables among those 16  non-redundant ones are not just desirable for consistency check purposes, but they are a necessity.  

Also, certain combinations of the observables have been shown to be useful in isolating some (low) partial-wave states to learn about the reaction dynamics. Motivated by these findings, we have carried out a model independent analysis of the peculiar angular behavior of the beam asymmetry observed in $\eta'$ photoproduction very close to threshold \cite{GRAAL15}. 
We have shown that the $P_{13}-D_{13}$, $S_{11}-F_{15}$, as well as the $P_{11}-D_{15}$ interferences - which give rise to the $\cos\theta\sin^2\theta$ angular dependence - are consistent with the extracted coefficients of the $\cos\theta\sin^2\theta$ terms from the existing data on cross section, beam asymmetry and their combinations. Moreover, the latter two interferences are shown to exhibit the identical feature as far as the $\cos\theta\sin^2\theta$ behavior is concerned. A way of distinguishing these interferences is through their energy dependences which depend on the underlying dynamics provided by some theoretical model. The present analysis rules out the $S_{11}-F_{17}$ interference as an alternative mechanism.
Recently,  the Bonn-Gatchina group \cite{BnGa18} introduced the $P_{13}-D_{13}$ interference with narrow $D_{13}(1900)$ resonance to describe both the cross section and beam asymmetry data. The Mainz group   \cite{TGKNOHOOSS18}, on the other hand, described the same data as an $S_{11}-F_{15}$ interference with
narrow $S_{11}(1900)$ resonance.
It would be very interesting for the authors of \cite{BnGa18,TGKNOHOOSS18}
to verify how well their models describe the combinations $d\sigma_o/d\Omega(1 \pm \Sigma)$,  as these provide additional information to help understand the underlying reaction dynamics.  Note, in particular, that  for the $P_{13}-D_{13}$ interference, there are three independent products of multipole amplitudes that cannot be fixed uniquely from the cross section and beam asymmetry alone. For this, one of the combinations $d\sigma_o/d\Omega(1 \pm \Sigma)$ is required.
Furthermore, we have also shown that at energies close to threshold, even for cases where the number of partial-waves are restricted to $l \le 2$, the unpolarized cross section and single-polarization observables alone are not sufficient to constrain the reaction dynamics model independently in the strict sense.  For this, double-polarization observables are required.

 The present work should be useful, especially, for newcomers in the field of baryon spectroscopy, where the photoproduction reaction processes are a major tool for probing the baryon spectra.

 \section*{Acknowledgement}
 
 We thank Igor Strakovsky for the encouragement to prepare this note and for valuable suggestions. 

\vskip 1cm

\appendix

\section{ }\label{app:1}

In this appendix we give the details how to identify all the possible sets of the minimum number of observables that determine the magnitudes and relative phases of the amplitudes $M_i\ (i=0-3)$ [cf.~Eq.~(\ref{eq:aux0})] expressed in Pauli-spin basis, apart from an arbitrary overall phase. To this end, we 
rewrite all  the 16 non-redundant observables given in Sec.~\ref{sec:NonRedundantObserv}
[cf. Eqs.~(\ref{LX-M},\ref{LS-M},\ref{T-M},\ref{P-M},\ref{Kii-M},\ref{Kij-M},\ref{MLG-BT},\ref{MLG-CT}) ] 
and group them as follows:
\begin{eqnarray}
\frac{d\sigma}{d\Omega} &  = \frac12 \left[ |M_0|^2  + |M_1|^2  + |M_2|^2 + |M_3|^2 \right]  \ ,  \nonumber \\
\Sigma &  = \frac12 \left[ |M_0|^2  - |M_1|^2  + |M_2|^2 - |M_3|^2 \right]  \ ,   \nonumber \\
T_x &  = \frac12 \left[ |M_0|^2  + |M_1|^2  - |M_2|^2 - |M_3|^2 \right]  \ ,   \nonumber \\
L_z &  = \frac12 \left[ |M_0|^2  - |M_1|^2  - |M_2|^2 + |M_3|^2 \right]  \ ,  \nonumber \\
\label{eq:comb0a}
\end{eqnarray}
\begin{eqnarray}
O^a_{1+} & \equiv  T & =  B_{02} \cos\phi_{02} + B_{13} \sin\phi_{13} \ ,   \nonumber \\
O^a_{1-} & \equiv  P & =  B_{02} \cos\phi_{02} -  B_{13} \sin\phi_{13} \ ,  \nonumber \\
O^a_{2+} & \equiv L_x & =  B_{13} \cos\phi_{13} + B_{02} \sin\phi_{02} \ ,  \nonumber \\
O^a_{2-} & \equiv  T_z & =  B_{13} \cos\phi_{13} -  B_{02} \sin\phi_{02} \ ,  \nonumber \\
\label{eq:comb1a}
\end{eqnarray}
\begin{eqnarray}
O^b_{1+} & \equiv -G & =  B_{03} \cos\phi_{03} + B_{12} \sin\phi_{12} \ ,  \nonumber \\
O^b_{1-} & \equiv -O_z & =  B_{03} \cos\phi_{03} -  B_{12} \sin\phi_{12} \ , \nonumber \\
O^b_{2+} & \equiv \ \ \ E & =  B_{12} \cos\phi_{12} + B_{03} \sin\phi_{03} \ , \nonumber \\
O^b_{2-} & \equiv  \ \ C_z & =  B_{12} \cos\phi_{12} -  B_{03} \sin\phi_{03} \ , \nonumber \\
\label{eq:comb1c}
\end{eqnarray}
\begin{eqnarray}
O^c_{1+}   & \equiv -H & =  B_{01} \cos\phi_{01} + B_{23} \sin\phi_{23} \ ,  \nonumber \\
O^c_{1-}  & \equiv -O_x & =  B_{01} \cos\phi_{01} -  B_{23} \sin\phi_{23} \ ,  \nonumber \\
O^c_{2+}  & \equiv -C_x & =  B_{23} \cos\phi_{23} + B_{01} \sin\phi_{01} \ , \nonumber \\
O^c_{2-}  & \equiv\ \ \  F &  =  B_{23} \cos\phi_{23} -  B_{01} \sin\phi_{01} \ , \nonumber \\
\label{eq:comb1b}
\end{eqnarray}
where, in Eqs.~(\ref{eq:comb1a},\ref{eq:comb1c},\ref{eq:comb1b}), the Pauli-spin amplitudes are expressed in the form given by Eq.~(\ref{eq:aux0}) and $B_{ij} \equiv B_i B_j$ and $\phi_{ij} \equiv \phi_i - \phi_j$. Also, all the spin observables above are multiplied by the unpolarized cross section. For example, $\Sigma$ should be understood as $(d\sigma / d\Omega) \Sigma$, and so on.  

The four observables in Eq.~(\ref{eq:comb0a}), the unpolarized cross section $d\sigma / d\Omega$, the single-spin observables $\Sigma$, and the double-spin observables $T_x$ and $L_z$, determine the magnitudes of the Pauli-spin amplitudes. We now search for all possible sets of four observables from the remaining three groups $\{a, b, c\}$ given by  Eqs.~(\ref{eq:comb1a},\ref{eq:comb1b},\ref{eq:comb1c}), which determine the phases of the Pauli-spin amplitudes up to an overall phase.  
There exist the following possibilities of forming a set of four observables from groups  $\{a, b, c\}$: 
\begin{itemize}
\item[1)]
Two from a given group and two from another group: $2 + 2$ case.

\item[2)]
Two from a given group and one from each of the remaining two groups: $2 + 1 + 1$ case.

\item[3)]
Three from a given group and one from another group: $3 + 1$ case.

\item[4)]   
All four observables from one group: $4$ case.

\end{itemize}

Then, in complete analogy with the derivation in Ref~\cite{Nak18} for the transversity amplitudes, we determine all the sets of observables that resolve the phase ambiguity of the Pauli-spin amplitudes. 
The results are displayed in Tables.~\ref{tab:ab}, \ref{tab:ac}, \ref{tab:bc} and \ref{tab:211}.
No sets of observables in the cases of items (3) and (4) above can resolve the phase ambiguity.

\vskip 0.5cm
From Eqs.~(\ref{eq:comb1a},\ref{eq:comb1c},\ref{eq:comb1b})), all the observables are of the form
\begin{equation}
O^m_{n\pm} = B_{ij} \cos\phi_{ij} \pm B_{kl} \sin\phi_{kl} \ .
\label{eq:PauliSpinform} 
\end{equation}
Thus, 
a pair of observables $(O^m_{n+}, O^m_{n-})$ determines $\cos\phi_{ij}$ and $\sin\phi_{kl}$ as
\begin{equation}
\cos\phi_{ij} = \frac{O^m_{n+} + O^m_{n-}}{2B_{ij}} \ , \qquad 
\sin\phi_{kl} = \frac{O^m_{n+} - O^m_{n-}}{2B_{kl}} \ ,
\label{eq:A4a}
\end{equation}
which implies that the relative phases $\phi_{ij}$ and  $\phi_{kl}$ are subject to a 2-fold ambiguity each:
\begin{equation}
\phi_{ij} = \pm \beta_{ij} , \qquad {\rm and} \qquad 
\phi_{kl} = \left\{
\begin{array}{l}
\alpha_{kl} \ , \\
\pi - \alpha_{kl} \ ,
\end{array} \right.
\label{eq:A4}
\end{equation}
where $\beta_{ij}$ and $\alpha_{kl}$ are uniquely defined with $0 \le \beta_{ij} \le \pi$ and $-\pi/2 \le \alpha_{kl} \le +\pi/2$. 
They result in 4-fold ambiguities
\begin{equation}
\fl \phi_{ij} - \phi_{kl}  =  \left\{
\begin{array}{l}
\ \ \, (\beta_{ij} - \alpha_{kl}) \ , \\
\ \ \, (\beta_{ij} + \alpha_{kl}) - \pi \ , \\
- (\beta_{ij} + \alpha_{kl}) \ , \\
- (\beta_{ij} - \alpha_{kl}) - \pi \ , 
\end{array} \right.
\quad {\rm and} \quad
\phi_{ij} + \phi_{kl}  =  \left\{
\begin{array}{l}
\ \ \, (\beta_{ij} + \alpha_{kl}) \ , \\
\ \ \, (\beta_{ij} - \alpha_{kl}) + \pi \ , \\
- (\beta_{ij} - \alpha_{kl}) \ , \\
- (\beta_{ij} + \alpha_{kl}) + \pi \ .
\end{array} \right.
\label{eq:A5}
\end{equation}

\vskip 0.5cm
For the pair of observables of the form $(O^m_{1\pm}, O^m_{2\pm})$ (here, the upper sign goes with upper sign and lower sign with lower sign),
\begin{equation}
\left\{
\begin{array}{l}
O^m_{1\pm}  = B_{ij} \cos\phi_{ij} \pm B_{kl} \sin\phi_{kl} \ ,  \\
 O^m_{2\pm} = B_{kl} \cos\phi_{kl} \pm B_{ij} \sin\phi_{ij} \ ,
 \end{array} \right.
\label{eq:PauliSpinform-a} 
\end{equation}
we have the 2-fold ambiguity 
\begin{equation}
\fl \phi_{ij} - \phi_{kl}  = \pm \left\{
\begin{array}{l}
 2\zeta -  (\beta_{ij} + \alpha_{kl})  \ , \\
 2\zeta + (\beta_{ij} + \alpha_{kl}) - \pi  \ ,
\end{array} \right. 
\qquad 
\phi_{ij} + \phi_{kl}  = \pm \left\{
\begin{array}{l}
- (\beta_{ij} - \alpha_{kl}) \ , \\
\ \ \, (\beta_{ij} - \alpha_{kl}) + \pi  \ , 
\end{array} \right. 
\label{eq:A12}
\end{equation}
where the angle $\zeta \equiv \zeta^m_{1\nu, 2\nu'}$ is uniquely defined through
\begin{equation}
\cos\zeta \equiv \frac{O^m_{1\nu}}{N}  \ , \quad\qquad \sin\zeta \equiv \frac{O^m_{2\nu'}}{N} \ , 
\label{eq:2-8a}
\end{equation}
with $N \equiv  N^m_{1\nu, 2\nu'} \equiv \sqrt{O^m_{1\nu}{^2} + O^m_{2\nu'}{^2}}$.
In the following we simply use $\zeta$ and $N$ to avoid the heavy notation, but it should be kept in mind that they depend on the given pair of observables. For the pair of observables considered above, $\nu=\nu'=\pm$. 
Here, $\beta_{ij}$ and $\alpha_{kl}$ are uniquely defined by
\begin{equation}
\cos\beta_{ij} = \frac{N^2 + B_{ij}^2 - B_{kl}^2}{2\, B_{ij}\, N} \ , \qquad 
\sin\alpha_{kl} = \frac{N^2 + B_{kl}^2 - B_{ij}^2}{2\, B_{kl}\, N} \  ,
\label{eq:2-8b}
\end{equation}
with $0 \le \beta_{ij} \le \pi$ and $-\pi/2 \le \alpha_{kl} \le +\pi/2$.

Analogously, for the pair of observables of the form $(O^m_{1\pm}, O^m_{2\mp})$, we have
\begin{equation}
\fl \phi_{ij} - \phi_{kl}  = \pm \left\{
\begin{array}{l}
\ \ \, (\beta_{ij} - \alpha_{kl}) \ , \\
- \left[ (\beta_{ij} - \alpha_{kl}) + \pi \right]  \ , 
\end{array} \right. 
\quad 
\phi_{ij} + \phi_{kl}  = \pm \left\{
\begin{array}{l}
- 2\zeta + (\beta_{ij} + \alpha_{kl})  \ , \\
- 2\zeta - \left[ (\beta_{ij} + \alpha_{kl}) - \pi \right]  \ .
\end{array} \right. 
\label{eq:A12a}
\end{equation}

\vskip 0.5cm
The determination of the phases of the photoproduction amplitude rests on whether or not the relative phase ambiguity discussed above can be resolved through one of the relations below \cite{ChT97,Nak18}:
\begin{eqnarray}
\phi_{02} - \phi_{13} & = \phi_{01} - \phi_{23} \qquad (a \leftrightarrow c)  \  ,  \nonumber \\
\phi_{02} + \phi_{13} & = \phi_{03} + \phi_{12} \qquad (a \leftrightarrow b) \  , \nonumber \\
\phi_{01} + \phi_{23} & = \phi_{03} - \phi_{12} \qquad (c \leftrightarrow b)  \  .
\label{eq:A6}
\end{eqnarray}

\vskip 0.5cm
The sets of four spin observables that resolve the phase ambiguity of the Pauli-spin amplitudes up to an arbitrary overall phase are displayed in Tables.~\ref{tab:ab}, \ref{tab:ac} and \ref{tab:bc} for the $2 + 2$ cases  mentioned in item(1) earlier in this appendix.

\vskip 0.5cm
The pair of observables of the form $(O^m_{n\nu}, O^{m'}_{n'\nu'})\ (m \ne m')$ enter in the $2 + 1 + 1$ case mentioned in item (2) in this appendix. Here, we use the same notation introduced in Ref.~\cite{Nak18}. Then,
consider, for example, the set of four observables of the form $[(O^{m''}_{1+}, O^{m''}_{1-}) , (O^m_{1+}, O^{m'}_{1+})]\ (m'' \ne m, m'\ {\rm and}\ m \ne m')$, with 
\begin{equation}
\fl \left\{
\begin{array}{l}
O^{m''}_{1+}  =  B_{ij} \cos\phi_{ij} + B_{kl} \sin\phi_{kl} \ , \\
O^{m''}_{1-}  = B_{ij} \cos\phi_{ij} - B_{kl} \sin\phi_{kl} \ , 
\end{array} \right. \nonumber \\
\qquad 
\left\{
\begin{array}{l}
O^m_{1+}   =  B_{ik} \cos\phi_{ik} + B_{jl} \sin\phi_{jl} \ , \\
O^{m'}_{1+}  = B_{il} \cos\phi_{il} + B_{kj} \sin\phi_{kj} \ .
\end{array} \right. 
\label{eq:A13}
\end{equation}
Using the relations
\begin{equation}
\phi_{il} = \phi_{ij} + \phi_{jl}  \qquad {\rm and} \qquad \phi_{kj} = \phi_{kl} - \phi_{jl} \ ,
\label{eq:A14}
\end{equation}
we obtain, from the pair $(O^m_{1+}, O^{m'}_{1+})$,
\begin{eqnarray}
\sin\phi^{\lambda \lambda'}_{jl}(\eta) & = \frac{- A^{\lambda \lambda'}_s O^{m'}_{1+} + \eta\, A_c \sqrt{D^{\lambda \lambda'\, 2}  - O^{m'}_{1+}{^2}}}{D^{\lambda \lambda'\, 2}} \ , \nonumber \\
\cos\phi^{\lambda\lambda'}_{ik}(\eta) & = \frac{O^{m}_{1+} - B_{jl} \sin\phi^{\lambda\lambda'}_{jl}(\eta)}{B_{ik}} \ ,
\label{eq:A15}
\end{eqnarray}
where $\eta$ takes the values $\pm 1$ and
\begin{eqnarray}
A_c & \equiv B_{il} \cos\phi_{ij} + B_{kj} \sin\phi_{kl}   \ , \nonumber \\  
A^{\lambda \lambda'}_s  &\equiv B_{il} \sin\phi^\lambda_{ij} + B_{kj} \cos\phi^{\lambda'}_{kl}  \ ,  \nonumber \\
D^{\lambda\lambda'\, 2} &\equiv A_c^2 + A^{\lambda\lambda'}_s{^2} =
B^2_{il} + B^2_{jk}  + 2 B_{il} B_{jk} \sin(\phi^\lambda_{ij} + \phi^{\lambda'}_{kl}) \ .
\label{eq:A16}
\end{eqnarray}
Note that, 
\begin{equation}
\phi^\lambda_{ij} = \lambda\, \beta_{ij} \ , \lambda=\pm \qquad 
{\rm and} \qquad
\phi^{\lambda'}_{kl} = \left\{
\begin{array}{ll}
\alpha_{kl} \ ,        &\ \lambda'=+ \  \\
\pi - \alpha_{kl} \ , &\  \lambda'=- \ . 
\end{array} \right.
\label{eq:A17}
\end{equation}
Also, note that the quantity $A_c$ introduced in Eq.~(\ref{eq:A16}) is independent on the indices $\lambda,\lambda'$ because $\cos\phi_{ij}$ and $\sin\phi_{kl}$ are uniquely defined by the pair of observables  $(O^{m''}_{1+}, O^{m''}_{1-})$.

From Eq.~(\ref{eq:A15}), we see that $\phi^{\lambda\lambda'}_{jl}(\eta)$ and $\phi^{\lambda\lambda'}_{ik}(\eta)$, each has a 2-fold ambiguity 
\begin{equation}
\phi^{\lambda \lambda'}_{jl}(\eta) = \left\{
\begin{array}{l}
\alpha^{\lambda \lambda'}_{jl}(\eta)  \ ,   \\
\pi - \alpha^{\lambda \lambda'}_{jl}(\eta)  \ ,
\end{array} \right.  
 \qquad {\rm and} \qquad 
\phi^{\lambda\lambda'}_{ik}(\eta) = \pm \beta^{\lambda\lambda'}_{ik}(\eta) \ ,
\label{eq:A18}
\end{equation}
with $-\pi/2 \le \alpha^{\lambda \lambda'}_{jl}(\eta) \le +\pi/2$ and $0 \le \beta^{\lambda \lambda'}_{ik}(\eta) \le \pi$.
Hence, we arrive at the possible solutions
\begin{equation}
\phi^\lambda_{ij} - \phi^{\lambda'}_{kl} = \phi^{\lambda\lambda'}_{ik}(\eta) - \phi^{\lambda\lambda'}_{jl}(\eta) = \left\{
\begin{array}{l}
\vspace{0.2cm} 
\ \ \, \left[ \beta^{\lambda\lambda'}_{ik}(\eta) - \alpha^{\lambda\lambda'}_{jl}(\eta) \right]  \ , \\
\vspace{0.2cm}
\ \ \, \left[ \beta^{\lambda\lambda'}_{ik}(\eta) + \alpha^{\lambda\lambda'}_{jl}(\eta) \right]  - \pi \ , \\
\vspace{0.2cm}
- \left[ \beta^{\lambda\lambda'}_{ik}(\eta) + \alpha^{\lambda\lambda'}_{jl}(\eta) \right] \ , \\
- \left[ \beta^{\lambda\lambda'}_{ik}(\eta) - \alpha^{\lambda\lambda'}_{jl}(\eta) \right]  - \pi \ , \\
\end{array} \right.
\label{eq:A19}
\end{equation}
for each set of $\{\lambda, \lambda', \eta\}$. For the pair $(O^{m''}_{1+}, O^{m''}_{1-})$, the difference $\phi^\lambda_{ij} - \phi^{\lambda'}_{kl}$, with the notation specified in Eq.~(\ref{eq:A17}), is given by Eq.~(\ref{eq:A5}). 
In Eq.~(\ref{eq:A19}), $\alpha^{\lambda\lambda'}_{jl}(\eta)$'s are all distinct from each other. They have the symmetries: $\alpha^{++}_{jl}(\pm) = - \alpha^{- -}_{jl}(\mp)$ and  $\alpha^{+-}_{jl}(\pm) = - \alpha^{- +}_{jl}(\mp)$. As a consequence, all $\beta^{\lambda\lambda'}_{ik}(\eta)$'s are distinct from each other.  
The results for other sets of the form $[(O^{m''}_{1+}, O^{m''}_{1-}) , (O^m_{1\nu}, O^{m'}_{1\nu'})]\ (\nu,\nu' =\pm; m'' \ne m, m'\ {\rm and}\ m \ne m')$, can be obtained from the above result for  $[(O^{m''}_{1+}, O^{m''}_{1-}) , (O^m_{1+}, O^{m'}_{1+})]$ by appropriate sign change of $B_{kl}$. 

For the set of four observables $[(O^{m''}_{2+}, O^{m''}_{2-}) , (O^m_{1\nu}, O^{m'}_{1\nu'})]\ (\nu,\nu'=\pm; m'' \ne m, m'\ {\rm and}\ \\ m \ne m')$,
we obtain the same result as for the set $[(O^{m''}_{1+}, O^{m''}_{1-}) , (O^m_{1\nu}, O^{m'}_{1\nu'})]$ given above, except that, in this case, the quantity $A_c$ defined in Eq.~(\ref{eq:A16}) becomes dependent on the indices  $\lambda, \lambda'$, while $A^{\lambda\lambda'}_s$ becomes independent on these indices ($A_c \to A^{\lambda\lambda'}_c$ and $A^{\lambda\lambda'}_s \to A_s$), since the pair of observables $(O^{m''}_{2+}, O^{m''}_{2-})$ defines $\sin\phi_{ij}$ and $\cos\phi_{kl}$ uniquely. As a consequence, in Eq.~(\ref{eq:A19}), the symmetries exhibited by the angles $\alpha^{\lambda\lambda'}_{jl}(\eta)$ are: $\alpha^{++}_{jl}(\pm) = \alpha^{- -}_{jl}(\mp)$ and  $\alpha^{+-}_{jl}(\pm) = \alpha^{- +}_{jl}(\mp)$, which implies that - unlike in the previous case discussed above - the angles $\beta^{\lambda\lambda'}_{ik}(\eta)$'s are not distinguishable from each other, i.e., $\beta^{++}_{ik}(\pm) = \beta^{- -}_{ik}(\mp)$ and  $\beta^{+ -}_{ik}(\pm) = \beta^{- +}_{ik}(\mp)$. For the pair $(O^{m''}_{2+}, O^{m''}_{2-})$, the difference $\phi^\lambda_{ij} - \phi^{\lambda'}_{kl}$ in Eq.~(\ref{eq:A19}), with the notation specified in Eq.~(\ref{eq:A17}), is given by Eq.~(\ref{eq:A5}). 

For a  set of four observables of the form $[(O^{m''}_{1\nu'''}, O^{m''}_{2\nu''}) , (O^m_{1\nu}, O^{m'}_{1\nu'})]\ (\nu''', \nu'', \nu', \nu= \pm; \\ m'' \ne m, m'\ {\rm and}\ m \ne m')$, we obtain the same results as for $[(O^{m''}_{1+}, O^{m''}_{1-}) , (O^m_{1\nu}, O^{m'}_{1\nu'})]$, except for the fact that 
the angles $\alpha^{\lambda\lambda'}_{jl}(\eta)$ and $\beta^{\lambda\lambda'}_{ik}(\eta)$ are all distinct from each other, since both $A_c$ and $A_s$ depend on the indices $\lambda,\lambda'$.  We also obtain the same feature for $\alpha^{\lambda\lambda'}_{jl}(\eta)$ and $\beta^{\lambda\lambda'}_{ik}(\eta)$ in the case of the sets $[(O^{m''}_{2\nu'''}, O^{m''}_{1\nu''}) , (O^m_{1\nu}, O^{m'}_{1\nu'})]\ (\nu''', \nu'', \nu', \nu= \pm; \\ m'' \ne m, m'\ {\rm and}\ m \ne m')$. Note that for the pairs $(O^{m''}_{n\nu}, O^{m''}_{n',\nu'})\ (n \ne n'; \nu,\nu'=\pm)$, the difference $\phi^\lambda_{ij} - \phi^{\lambda'}_{kl}$ in Eq.~(\ref{eq:A19}), with the notation specified in Eq.~(\ref{eq:A17}), are given by Eqs.~(\ref{eq:A12},\ref{eq:A12a}).

\vskip 0.5cm
Analogously, for the set of four observables of the form $[(O^{m''}_{1+}, O^{m''}_{1-}) , (O^m_{2+}, O^{m'}_{2+})]\ (m'' \ne m, m'\ {\rm and}\ m \ne m')$ with
\begin{equation}
\fl \left\{
\begin{array}{l}
O^{m''}_{1+}  =  B_{ij} \cos\phi_{ij} + B_{kl} \sin\phi_{kl} \ , \\
O^{m''}_{1-}  = B_{ij} \cos\phi_{ij} - B_{kl} \sin\phi_{kl} \ , 
\end{array} \right. \nonumber \\
\qquad 
\left\{
\begin{array}{l}
O^m_{2+}  = B_{jl} \cos\phi_{jl} + B_{ik} \sin\phi_{ik} \ , \\
O^{m'}_{2+}   =  B_{kj} \cos\phi_{kj} + B_{il} \sin\phi_{il} \ ,
\end{array} \right. 
\label{eq:A13a}
\end{equation}
we obtain 
\begin{eqnarray}
\cos\phi^{\lambda \lambda'}_{jl}(\eta) & = \frac{A^{\lambda \lambda'}_s O^{m'}_{2+} + \eta\, A_c \sqrt{D^{\lambda \lambda'\, 2}  - O^{m'}_{2+}{^2}}}{D^{\lambda \lambda'\, 2}} \ , \nonumber \\
\sin\phi^{\lambda\lambda'}_{ik}(\eta) & = \frac{O^{m}_{2+} - B_{jl} \cos\phi^{\lambda\lambda'}_{jl}(\eta)}{B_{ik}} \ ,
\label{eq:A15a}
\end{eqnarray}
where $A_c, A^{\lambda\lambda'}_s$ and $D^{\lambda\lambda'}{^2}$ are given by Eq.~(\ref{eq:A16});
$\eta$ takes the values $\pm 1$.

From the above equation, we have
\begin{equation}
\phi^{\lambda \lambda'}_{ik}(\eta) = \left\{
\begin{array}{l}
\alpha^{\lambda \lambda'}_{ik}(\eta)  \ ,   \\
\pi - \alpha^{\lambda \lambda'}_{ik}(\eta)  \ ,
\end{array} \right.  
 \qquad {\rm and} \qquad 
\phi^{\lambda\lambda'}_{jl}(\eta) = \pm \beta^{\lambda\lambda'}_{jl}(\eta) \ ,
\label{eq:A18a}
\end{equation}
with $-\pi/2 \le \alpha^{\lambda \lambda'}_{ik}(\eta) \le +\pi/2$ and $0 \le \beta^{\lambda \lambda'}_{jl}(\eta) \le \pi$. 
Hence, we arrive at the possible solutions
\begin{equation}
\phi^\lambda_{ij} - \phi^{\lambda'}_{kl} = \phi^{\lambda\lambda'}_{ik}(\eta) - \phi^{\lambda\lambda'}_{jl}(\eta) = \left\{
\begin{array}{l}
\vspace{0.2cm} 
\ \ \ \left[ \alpha^{\lambda\lambda'}_{ik}(\eta) - \beta^{\lambda\lambda'}_{jl}(\eta) \right]  \ , \\
\vspace{0.2cm}
\ \ \ \left[ \alpha^{\lambda\lambda'}_{ik}(\eta) + \beta^{\lambda\lambda'}_{jl}(\eta) \right] \ , \\
\vspace{0.2cm}
- \left[ \alpha^{\lambda\lambda'}_{ik}(\eta) + \beta^{\lambda\lambda'}_{jl}(\eta) \right]  + \pi \ , \\
- \left[ \alpha^{\lambda\lambda'}_{ik}(\eta) - \beta^{\lambda\lambda'}_{jl}(\eta) \right]  + \pi \ , \\
\end{array} \right.
\label{eq:A19a}
\end{equation}
for each set of $\{\lambda, \lambda', \eta\}$. The difference $\phi^\lambda_{ij} - \phi^{\lambda'}_{kl}$, with the notation specified in Eq.~(\ref{eq:A17}), is given by Eq.~(\ref{eq:A5}). 
The symmetries of the angles in the above equation are: $\beta^{++}_{jl}(\pm) = \pi - \beta^{- -}_{jl}(\mp)$, $\beta^{+-}_{jl}(\pm) = \pi - \beta^{+ -}_{jl}(\mp)$, and the distinct $\alpha^{\lambda\lambda'}_{ik}(\eta)$'s. 

The results for the sets $[(O^{m''}_{n\nu'''}, O^{m''}_{n'\nu''}) , (O^m_{2\nu}, O^{m'}_{2\nu'})]\ (n \ne n'; \nu''', \nu'', \nu', \nu=\pm; m'' \ne m, m'\ {\rm and}\ m \ne m')$ can be obtained from the results for $[(O^{m''}_{1+}, O^{m''}_{1-}) , (O^m_{2+}, O^{m'}_{2+})]\ (m'' \ne m, m'\ {\rm and}\ m \ne m')$ following - in complete analogy - what have been done  for obtaining the results for the corresponding sets $[(O^{m''}_{n\nu'''}, O^{m''}_{n'\nu''}) , (O^m_{1\nu}, O^{m'}_{1\nu'})]$.

\vskip 0.5cm
The remaining pairs of two spin observables are of the form $(O^m_{n\nu}, O^{m'}_{n'\nu'})\ (n \ne n'; m \ne m'\ {\rm and}\ \nu, \nu'=\pm)$. For example, consider the pair $[(O^{m''}_{1+}, O^{m''}_{1-}) , (O^m_{1+}, O^{m'}_{2+})]$
\begin{equation}
\fl \left\{
\begin{array}{l}
O^{m''}_{1+}  =  B_{ij} \cos\phi_{ij} + B_{kl} \sin\phi_{kl} \ , \\
O^{m''}_{1-}  = B_{ij} \cos\phi_{ij} - B_{kl} \sin\phi_{kl} \ , 
\end{array} \right. \nonumber \\
\qquad 
\left\{
\begin{array}{l}
O^m_{1+}   =  B_{ik} \cos\phi_{ik} + B_{jl} \sin\phi_{jl} \ , \\
O^{m'}_{2+}  = B_{kj} \cos\phi_{kj} + B_{il} \sin\phi_{il} \ .
\end{array} \right. 
\label{eq:A20}
\end{equation}
In complete analogy to the previous cases above, we obtain
\begin{eqnarray}
\sin\phi^{\lambda \lambda'}_{jl}(\eta) & = \frac{A_c  O^{m'}_{2+} + \eta\, A^{\lambda \lambda'}_s \sqrt{D^{\lambda \lambda'\, 2}  - O^{m'}_{2+}{^2}}}{D^{\lambda \lambda'\, 2}} \ , \nonumber \\
\cos\phi^{\lambda\lambda'}_{ik}(\eta) & = \frac{O^{m}_{1+} - B_{jl} \sin\phi^{\lambda\lambda'}_{jl}(\eta)}{B_{ik}} \ .
\label{eq:A21}
\end{eqnarray}
The above results lead to the same form of the solutions as given Eq.~(\ref{eq:A19}). 
 Here, the symmetries of the corresponding phases are the same to those for the set $[(O^{m''}_{2+}, O^{m''}_{2-}) , (O^m_{1\nu}, O^{m'}_{1\nu'})]$ discussed previously, i.e., 
  $\alpha^{++}_{jl}(\pm) = \alpha^{- -}_{jl}(\mp)$, $\alpha^{+-}_{jl}(\pm) = \alpha^{-+}_{jl}(\mp)$, $\beta^{++}_{ik}(\pm) = \beta^{- -}_{ik}(\mp)$ and $\beta^{+-}_{ik}(\pm) = \beta^{-+}_{ik}(\mp)$.

The results for other sets of the form $[(O^{m''}_{n\nu'''}, O^{m''}_{n'\nu''}), (O^m_{1\nu}, O^{m'}_{2\nu'})]\ (n \ne n'; \nu''', \nu'', \nu', \nu \\=\pm; m \ne m'\ {\rm and}\ \nu,\nu'=\pm)$ can be obtained from the above results for $[(O^{m''}_{1+}, O^{m''}_{1-}), \\(O^m_{1+}, O^{m'}_{2+})]$, following - in complete analogy - what have been done  for obtaining the results for the corresponding sets $[(O^{m''}_{n\nu'''}, O^{m''}_{n'\nu''}) , (O^m_{1\nu}, O^{m'}_{1\nu'})]$.

\vskip 0.5cm
Finally, for the set of the form $[(O^{m''}_{1+}, O^{m''}_{1-}) , (O^m_{2+}, O^{m'}_{1+})]$ with
\begin{equation}
\fl \left\{
\begin{array}{l}
O^{m''}_{1+}  =  B_{ij} \cos\phi_{ij} + B_{kl} \sin\phi_{kl} \ , \\
O^{m''}_{1-}  = B_{ij} \cos\phi_{ij} - B_{kl} \sin\phi_{kl} \ , 
\end{array} \right. \nonumber \\
\qquad 
\left\{
\begin{array}{l}
O^m_{1+}  = B_{jl} \cos\phi_{jl} + B_{ik} \sin\phi_{ik} \ , \\
O^{m'}_{2+}   =  B_{il} \cos\phi_{il} + B_{kj} \sin\phi_{kj} \ ,
\end{array} \right. 
\label{eq:A20a}
\end{equation}
we have
\begin{eqnarray}
\cos\phi^{\lambda \lambda'}_{jl}(\eta) & = \frac{A_c  O^{m'}_{2+} + \eta\, A^{\lambda \lambda'}_s \sqrt{D^{\lambda \lambda'\, 2}  - O^{m'}_{2+}{^2}}}{D^{\lambda \lambda'\, 2}} \ , \nonumber \\
\sin\phi^{\lambda\lambda'}_{ik}(\eta) & = \frac{O^{m}_{1+} - B_{jl} \sin\phi^{\lambda\lambda'}_{jl}(\eta)}{B_{ik}} \ .
\label{eq:A21a}
\end{eqnarray}
The above results lead to the same form of the solutions as given Eq.~(\ref{eq:A19a}). 
 Here, the symmetries of the corresponding phases are the same to those for the set $[(O^{m''}_{2+}, O^{m''}_{2-}) , (O^m_{1\nu}, O^{m'}_{1\nu'})]$ discussed previously, i.e., 
  $\alpha^{++}_{jl}(\pm) = \alpha^{- -}_{jl}(\mp)$, $\alpha^{+-}_{jl}(\pm) = \alpha^{-+}_{jl}(\mp)$, $\beta^{++}_{ik}(\pm) = \beta^{- -}_{ik}(\mp)$ and $\beta^{+-}_{ik}(\pm) = \beta^{-+}_{ik}(\mp)$.

The results for other sets of the form $[(O^{m''}_{n\nu'''}, O^{m''}_{n'\nu''}), (O^m_{2\nu}, O^{m'}_{1\nu'})]\ (n \ne n'; \nu''', \nu'', \nu', \nu \\=\pm; m \ne m'\ {\rm and}\ \nu,\nu'=\pm)$ can be obtained from the above results for $[(O^{m''}_{1+}, O^{m''}_{1-}), \\(O^m_{2+}, O^{m'}_{1+})]$, following - in complete analogy - what have been done  for obtaining the results for the corresponding sets $[(O^{m''}_{n\nu'''}, O^{m''}_{n'\nu''}) , (O^m_{1\nu}, O^{m'}_{1\nu'})]$.

\vskip 0.5cm
The sets of four spin observables that resolve the phase ambiguity of the Pauli-spin amplitudes up to an arbitrary overall phase are displayed in Table.~\ref{tab:211} for the case of $2(a) + 1(b) + 1(c)$. The other sets can be obtained by appropriate permutations of $a, b$ and $c$.

It should be mentioned that, as has been pointed out in Ref.~\cite{Nak18}, there are some kinematic restrictions 
 on the sets of carefully chosen four observables to be able to resolve the phase ambiguity of the Pauli-spin amplitudes - as indicated in Tables.~\ref{tab:ab},\ref{tab:ac},\ref{tab:bc} and \ref{tab:211}. These restrictions wouldn't be much of a concern if their violations were rare events. Unfortunately, we have no reason \textit{a priori} to expect the violations to be rare.  In the following \ref{app:2}, we provide a way to gauge when such restrictions are violated.

\section{}{\label{app:2}
 
There are two different levels of kinematic restrictions in the relative phase angles;  one is at  the level of a chosen pair of observables and the other is at the level of a set of two pairs of observables. First, we discuss the restrictions at the level of a pair of observables and, then, at the level of a set of two pairs of observables.  

Let's start with the pair of observables of the form $(O^m_{n+}, O^m_{n-})$. From Eq.~(\ref{eq:A5}), we see that the 4-fold ambiguities become degenerated when
$(\beta_{ij} \pm \alpha_{kl}) = \pm \pi/2$ or $\alpha_{kl}= \pi/2$.  These translate to the conditions
\begin{equation}
\frac{O^m_{n-}}{O^m_{n+}} = \frac{B_{kl} - B_{ij}}{B_{kl} + B_{ij}} \qquad {\rm or} \qquad  (O^m_{n+} - O^m_{n-}) = 2 B_{kl} \ .
\label{eq:B1}
\end{equation}
Thus, by measuring the cross section, $d\sigma/d\Omega$, and the spin observables $\Sigma$, $T_x$ and $L_z$ - which determine the magnitudes of the Pauli-spin amplitudes (cf.~Eqs.~(\ref{eq:aux0}, \ref{eq:comb0a})) - together with the observables $O^m_{n+}$ and $O^m_{n-}$, we will be able to gauge the restriction condition.

Similarly, from Eqs.~(\ref{eq:A12},\ref{eq:A12a}), for the observables of the form $(O^m_{n\nu}, O^m_{n'\nu'})\ (n,n'=1,2; \nu, \nu' = \pm)$, the degeneracies in the ambiguities occur when $(\beta_{ij} \pm \alpha_{kl}) = \pm \pi/2$. In terms of the observables, with the help of Eq.~(\ref{eq:2-8b}),  this leads to
\begin{equation}
N^m_{n\nu, n'\nu'} = B_{ij} + B_{kl} \ ,
\label{eq:B2}
\end{equation}
where the indices on which $N$ depends are restored. 

For the observables of the form $(O^m_{1\nu}, O^{m'}_{1\nu'})\ (m \ne m'\ {\rm and}\ \nu,\nu'=\pm)$, the degeneracy in the 4-fold ambiguity - according to Eq.~(\ref{eq:A19}) -
occurs when $(\beta^{\lambda\lambda'}_{ik}(\eta) \pm \alpha^{\lambda\lambda'}_{jl}(\eta)) = \pm \pi/2$. With the help of Eq.~(\ref{eq:A15}), this translates to
\begin{equation}
\frac{O^{m}_{1\nu}}{B_{jl} + B_{ik}} =
\frac{- A^{\lambda \lambda'}_s O^{m'}_{1\nu'} + \eta\, A_c \sqrt{D^{\lambda \lambda'\, 2}  - O^{m'}_{1\nu'}{^2}}}{D^{\lambda \lambda'\, 2}} \ .
\label{eq:B3}
\end{equation}

For all the other pairs of observables of the form $(O^m_{n\nu}, O^{m'}_{n'\nu'})\ (m \ne m'\ {\rm and}\ \nu,\nu'=\pm)$, the occurrence of the degeneracy in the 4-fold ambiguity can be verified using Eq.~(\ref{eq:B3}), with the left-hand-side of the equation replaced by the corresponding term - given by Eqs.~(\ref{eq:A15a},\ref{eq:A21},\ref{eq:A21a}) - according to the particular pair $(O^m_{n\nu}, O^{m'}_{n\nu'})$ as explained in \ref{app:1}.
Also, $O^m_{1\nu}$ on the right-hand-side of Eq.~(\ref{eq:B3}) should be replaced by the corresponding observable $O^m_{n\nu}$.

\vskip 0.5cm
We now discuss the kinematic restrictions that arise at the level of a set of two pairs of observables. 
As a concrete example, let's take the set of four observables $[(O^a_{1+}, O^a_{2+}) , (O^b_{1+}, O^b_{2-})]$. From Eqs.~(\ref{eq:A12},\ref{eq:A12a}), and using the relation for $(a \leftrightarrow b)$ in Eq.~(\ref{eq:A6}), we have a 4-fold ambiguity (2-fold for each pair of observables) leading to the possible solutions
\begin{eqnarray}
\ \ \ - (\beta_{02} - \alpha_{13}) & = - 2\zeta + (\beta_{03} + \alpha_{12})  \ , \nonumber \\
\ \ \ - (\beta_{02} - \alpha_{13}) & = - 2\zeta - (\beta_{03} + \alpha_{12}) + \pi \ , \nonumber \\
\pi +  (\beta_{02} - \alpha_{13}) & = - 2\zeta + (\beta_{03} + \alpha_{12})  \ , \nonumber \\
\pi +  (\beta_{02} - \alpha_{13}) & = - 2\zeta - (\beta_{03} + \alpha_{12}) + \pi \ .
\label{eq:B4}
\end{eqnarray}
It is then immediate that this set of four observables cannot resolve the phase ambiguity of the Pauli-spin amplitudes when  $\beta_{02}=\alpha_{13}$ \textit{and} $\beta_{03} = - \alpha_{12}$, simultaneously, in which case there are degenerated solutions that prevent resolving the phase ambiguity of the reaction amplitude. 
From Eq.~(\ref{eq:2-8b}), the two equalities in relative phase angles imply that 
\begin{eqnarray}
\left[ \frac{(N^a_{1+, 2+})^2 + B_{02}^2 - B_{13}^2}{2 (N^a_{1+, 2+}) B_{02}} \right]^2 + \left[\frac{(N^a_{1+, 2+})^2 + B_{13}^2 - B_{02}^2}{2 (N^a_{1+, 2+}) B_{13}} \right]^2 = 1  \  , \nonumber \\
\fl {\rm and} \\
\left[ \frac{(N^b_{1+, 2-})^2 + B_{03}^2 - B_{12}^2}{2 (N^b_{1+, 2-}) B_{03}} \right]^2 + \left[\frac{(N^b_{1+, 2-})^2 + B_{12}^2 - B_{03}^2}{2 (N^b_{1+, 2-}) B_{12}} \right]^2 = 1  \  .  \nonumber
\label{eq:B5}
\end{eqnarray}
The above relations provide a means of gauging whether the restriction condition has been violated or not.  
Of course - as mentioned above in connection to the degeneracy of the phase ambiguity at the level of a pair of observables - here, the phase ambiguity cannot be resolved also when $(\beta_{02}-\alpha_{13})= -\pi/2$ or $(\beta_{03}+\alpha_{12})=\pi/2$. These two later conditions translate to (cf.~Eq.~(\ref{eq:B2}))
\begin{equation}
N_{1+, 2+}^a  = B_{02} + B_{13}  \qquad {\rm or} \qquad  N_{1+, 2-}^b = B_{03} + B_{12} \ ,
\label{eq:B6}
\end{equation}
respectively.

The kinematic restrictions for all other sets of four observables can be found in an analogous way to what has been done for the set $[(O^a_{1+}, O^a_{2+}) , (O^b_{1+}, O^b_{2-})]$ above. The results are indicated in Tables.~\ref{tab:ab},\ref{tab:ac},\ref{tab:bc} and \ref{tab:211}. For the sets involving the pair of observables of the form $(O^m_{n\nu}, O^m_{n\nu'})$, the condition $\beta_{ij} = |\alpha_{kl}|$ can be gauged by verifying the relation (from Eq.~(\ref{eq:A4a}))
\begin{equation}
\left[\frac{O^m_{n\nu}+O^m_{n\nu'}}{B_{ij}}\right]^2 + \left[\frac{O^m_{n\nu}-O^m_{n\nu'}}{B_{kl}}\right]^2 = 1 \ .
\label{eq:B7}
\end{equation}

Analogously, for the sets of four observables involving the pair of the form $(O^m_{n\nu}, O^m_{n'\nu'})\ (n \ne n')$, the condition $\beta_{ij}= |\alpha_{kl}|$ leads to  (from Eq.~(\ref{eq:2-8b}))
\begin{equation}
\left[ \frac{N^2 + B_{ij}^2 - B_{kl}^2}{2 N B_{ij}} \right]^2 + \left[\frac{N^2 + B_{kl}^2 - B_{ij}^2}{2 N B_{kl}} \right]^2 = 1  \  .
\label{eq:B8}
\end{equation}

For the pair of observables $(O^m_{1+}, O^{m'}_{1+})$ in the set of four observables $[(O^{m''}_{1+}, O^{m''}_{1-}) , (O^m_{1+}, O^{m'}_{1+})]\ (m'' \ne m, m'\ {\rm and}\  m \ne m')$, the condition $\beta_{ik}= |\alpha_{jl}|$ for a given set $\{\lambda, \lambda', \eta \}$ leads to 
\begin{equation}
 \cos^2\beta_{ik} + \sin^2\alpha_{jl} = 1 \ ,
\label{eq:B9}
\end{equation}
where $\cos\beta_{ik}$ and $\sin\alpha_{jl}$ are given by Eqs.~(\ref{eq:A15}) with $\phi_{ik} = \beta_{ik}$ and $\phi_{jl} = \alpha_{jl}$.
For the pairs of the form $(O^m_{n\nu}, O^{m'}_{n'\nu'})\ (m \ne m'\ {\rm and}\ \nu,\nu'=\pm)$ involved in the sets of four observables in general, the condition   
$\beta_{ik}= |\alpha_{jl}|$ for a given set $\{\lambda, \lambda', \eta \}$ leads to the relation (\ref{eq:B9}), with $\cos\beta_{ik}$ and $\sin\alpha_{jl}$ 
given by the corresponding relation (\ref{eq:A15}, \ref{eq:A15a}, \ref{eq:A21}, \ref{eq:A21a}), according to the particular pair $(O^m_{n\nu}, O^{m'}_{n\nu'})$
as explained in \ref{app:1}.

\vskip 1cm
We now turn our attention to some of the features of the Pauli-spin amplitudes and transversity amplitudes in
resolving the phase ambiguity of the photoproduction  amplitude. 
We note that in Pauli-spin 
representation, all the observables are of the form given by Eq.~(\ref{eq:PauliSpinform}) involving $\cos\phi_{ij}$ and $\sin\phi_{kl}$,
while in transversity representation, the observables are of the forms \cite{ChT97,Nak18}
\begin{equation}
\fl O^m_{1\pm}  = B_{ij} \sin\phi_{ij} \pm B_{kl} \sin\phi_{kl}  \quad {\rm or} \quad 
O^m_{2\pm}  = B_{ij} \cos\phi_{ij} \pm B_{kl} \cos\phi_{kl} \ ,
\label{eq:Transform} 
\end{equation}
involving either $\sin\phi_{ij}$ and $\sin\phi_{kl}$ or $\cos\phi_{ij}$ and $\cos\phi_{kl}$.
In helicity representation, the observables are also of the form given by Eq.~(\ref{eq:Transform}) above \cite{ChT97}. 

The above mentioned difference in expressing the observables in different representations leads 
to different kinematic constraints for the possible solutions to resolve the phase ambiguities of the corresponding amplitudes. 
For example, consider the set of four observables $[(O^a_{1+}, O^a_{2+}) , (O^b_{1+}, O^b_{2-})]$ we have discussed above.
This set of observables leads to the possible solutions as given by Eq.~(\ref{eq:B4}) which, as has been pointed out, cannot 
resolve the phase ambiguity of the Pauli-spin amplitudes when  $\beta_{02}=\alpha_{13}$ \textit{and} $\beta_{03} = - \alpha_{12}$, simultaneously.
Note that, due to the distinct domains of the angles $\beta_{ij}$'s and $\alpha_{kl}$'s, the first equality can happen only when $\beta_{02}$ and $\alpha_{13}$ are both on the first quadrant, while the second equality happens only if $\beta_{03}$ is in the first quadrant and $\alpha_{12}$ in the fourth quadrant. 
In contrast, the analog for transversity amplitudes yields (from Eqs.~(37,38) of \cite{Nak18})
\begin{eqnarray}
- (\alpha_{14} - \alpha_{23}) &= -2\zeta + [(\alpha_{13} + \alpha_{24}) - \pi]   \ , \nonumber \\
- (\alpha_{14} - \alpha_{23}) &= -2\zeta - [(\alpha_{13} + \alpha_{24}) - \pi]  \ , \nonumber \\
\ \ \,  (\alpha_{14} - \alpha_{23}) &= - 2\zeta + [(\alpha_{13} + \alpha_{24}) - \pi] \ , \nonumber \\
\ \ \,  (\alpha_{14} - \alpha_{23}) &= -2\zeta - [(\alpha_{13} + \alpha_{24}) - \pi] \ , 
\label{eq:A24}
\end{eqnarray}
from which we see that the phase ambiguity of the transversity amplitudes cannot be resolved if 
$\alpha_{14} = \alpha_{23}$ \textit{or} $\alpha_{13} = - \alpha_{24}$. Note that in contrast to the 
amplitude in Pauli-spin representation, here, the domains of all $\alpha_{ij}$'s are the same. Thus,
the equality can happen if $\alpha_{14}$ and $\alpha_{23}$ are either in the first or fourth quadrant.
Analogously, $\alpha_{13}$ in the first(fourth) and $\alpha_{24}$ in the fourth(first) quadrant.      

Similar situations as presented in the above example occur with other sets of four observables. 

From the above considerations, we conclude that the kinematic restrictions to be satisfied to resolve the phase ambiguity of the photoproduction amplitude is more severe in the transversity representation than in the Pauli-spin representation,  in general. This kind of differences is a direct consequence of the fact that the observables are bilinear combinations of the basic amplitudes that constitute the reaction amplitude.

\section{}\label{app:3}

In the literature \cite{GDH06}, one does introduce the cross section difference of the parallel and anti-parallel helicity states of the initial state particles, i.e., beam photon and target nucleon. Explicitly,
\begin{equation}
\frac{d\sigma_{31}}{d\Omega}  \equiv \frac{d\sigma_{3/2}}{d\Omega} -  \frac{d\sigma_{1/2}}{d\Omega} \ , 
\label{eq:D32}
\end{equation}
where  $\sigma_{3/2}$  and $\sigma_{1/2}$ stand for the cross sections with the parallel ($\lambda_{N_i} - \lambda_\gamma = \pm 3/2$) and the anti-parallel  ($\lambda_{N_i} - \lambda_\gamma = \pm 1/2$) initial state helicities, respectively. 

Introducing now the helicity matrix elements (in the uncoupled basis) of the reaction amplitude $\hat{M}$ (cf. Eq.~(\ref{Jampl-NL}))
\begin{equation}
 H_{\lambda_f, \lambda_i} \equiv  \langle{\frac12 \lambda_f} \big|  \hat{M} \big| {\frac12 \lambda_{N_i}, 1-\lambda_\gamma} \rangle \equiv  \frac{1}{4\pi} \sum_J (2J+1) d^{J\, *}_{\lambda_i\lambda_f}(\theta) \langle{\lambda_f} \big| M^J \big| {\lambda_i} \rangle \ , 
 \label{eq:Hampl}
\end{equation}
with $\lambda_f \equiv \lambda_{N_f}$  and $\lambda_i \equiv \lambda_{N_i} - \lambda_\gamma$,  we can express the observable $T^c_z \equiv E$ defined in 
Eq.~(10$f$) directly in terms of these helicity matrix elements by evaluating the trace over the photon, target nucleon and recoil nucleon helicities. This gives
 \begin{eqnarray}
  \frac{d\sigma}{d\Omega} E  = \frac14 \sum_{\lambda's} H_{\lambda_{N'_f},\lambda_{N'_i}-\lambda_\gamma'} (\sigma^\gamma_z)_{\lambda'_\gamma\lambda_\gamma} (-\sigma_z)_{\lambda_{N'_i}\lambda_{N_i}} H^\dagger_{\lambda_{N_i}-\lambda_\gamma,\lambda_{N_f}}  \nonumber \\
\qquad\   =  \frac14 \sum_{\lambda's} H_{\lambda_{N_f},\lambda_{N'_i}-\lambda_\gamma} (\lambda_\gamma \delta_{\lambda'_\gamma \lambda_\gamma} -2\lambda_{N_i} \delta_{\lambda_{N'_i}\lambda_{N_i}}) H^*_{\lambda_{N_f}, \lambda_{N_i}-\lambda_\gamma}  \nonumber \\
\qquad\  = \frac14 \sum_{\lambda's} | H_{\lambda_{N_f},\lambda_{N_i}-\lambda_\gamma} |^2  (-2\lambda_{N_i}) \lambda_\gamma \nonumber \\
\qquad\   = \frac12 \left[|H_{\frac12, \frac32} |^2 + |H_{\frac12, -\frac32} |^2 - |H_{\frac12, \frac12} |^2 - |H_{\frac12, -\frac12} |^2\right] \ , 
\label{eq:Ehel}
\end{eqnarray}
 where we have used the fact that $(\sigma^\gamma_z = \hat{P}^+ - \hat{P}^-) $ for a real photon (cf.~Sec.~\ref{sec:SpinObservLinear}, below Eq.~(\ref{LS-M})).  The minus sign appearing in $(-\sigma_z)_{\lambda_{N'_i}\lambda_{N_i}}$ is  due to our definition of $E$, where $\hat{z} = \hat{k}$, while for the nucleon helicity, $\hat{z} = - \hat{k}$. 
  Note that, in terms of the helicity amplitudes, $d\sigma/d\Omega = \frac12 \left[|H_{\frac12, \frac32} |^2 + |H_{\frac12, -\frac32} |^2 + |H_{\frac12, \frac12} |^2 + |H_{\frac12, -\frac12} |^2\right]$.

\vskip 0.3cm
Comparing Eqs.~(\ref{eq:D32},\ref{eq:Ehel}) and remembering that the proper cross section is given by Eq.(\ref{xsc1}), we obtain
\begin{equation}
\frac{d\sigma_o}{d\Omega}E = \frac12 \frac{d\sigma_{31}}{d\Omega} \ .
\label{eq:EDs31}
\end{equation}
where the factor 1/2 is due to the fact that $d\sigma_o/d\Omega$ contains the initial spin averaging factor of 1/4, while 
$d\sigma_{3/2}/d\Omega$ and $d\sigma_{1/2}/d\Omega$ contain the spin averaging factor of 1/2. 

\vskip 1cm
\fl \qquad\qquad

\end{document}